\pdfoutput=1

\documentclass[11pt,twoside,a4paper,cmspaper,final,collab]{cms-tdr}
\def\svnVersion{491559P}\def\cmsCernNoTag{CERN-EP-2018-307}\def\cmsCernDate{\today}\def\cmsMessage{Published in the Journal of High Energy Physics as \href{http://dx.doi.org/10.1007/JHEP03(2019)031}{\doi{10.1007/JHEP03(2019)031.}}}

\begin{document}\cmsNoteHeader{SUS-16-017}

\hyphenation{had-ron-i-za-tion}
\hyphenation{cal-or-i-me-ter}
\hyphenation{de-vices}
\RCS$Revision: 488909 $
\RCS$HeadURL: svn+ssh://svn.cern.ch/reps/tdr2/papers/SUS-16-017/trunk/SUS-16-017.tex $
\RCS$Id: SUS-16-017.tex 488909 2019-02-12 21:38:13Z alverson $

\newlength\cmsFigWidth
\ifthenelse{\boolean{cms@external}}{\setlength\cmsFigWidth{0.85\columnwidth}}{\setlength\cmsFigWidth{0.4\textwidth}}
\ifthenelse{\boolean{cms@external}}{\providecommand{\cmsLeft}{top\xspace}}{\providecommand{\cmsLeft}{left\xspace}}
\ifthenelse{\boolean{cms@external}}{\providecommand{\cmsRight}{bottom\xspace}}{\providecommand{\cmsRight}{right\xspace}}
\ifthenelse{\boolean{cms@external}}{\providecommand{\CL}{C.L.\xspace}}{\providecommand{\CL}{CL\xspace}}
\ifthenelse{\boolean{cms@external}}{\providecommand{\cmsUpperLeft}{top}}{\providecommand{\cmsUpperLeft}{upper left}}
\ifthenelse{\boolean{cms@external}}{\providecommand{\cmsUpperRight}{middle}}{\providecommand{\cmsUpperRight}{upper right}}

\newcommand{\MR}{\ensuremath{M_\mathrm{R}}\xspace}
\newcommand{\dPhiR}{\ensuremath{\Delta\phi_\mathrm{R}}\xspace}
\newcommand{\Rtwo}{\ensuremath{\mathrm{R}^2}\xspace}
\newcommand{\R}{\ensuremath{\mathrm{R}}\xspace}
\newcommand{\MRT}{\ensuremath{M_\mathrm{T}^\mathrm{R}}\xspace}
\providecommand{\cPV}{\ensuremath{\cmsSymbolFace{V}}\xspace}
\newlength\cmsTabSkip\setlength{\cmsTabSkip}{1ex}

\cmsNoteHeader{SUS-16-017}
\title{Inclusive search for supersymmetry in pp collisions at \texorpdfstring{$\sqrt{s}=13\TeV$}{sqrt(s) = 13 TeV} using razor variables and boosted object identification in zero and one lepton final states }

\date{\today}

\abstract{
An inclusive search for supersymmetry (SUSY) using the razor variables is performed using a data sample of proton-proton collisions corresponding to an integrated luminosity of $35.9\fbinv$, collected with the CMS experiment in 2016 at a center-of-mass energy of $\sqrt{s}=13\TeV$. The search looks for an excess of events with large transverse energy, large jet multiplicity, and large missing transverse momentum. The razor kinematic variables are sensitive to large mass differences between the parent particle and the invisible particles of a decay chain and help to identify the presence of SUSY particles. The search covers final states with zero or one charged lepton and features event categories divided according to the presence of a high transverse momentum hadronically decaying \PW~boson or top quark, the number of jets, the number of \PQb-tagged jets, and the values of the razor kinematic variables, in order to separate signal from background for a broad range of SUSY signatures. The addition of the boosted $\PW$~boson and top quark categories within the analysis further increases the sensitivity of the search, particularly to signal models with large mass splitting between the produced gluino or squark and the lightest SUSY particle. The analysis is interpreted using simplified models of \textit{R}-parity conserving SUSY, focusing on gluino pair production and top squark pair production. Limits on the gluino mass extend to $2.0\TeV$, while limits on top squark mass reach $1.14\TeV$.
}

\hypersetup{
pdfauthor={CMS Collaboration},
pdftitle={Inclusive search for supersymmetry in pp collisions at sqrt(s) = 13 TeV using razor variables in zero or one lepton final states with boosted object identification},
pdfsubject={CMS},
pdfkeywords={CMS, physics,  supersymmetry}}

\maketitle
\section{Introduction}
\label{sec:intro}

We present an inclusive search for supersymmetry (SUSY) using the razor
variables~\cite{Rogan:2010kb,razor2015,razorboost}
on data collected by the CMS experiment in 2016.
Supersymmetry extends space-time symmetry such that every fermion (boson) in the
standard model (SM) has a bosonic (fer\-mi\-onic) partner~\cite{Wess,Golfand,Volkov,Chamseddine,Kane,Fayet,Barbieri,Hall,Ramond}.
Supersymmetric extensions of the SM yield solutions to the gauge hierarchy
problem without the need for large fine tuning of fundamental
parameters~\cite{Witten:1981nf,Dimopoulos:1981zb,Dine:1981za,Dimopoulos:1981au,Sakai:1981gr,Kaul:1981hi},
exhibit gauge coupling unification~\cite{Dimopoulos:1981yj,Marciano:1981un,Einhorn:1981sx,Ibanez:1981yh,Amaldi:1991cn,Langacker:1995fk},
and can provide weakly interacting particle candidates for dark matter~\cite{Ellis:1983ew,Jungman:1995df}.

The search described in this paper is an extension of previous work presented in Refs.~\cite{razor2015,razorboost}.
The search is inclusive in scope, covering final states with zero or one charged lepton. To enhance sensitivity to specific types
of SUSY signatures, the events are categorized according to the presence of
jets consistent with high transverse momentum ($\pt$) hadronically decaying \PW~bosons or top quarks,
the number of identified charged leptons, the number of jets, and the number of \PQb-tagged jets.
The search is performed in bins of the razor variables $\MR$ and $\Rtwo$~\cite{Rogan:2010kb,razor2015,razorboost}.
The result presented in this paper is the first search for SUSY from the CMS experiment that
incorporates both Lorentz-boosted and ``non-boosted'' (resolved) event categories.
This search strategy provides broad sensitivity to gluino and squark
pair production in \textit{R}-parity~\cite{Farrar:1978xj} conserving scenarios for a large variety of decay modes
and branching fractions. The prediction of the SM background in the search
regions (SRs) is obtained from Monte Carlo (MC) simulation calibrated with data control regions (CRs) that isolate
the major background components. Additional validation of the assumptions made by
the background estimation method yields estimates of the systematic uncertainties.

Other searches for SUSY by the
CMS~\cite{Sirunyan:2018vjp,Sirunyan:2017xse,Sirunyan:2017kqq,Sirunyan:2017fsj,Sirunyan:2017cwe,Sirunyan:2017uyt,Khachatryan:2017rhw,Sirunyan:2017pjw,Sirunyan:2016jpr}
and ATLAS
~\cite{Aaboud:2017ayj,Aaboud:2017wqg,Aaboud:2017bac,Aaboud:2017nfd,Aaboud:2017hdf,Aaboud:2017dmy,Aaboud:2017faq} Collaborations
have been performed using similar data sets and yield complementary sensitivity.
Compared to those searches, the razor kinematic variables explore alternative
signal-sensitive phase space and add robustness to the understanding of
the background composition and the potential systematic uncertainties in
the background models. To give a characteristic example, for squark
pair production with a squark mass of $1000\GeV$
and a neutralino mass of $100\GeV$, we find that the overlap of signal events falling in the most
sensitive tail regions of the razor kinematic variables and of other kinematic variables used in
alternative searches described in Ref.~\cite{Sirunyan:2017cwe} is $50$--$70\%$.

\begin{figure}[!tbp] \centering
\includegraphics[width=0.32\textwidth]{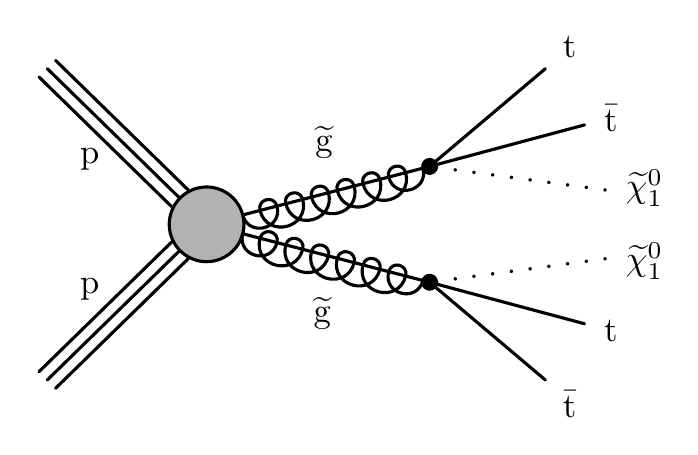}
\includegraphics[width=0.32\textwidth]{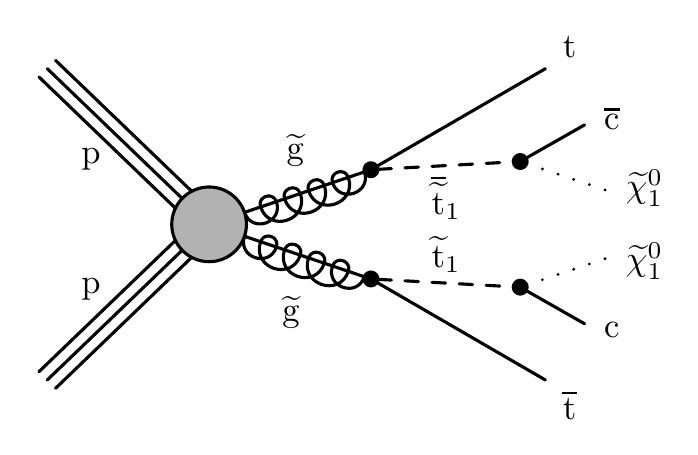}
\includegraphics[width=0.32\textwidth]{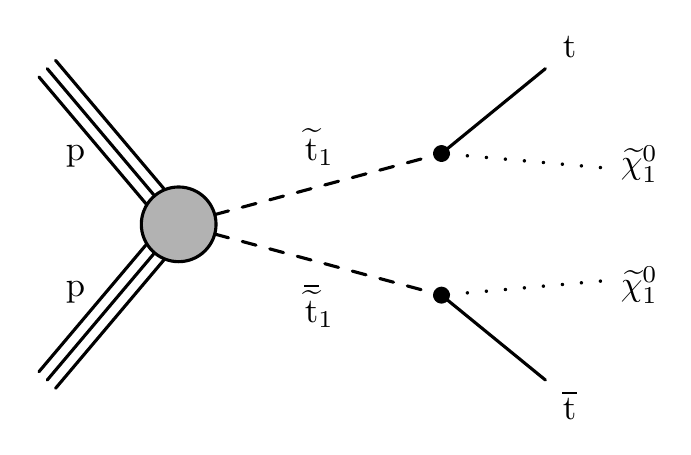}
\caption{ Diagrams for the simplified models considered in this analysis:
(left) pair-produced gluinos, each decaying to two top quarks and the LSP, denoted T1tttt;
(middle) pair-produced gluinos, each decaying to a top quark and a low mass top squark that subsequently
decays to a charm quark and the LSP, denoted T5ttcc;
(right) pair-produced top squarks, each decaying to a top quark and the LSP, denoted T2tt.
In the diagrams, the gluino is denoted by \PSg, the top squark is denoted by \PSQt,
and the lightest neutralino is denoted by \PSGczDo and is the LSP.
}
\label{fig:FeynmanDiagrams}
\end{figure}

We present interpretations of the results in terms of production cross section limits
for several simplified models~\cite{bib-sms-1,bib-sms-2,bib-sms-3,bib-sms-4}
for which this search has enhanced sensitivity. The simplified
models considered include gluino pair production, with each gluino decaying to a pair of top quarks
and the lightest SUSY particle (LSP), referred to as ``T1tttt'';
gluino pair-production, with each gluino decaying to a
top quark and a low-mass top squark that subsequently decays to a charm quark and the LSP,
referred to as ``T5ttcc''; and top squark pair production, with each top squark
decaying to a top quark and the LSP, referred to as ``T2tt''.
The corresponding diagrams for these simplified models are shown in Fig.~\ref{fig:FeynmanDiagrams}.
Although we only interpret the search results in a limited set of simplified models,
the search can be sensitive to other simplified models that are not explicitly considered
in this paper.

This paper is organized as follows.
Details of the detector, trigger, and object reconstruction and identification
are described in Section~\ref{sec:CMSDetector}. The MC simulation samples used
to model background and signal processes are described in Section~\ref{sec:simulation}.
The analysis strategy and event categorization are discussed in Section~\ref{sec:StrategySelection},
and the background modeling is discussed in Section~\ref{sec:Background}.
Systematic uncertainties are discussed in Section~\ref{sec:Systematics},
and finally the results and interpretations are presented in Section~\ref{sec:Results}.
We summarize the paper in Section~\ref{sec:Summary}.

\section{The CMS detector and object reconstruction}
\label{sec:CMSDetector}
The CMS detector consists of a superconducting solenoid of
6\unit{m} internal diameter, providing a magnetic field of
3.8\unit{T}. Within the solenoid volume there are a silicon pixel and a silicon strip tracker, a
lead tungstate crystal electromagnetic calorimeter (ECAL), and a
brass and scintillator hadron calorimeter (HCAL), each composed of a barrel and
two endcap sections. Extensive forward calorimetry complements the coverage
provided by the barrel and endcap detectors. Muons are measured in gas-ionization detectors
embedded in the magnet steel flux-return yoke outside the
solenoid. Events are selected by a two-level trigger system. The first
level is based on a hardware
filter, and the second level, the high level trigger,
is implemented in software.
A more detailed description of the CMS detector, together with a definition
of the coordinate system used and the relevant kinematic variables,
can be found in Ref.~\cite{Adolphi:2008zzk}.

Physics objects are defined using the particle-flow (PF)
algorithm~\cite{Sirunyan:2017ulk}, which aims to
reconstruct and identify each individual particle in an event using
an optimized combination of information from the various
elements of the CMS detector. Jets are clustered from PF candidates
using the anti-$\kt$ algorithm~\cite{Cacciari:2008gp, Cacciari:2011ma}
with a distance parameter of 0.4. Jet energy corrections are derived from simulation and
confirmed by in-situ measurements of the energy balance in dijet, multijet,
photon+jet, and leptonically decaying \cPZ+jet events~\cite{Khachatryan:2016kdb}.
Further details of the performance of the jet reconstruction can
be found in Ref.~\cite{CMS-PAS-JME-16-003}.  Jets used in any selection of
this analysis are required to have $\pt > 30\GeV$ and pseudorapidity $\abs{\eta} < 2.4$.
To identify jets originating from \cPqb~quarks,
we use the ``medium'' working point of the
combined secondary vertex (CSVv2) \cPqb~jet tagger, which uses an inclusive
vertex finder to select \cPqb~jets~\cite{Sirunyan:2017ezt}. The efficiency
to identify a bottom jet is in the range of $50$--$65\%$ for jets with
$\pt$ between $20$ and $400\GeV$, while the misidentification
rate for light-flavor quark and gluon jets (charm jets) is about $1$~$(10)\%$.
We also use the ``loose" working point of the CSVv2 \cPqb~jet tagger to identify \cPqb~jets
to be vetoed in the definition of various CRs. The loose \cPqb~jet tagging working point
has an efficiency of $80\%$ and a misidentification rate for light-flavor and gluon jets
of $10\%$.

Large-radius jets used for identifying Lorentz-boosted \PW~bosons and top quarks are clustered using the
anti-$\kt$ algorithm with a distance parameter of 0.8.  The subset of these jets having
$\abs{\eta} < 2.4$ and $\pt > 200$ (400)\GeV are used to identify \PW~bosons (top quarks).
Identification is done using jet mass, the $N$-subjettiness variables~\cite{Thaler:2010tr},
and subjet \cPqb~tagging for top quarks.  Jet mass is computed using the soft-drop
algorithm~\cite{Larkoski:2014wba}, and is required to be between 65--105 and 105--210\GeV
for $\PW$~bosons and top quarks, respectively.  The $N$-subjettiness variables:
\begin{equation}
\tau_N = \frac{1}{d_0} \sum_k p_{\text{T},k} \min\left( \Delta R_{1,k}, \Delta R_{2,k}, \cdots , \Delta R_{N,k} \right),
\end{equation}
where $N$ denotes candidate axes for subjets, $k$ runs over all constituent particles,
and $d_0 = R_0 \sum_k p_{\text{T},k} $. $R_{0}$ is the clustering parameter of the original jet,
and $\Delta R_{n,k}$ is the distance from constituent particle $k$ to subjet $n$.
The $N$-subjettiness variable is used to evaluate the consistency of a jet with
having $N$ subjets.  To enhance discrimination, the ratios
$\tau_{21} = \tau_2 / \tau_1$ and $\tau_{32} = \tau_3 / \tau_2$ are used for the $\PW$~boson
and top quark tagging, respectively, with the criteria of $\tau_{21} < 0.40$ and $\tau_{32} < 0.65$.
For tagging top quarks (``\cPqt~tagging''), an additional
requirement is imposed on the subjet \cPqb~tagging discriminant based on the multivariate CSVv2
algorithm~\cite{Sirunyan:2017ezt}.
The efficiencies for $\PW$~boson and top quark tagging are on
average 66~and~15\%, respectively, with mistagging rates of 4.0~and~0.1\%~\cite{CMS-PAS-JME-16-003}.

The missing transverse momentum vector \ptvecmiss
is defined as the projection of the negative vector sum of the
momenta of all reconstructed PF candidates on the plane perpendicular to the beams.
Its magnitude is referred to as \ptmiss. Events containing signatures consistent with
beam-induced background or anomalous noise in the calorimeters
sometimes results in events with anomalously large values of
\ptmiss and are rejected using dedicated filters~\cite{Chatrchyan:2011tn,Khachatryan:2014gga}.
The performance of the \ptmiss at CMS may be found in Ref.~\cite{CMS-PAS-JME-17-001}.

Electrons are reconstructed by
associating an energy cluster in the ECAL with a reconstructed
track~\cite{Khachatryan:2015hwa}, and are identified on the basis of
the electromagnetic shower shape, the ratio of energies deposited in the
ECAL and HCAL, the geometric matching of
the track and the calorimeter cluster, the track quality and impact
parameter, and isolation. To improve the efficiency for models
that produce a large number of jets, a so-called ``mini-isolation'' technique is used,
where the isolation cone shrinks as the momentum of the object increases.
Further details are discussed in Ref.~\cite{razor2015}.
Muons are reconstructed by combining
tracks found in the muon system with corresponding tracks in the
silicon tracking detectors~\cite{Sirunyan:2018fpa}, and are identified
based on the quality of the track fit, the number of detector hits used in the
tracking algorithm, the compatibility between track segments, and isolation.
Two types of selections are defined for electrons and muons: a ``tight''
selection with an average efficiency of about $70$--$75\%$,
and a ``loose'' selection with an efficiency of about $90$--$95\%$.
The loose selections are required to have $\pt > 5\GeV$, while the tight selections are required
to have $\pt > 30$ and $25\GeV$ for electrons and muons, respectively.
Similarly electrons (muons) are required to have $\abs{\eta} < 2.5$ ($2.4$), and
electrons with $\abs{\eta}$ (of $1.442$--$1.556$) in the transition region between the barrel and endcap
ECAL are not considered because of limited electron reconstruction capabilities in that region.

{\tolerance=800
Hadronically decaying \PGt~leptons ($\tauh$) are reconstructed using the
hadron-plus-strips algorithm~\cite{Khachatryan:2015dfa}, which identifies \PGt~lepton decay modes
with one charged hadron and up to two neutral pions or three charged hadrons, and
are required to be isolated. The ``loose'' selection used successfully reconstructs
\tauh decays with an efficiency of about $50\%$.
The reconstructed \tauh leptons have $\pt > 20\GeV$
and $\abs{\eta} < 2.4$.
\par}

Finally, photon candidates are reconstructed
from energy clusters in the ECAL~\cite{Khachatryan:2015iwa} and identified
based on the transverse shower width, the hadronic to electromagnetic
energy ratio in the HCAL and ECAL, and isolation. Photon candidates that share
the same energy cluster as an identified electron are vetoed.  Photons are used in the estimation of $\cPZ\to\nu\nu$+jets backgrounds, and are required to have $\abs{\eta} < 2.5$ and $\pt > 185$ or $80\GeV$ for the non-boosted or boosted categories, respectively.

\section{Simulation}
\label{sec:simulation}

{\tolerance=800
Monte Carlo simulated samples are used to predict the
SM backgrounds in the SRs and to calculate the selection efficiencies
for SUSY signal models. Events corresponding to the
$\cPZ$+jets, $\cPgg$+jets, and quantum chromodynamics (QCD)
multijet background processes, as well as the
SUSY signal processes, are generated at leading order with
\MGvATNLO~2.2.2~\cite{Alwall:2011uj,Alwall:2014hca} interfaced with
\PYTHIA V8.205~\cite{Sjostrand:2014zea} for fragmentation and parton
showering, and matched to the matrix element kinematic configuration using the MLM
algorithm~\cite{Hoche:2006ph, Alwall:2007fs}. The CUETP8M1 \PYTHIA8 tune~\cite{Khachatryan:2015pea} was used.
Other background processes are generated at next-to-leading order (NLO) with
\MGvATNLO~2.2.2~\cite{Alwall:2014hca} ($\PW$+jets, $s$-channel single top quark, $\ttbar\PW$, $\ttbar\cPZ$ processes) or
with \POWHEG v2.0~\cite{Frixione:2007nw, Alioli:2009je, Re:2010bp} ($\ttbar$+jets,
$t$-channel single top quark, and $\cPqt\PW$ production), both interfaced with \PYTHIA V8.205.
Simulated samples generated at LO (NLO) used the
NNPDF3.0LO (NNPDF3.0NLO)~\cite{Ball:2014uwa} parton distribution
functions. The SM background events are simulated using a \GEANTfour-based model~\cite{geant4}
of the CMS detector, while SUSY signal events are simulated using the CMS fast
simulation package~\cite{FastSim}. All simulated events include the effects of pileup,
multiple $\Pp\Pp$ collisions within the same or neighboring bunch crossings.
\par}

The SUSY particle production cross sections are calculated to NLO plus next-to-leading-log (NLL)
precision~\cite{NLONLL1,NLONLL2,NLONLL3,NLONLL4,NLONLL5,Borschensky:2014cia} with all
other sparticles assumed to be heavy and decoupled. The NLO+NLL cross sections and
their associated uncertainties from Ref.~\cite{Borschensky:2014cia} are taken as a
reference to derive the exclusion limit on the SUSY particle masses.

To improve on the \MGvATNLO modeling of the multiplicity of additional
jets from initial-state radiation (ISR), strongly produced SUSY signal samples
are reweighted as a function of the number of ISR jets ($N_{\mathrm{jets}}^{\mathrm{ISR}}$).
This correction is derived from a \ttbar enriched control sample such that
the jet multiplicity from the \MGvATNLO-generated \ttbar sample agrees with data.
The reweighting factors vary between 0.92 and 0.51 for $N_{\mathrm{jets}}^{\mathrm{ISR}}$ between one and six.
We take one half of the deviation from unity as the systematic uncertainty in these
reweighting factors.

\section{Analysis strategy and event categorization }
\label{sec:StrategySelection}

We perform the search in several event categories defined according to the presence of
jets tagged as originating from a boosted hadronic $\PW$~boson or top quark,
the number of identified charged leptons, jets,
and \cPqb-tagged jets. A summary of the categories used is shown in Table~\ref{tab:categoryRequirements} below.

\begin{table}[!tbp]
\centering
\topcaption{Summary of the search categories, their charged lepton and jet count requirements, and
the \cPqb~tag bins that define the subcategories.
Events passing the ``Lepton veto'' requirement must have no electron or muon passing
the loose selection, and no $\tauh$ candidate. }
\label{tab:categoryRequirements}
\begin{tabular}{cccc} \hline
Category & Lepton requirement   & Jet requirement & \cPqb~tag bins \\
\hline
Lepton multijet & 1 ``Tight'' electron or muon   &4--6 jets & 0, 1, 2, $\ge$3 \cPqb~tags   \\[\cmsTabSkip]

Lepton seven-jet  & 1 ``Tight'' electron or muon  & $\geq$7 jets & 0, 1, 2, $\ge$3 \cPqb~tags \\[\cmsTabSkip]

\multirow{2}{*}{Boosted $\PW$ 4--5 jet}  & \multirow{2}{*}{Lepton veto}    & $\ge$1 \PW-tagged jet & \multirow{2}{*}{$\ge$1 \cPqb~tags}  \\
&  &  4--5 jets &   \\[\cmsTabSkip]

\multirow{2}{*}{Boosted $\PW$ 6 jet}  & \multirow{2}{*}{Lepton veto}    & $\ge$1 \PW-tagged jet  & \multirow{2}{*}{$\ge$1 \cPqb~tags}  \\
&  &  $\geq$6 jets &   \\[\cmsTabSkip]

\multirow{3}{*}{Boosted top}  & \multirow{3}{*}{Lepton veto}    & 0 \PW-tagged jets  & \multirow{3}{*}{$\ge$0 \cPqb~tags} \\
&  &  $\ge$1 \cPqt-tagged jet &   \\
&  &  $\geq$6 jets &   \\[\cmsTabSkip]

\multirow{3}{*}{Dijet}  & \multirow{3}{*}{Lepton veto}    & 0 \PW-tagged jets & \multirow{3}{*}{0, 1, $\ge$2 \cPqb~tags}\\
&  &  0 \cPqt-tagged jets &   \\
&  &  2--3 jets &   \\[\cmsTabSkip]

\multirow{3}{*}{Multijet}  &  \multirow{3}{*}{Lepton veto}    & 0 \PW-tagged jets  & \multirow{3}{*}{0, 1, 2, $\ge$3 \cPqb~tags} \\
&  &  0 \cPqt-tagged jets &   \\
&  &  4--6 jets &   \\[\cmsTabSkip]

\multirow{3}{*}{Seven-jet}  & \multirow{3}{*}{Lepton veto}    & 0\PW-tagged jets  & \multirow{3}{*}{0, 1, 2, $\ge$3 \cPqb~tags} \\
&  &  0 \cPqt-tagged jets &   \\
&  &  $\geq$7 jets &   \\[\cmsTabSkip]
\hline
\end{tabular}
\end{table}

Events in the one-lepton category are required to have one and only one charged lepton (electron or muon),
with $\pt$ above 30\,(25)\GeV for electrons (muons) selected using the tight criteria,
while events in the zero-lepton category are required to have no electrons or muons passing the
loose selection criteria and no $\tauh$ candidates. One-lepton events are placed in the
``Lepton Multijet'' category if they have between 4 and 6 jets, and placed in the
``Lepton Seven-jet'' category if they have 7 or more jets. One-lepton events with fewer than 4
jets are not considered in the analysis.

Zero-lepton events with jets tagged as originating from a boosted hadronic $\PW$~boson or top quark decay are placed
in a dedicated ``boosted'' event category. Events in this ``boosted'' category are analyzed separately
with a set of CRs and validation tests specific for the analysis with boosted objects.
They are further classified into those having at least one tagged $\PW$~boson
and one tagged \cPqb~jet (``$\PW$" category), and those having at least one tagged
top quark (``Top" category).  Events in the $\PW$ category are further divided
into subcategories with 4--5 jets, and 6 jets or more.
Zero-lepton events not tagged as having boosted $\PW$~bosons or top quarks are placed into
the ``Dijet'' category if they have two or three jets, the ``Multijet'' category if they
have between 4 and 6 jets, and into the ``Seven-jet'' category if they have 7 or more
jets.

The Dijet category is further divided into subcategories
with zero, one, and two or more \cPqb-tagged jets, and all other non-boosted categories are divided
into subcategories with zero, one, two, and three or more \cPqb-tagged jets.

For each event in the above categories, we group the selected charged leptons
and jets in the event into two distinct hemispheres called megajets, whose four-momenta are
defined as the vector sum of the four-momenta of the physics objects in each hemisphere. The
clustering algorithm selects the grouping that minimizes the sum of the squared invariant masses
of the two megajets~\cite{razor2010}. We define the razor variables $\MR$ and $\MRT$ as:
\begin{align}
 \label{eq:MRstar}
 \MR &\equiv
 \sqrt{
(\abs{\vec{p}^{\mathrm{j}_{1}}}+\abs{\vec{p}^{\mathrm{j}_{2}}})^2 -({p}^{\mathrm{j}_1}_\mathrm{z}+{p}^{\mathrm{j}_2}_\mathrm{z})^2},\\
\MRT &\equiv \sqrt{ \frac{\ptmiss(\pt^{\mathrm{j}_1}+\pt^{\mathrm{j}_2}) -
\ptvecmiss \cdot
 (\ptvec^{\,\mathrm{j}_1}+\ptvec^{\,\mathrm{j}_2}) }{2}},
\end{align}
where $\vec{p}^{\mathrm{j}_i}$, $\ptvec^{\,\mathrm{j}_i}$, and
$p^{\mathrm{j}_\mathrm{i}}_\mathrm{z}$ are the momentum of the $i$-th megajet, its
transverse component with respect to the beam axis, and its
longitudinal component, respectively.  The dimensionless variable $\R$ is defined as:
\begin{equation}
\R \equiv \frac{\MRT}{\MR}.
\end{equation}
For pair-produced SUSY signals, the variable $\MR$ quantifies the mass splitting between the
pair-produced particle and the LSP, and exhibits a peaking structure, while for background it is distributed
as an exponentially decaying spectrum. The variable $\R$ quantifies the degree of imbalance between
the visible and invisible decay products and helps to suppress backgrounds which do not
produce any weakly interacting particles. The combination of the two variables provide
powerful discrimination between the SUSY signal and SM backgrounds.

Single-electron or single-muon triggers are used to collect
events in the one-lepton categories, with a total trigger efficiency
of about $80\%$ for reconstructed $\pt$ around $30\GeV$, growing
to $95\%$ for reconstructed $\pt$ above $50\GeV$.
Events in the boosted category are collected using
triggers that select events based on the $\pt$
of the leading jet and the scalar $\pt$ sum of all jets, $\HT$. The trigger
efficiency is about $50\%$ at the low range
of the $\MR$ and $\Rtwo$ kinematic variables and grows
to $100\%$ for $\MR>1.2\TeV$ and $\Rtwo>0.16$.
For the zero-lepton non-boosted event categories, dedicated triggers
requiring at least two jets with $\pt > 80\GeV$ and loose thresholds on the
razor variables $\MR$ and $\Rtwo$ are used to collect the events.
The trigger efficiency ranges from $95$--$100\%$ and increases
with $\MR$ and $\Rtwo$.

Preselection requirements
on the $\MR$ and $\Rtwo$ variables are made depending on the event category.
For events in the one-lepton categories, further
requirements are made on the transverse mass \mT defined as follows:
\begin{equation}
\mT = \sqrt{2 \ptmiss \pt^{\ell} [1 - \cos(\Delta\phi)] },\\
\end{equation}
where $\pt^{\ell}$ is the charged-lepton transverse momentum, and $\Delta\phi$
is the azimuthal angle (in radians) between the charged-lepton momentum and the \ptmiss.
For events in the zero-lepton categories, further
requirements are made on the azimuthal angle $\dPhiR$ between
the axes of the two razor megajets. These requirements are
summarized in Table~\ref{tab:categoryBins}.

\begin{table}[!tbp]
\centering
\topcaption{The baseline requirements on the razor variables $\MR$ and $\Rtwo$, additional requirements on
\mT and $\dPhiR$, and the trigger requirements are shown for each event category. }
\label{tab:categoryBins}
\begin{tabular}{cccc}
\hline
\multirow{2}{*}{Category} & \multirow{2}{*}{Preselection} & Additional   & Trigger          \\
                          &                               & requirements & requirement       \\
\hline
Lepton multijet    & $\MR > 550\GeV$ \& $\Rtwo > 0.20$        & $\mT>120\GeV$  & Single lepton \\

Lepton seven-jet    & $\MR > 550\GeV$ \& $\Rtwo > 0.20$       & $\mT>120\GeV$  & Single lepton \\

Boosted $\PW$ 4--5 jet  & $\MR > 800\GeV$ \& $\Rtwo > 0.08$       & $\dPhiR < 2.8$    & $\HT$, jet $\pt$ \\

Boosted $\PW$ 6 jet  & $\MR > 800\GeV$ \& $\Rtwo > 0.08$              & $\dPhiR < 2.8$          & $\HT$, jet $\pt$ \\

Boosted top   & $\MR > 800\GeV$ \& $\Rtwo > 0.08$             & $\dPhiR < 2.8$          & $\HT$, jet $\pt$ \\

Dijet         & $\MR > 650\GeV$ \& $\Rtwo > 0.30$             & $\dPhiR < 2.8$          & Hadronic razor \\

Multijet      & $\MR > 650\GeV$ \& $\Rtwo > 0.30$              & $\dPhiR < 2.8$          & Hadronic razor \\

Seven-jet     & $\MR > 650\GeV$ \& $\Rtwo > 0.30$              & $\dPhiR < 2.8$          & Hadronic razor  \\
\hline
\end{tabular}
\end{table}

Finally, in each event category, the search is performed in bins of the kinematic variables $\MR$ and $\Rtwo$
in order to take advantage of the varying signal-to-background ratio in the different bins. For one-lepton
categories, the SRs are composed of five bins in $\MR$, starting from $550\GeV$, and five bins in $\Rtwo$
starting from $0.20$. For the zero-lepton boosted categories, the SRs are composed of five bins in $\MR$,
starting from $800\GeV$, and five bins in $\Rtwo$, starting from $0.08$. Finally, for the zero-lepton non-boosted
categories, the SRs are composed of five bins in $\MR$, starting from $650\GeV$, and four bins in $\Rtwo$
starting from $0.30$. To match with the expected resolution, the bin widths in $\MR$ increases from $100$ to $300\GeV$
as the value of $\MR$ grows from $400$ to $1200\GeV$. In each category, to limit the impact of
statistical uncertainties due to the limited size of the MC simulation samples, bins are merged such that the expected
background in each bin is larger than about $0.1$ events. As a result, the SRs have a decreasing number
of bins as the number of jets, \cPqb-tagged jets, and $\MR$ increases.

\section{Background modeling}
\label{sec:Background}

The main background processes in the SRs considered are
$\PW(\ell\nu)$+jets (with $\ell=\Pe,\Pgm,\Pgt$), $\cPZ(\nu\overline{\nu})$+jets, $\ttbar$, and QCD multijet production. For event categories with
zero \PQb-tagged jets, the background is primarily composed of the $\PW(\ell\nu)$+jets and $\cPZ(\nu\overline{\nu})$+jets
processes, while for categories with two or more \PQb-tagged jets it is
dominated by the $\ttbar$ process. There are also small contributions
at the level of a few percent from single top quark production,
production of two or three electroweak bosons, and production of $\ttbar$ in
association with a $\PW$ or $\cPZ$~boson.

The background prediction strategy relies on the use of CRs
to isolate each background process, address deficiencies of the MC
simulation using control samples in data, and estimate systematic uncertainties
in the expected event yields. The CRs are defined such that they have no
overlap with any SRs. For the dominant backgrounds discussed above,
the primary sources of mismodeling come from inaccuracy in the MC prediction
of the hadronic recoil spectrum and the jet multiplicity.
Corrections to the MC simulation are applied
first in bins of $\MR$ and $\Rtwo$, and then subsequently in
the number of jets ($N_{\text{jets}}$)
to address these modeling inaccuracies.
The CR bins generally follow the bins of the
SRs described in Section~\ref{sec:StrategySelection},
but bins with limited statistical power are merged in order
to avoid large statistical fluctuations in the
background predictions.

For the boosted categories, the CR selection and categorization
are slightly adapted and the details are discussed further in
Section~\ref{sec:bgboost}. An additional validation of the background
prediction method is also performed for the boosted categories.

In what follows, all background MC samples are corrected for
known mismodeling of the jet energy response,
the trigger efficiency, and the selection efficiency
of electrons, muons, and \PQb-tagged jets.
These corrections are mostly in the range of $0$--$5\%$,
but can be as large as $10\%$ in bins with large $\MR$ and $\Rtwo$,
where the corrections have larger statistical uncertainties.

\subsection{The \texorpdfstring{$\ttbar$ and $\PW(\ell\nu)$+jets backgrounds}{ttbar and W(ell nu)+jets background}}
\label{sec:TTBarWJetsCR}
We predict the $\ttbar$ and $\PW(\ell\nu)$ backgrounds from the MC simulation corrected
for inaccuracies in the modeling of the hadronic recoil. The corrections are derived
in a CR consisting of events having at least one tight electron or muon.
In order to separate the CR from the SRs
and to reduce the QCD multijet background, the $\ptmiss$ is required to be
larger than $30\GeV$, and $\mT$ is required to be between $30$~and~$100\GeV$.

The tight lepton control sample is separated into $\PW(\ell\nu)$+jets-enriched and $\ttbar$-enriched
samples by requiring events to have zero (for $\PW(\ell\nu)$+jets), or one or more
(for $\ttbar$) \PQb-tagged jets, respectively. The purity of the
$\PW(\ell\nu)$+jets and $\ttbar$ dominated CRs are both about $80\%$.
In each sample, corrections to the MC prediction are derived in
two-dimensional bins in $\MR$ and $\Rtwo$.
The contribution from all other background processes estimated from
simulation in each bin in a given CR ($N^{\text{MC,bkg}}_{\text{CR\ bin}\ i}$)
is subtracted from the data yield in the corresponding bin in the
CR ($N^{\text{data}}_{\text{CR\ bin}\ i}$),
and compared to the MC prediction ($N^{\text{MC,\ttbar}}_{\text{CR\ bin}\ i}$)
to derive the correction factor:
\begin{equation}
C_{\text{bin}\ i}^{\ttbar} = \frac{N^{\text{data}}_{\text{CR\ bin}\ i} - N^{\text{MC,bkg}}_{\text{CR\ bin}\ i}}{N^{\text{MC,\ttbar}}_{\text{CR\ bin}\ i}}.
\end{equation}
Finally, the prediction for the \ttbar background in the SR ($N^{\text{\ttbar}}_{\text{SR\ bin}\ i}$) is:
\begin{equation}
N^{\text{\ttbar}}_{\text{SR\ bin}\ i} = N^{\text{MC,\ttbar}}_{\text{SR\ bin}\ i} C_{\text{bin}\ i}^{\ttbar},
\end{equation}
where $N^{\text{MC,\ttbar}}_{\text{SR\ bin}\ i}$ is the prediction for the SR from the MC simulation.

Because the $\ttbar$-enriched sample is the purer of the two, the corrections are first derived
in this sample.  These corrections are applied to the $\ttbar$ simulation in the $\PW(\ell\nu)$+jets-enriched
sample, and then analogous corrections and predictions for the $\PW(\ell\nu)$+jets background process are derived.

The corrections based on $\MR$ and $\Rtwo$ are measured and applied by averaging over all jet multiplicity bins.
As our SRs are divided according to the jet multiplicity, additional
corrections are needed in order to ensure correct background modeling for different numbers of jets.
We derive these corrections separately for the $\ttbar$ and $\PW(\ell\nu)$+jets samples,
obtaining correction factors for events with two or three jets, four to six jets, and seven or more jets.
The $\ttbar$ correction is derived prior to the $\PW(\ell\nu)$+jets correction
to take advantage of the slightly higher purity of the $\ttbar$ CR.

We also check for MC mismodeling that depends on the number of \PQb~jets
in the event.  To do this we apply the above-mentioned corrections
in bins of $\MR$, $\Rtwo$, and the number of jets and derive
an additional correction needed to make the predicted $\MR$ spectrum
match that in data for each \PQb~tag multiplicity. This correction is performed
separately for events with two or three, four to six, and seven or more jets.

A final validation of the MC modeling in this tight lepton CR
is completed by comparing the $\Rtwo$ spectrum in data with the MC prediction
in each jet multiplicity and \PQb~tag multiplicity category.
We do not observe any systematic mismodeling in the $\Rtwo$ spectra, and
we propagate the total uncertainty in the data-to-MC ratio in each bin of $\Rtwo$
as a systematic uncertainty in the $\ttbar$ and $\PW$+jets
backgrounds in the analysis SRs.

\begin{figure}[!tbp] \centering
\includegraphics[width=0.49\textwidth]{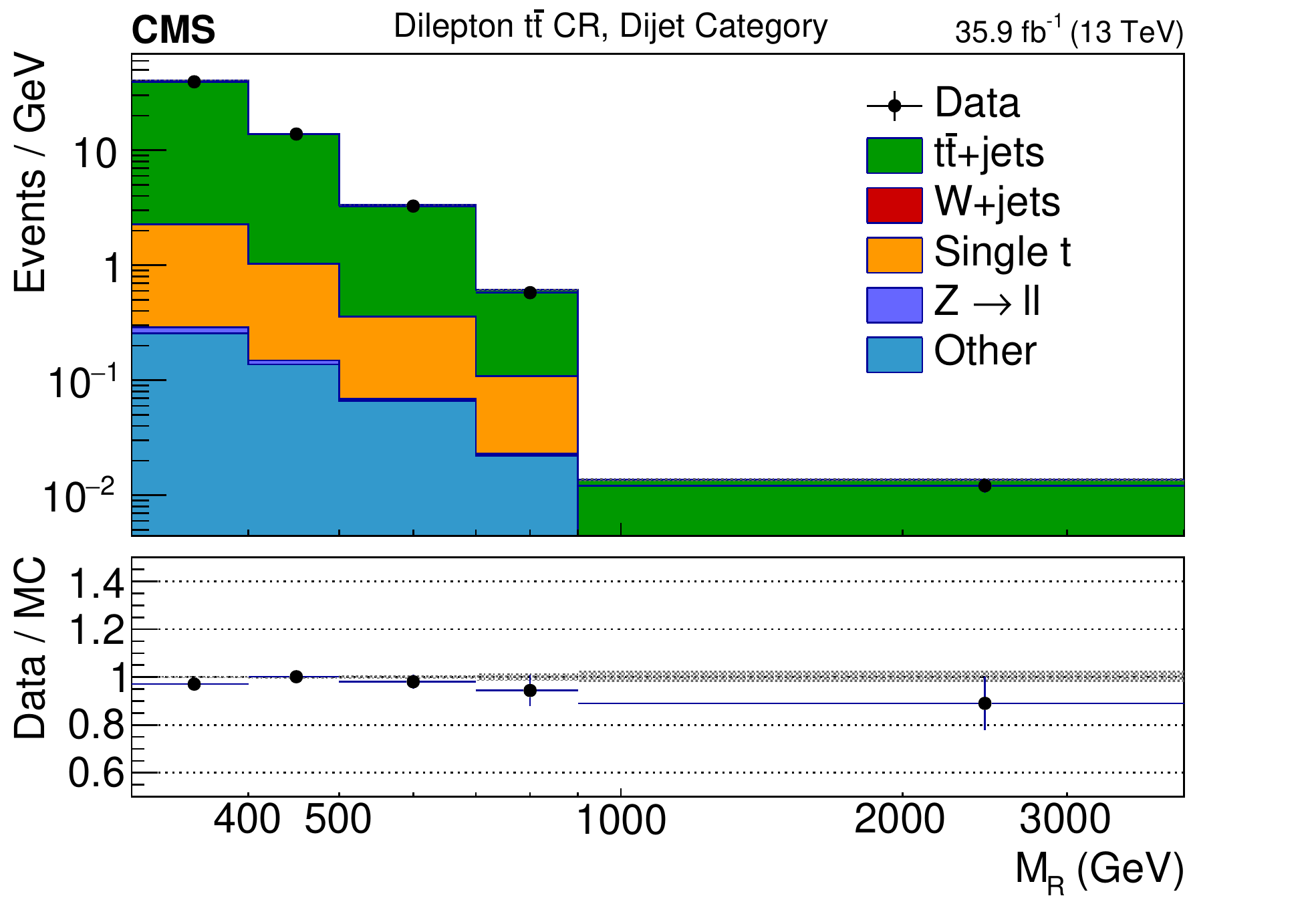}
\includegraphics[width=0.49\textwidth]{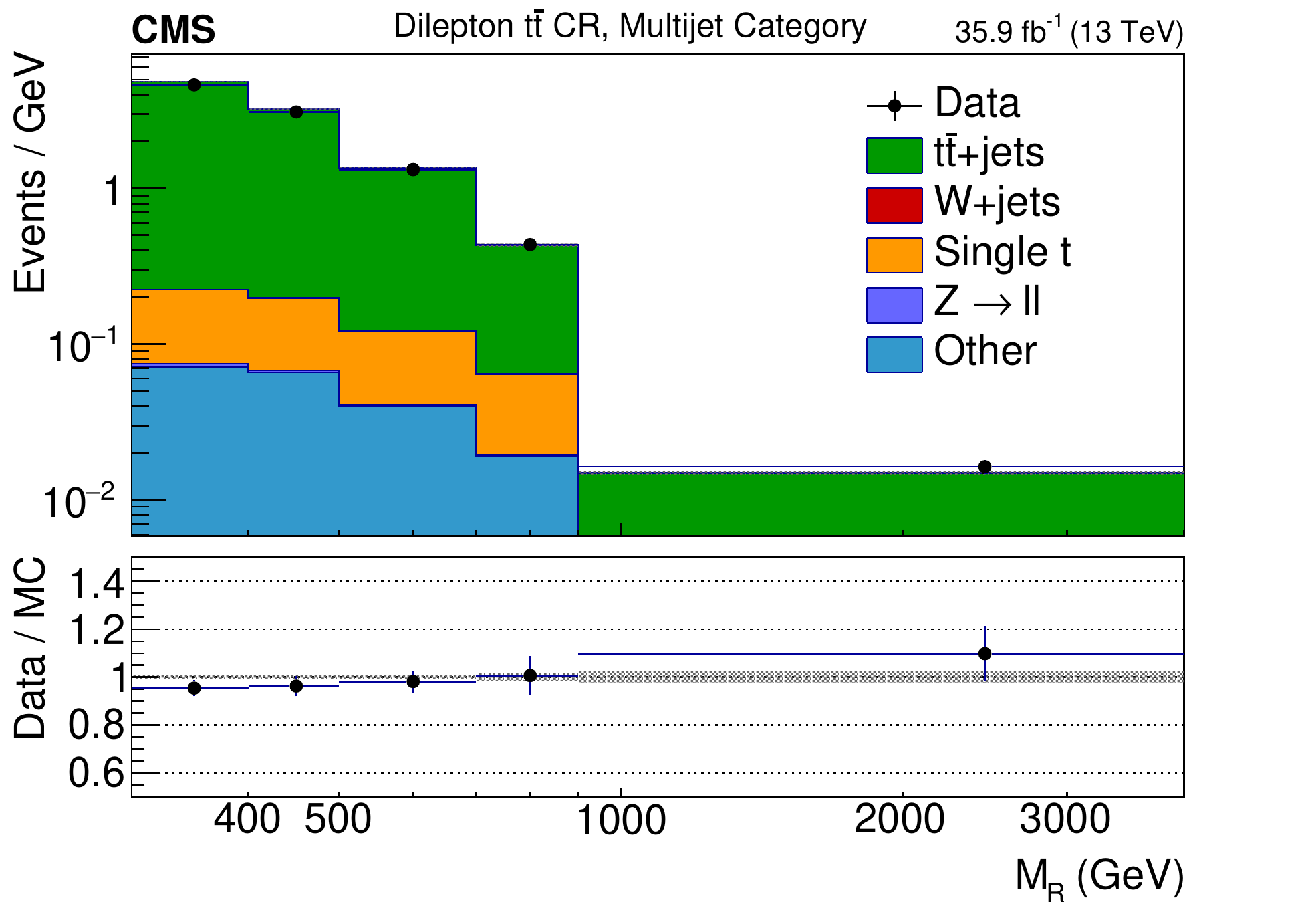} \\
\includegraphics[width=0.49\textwidth]{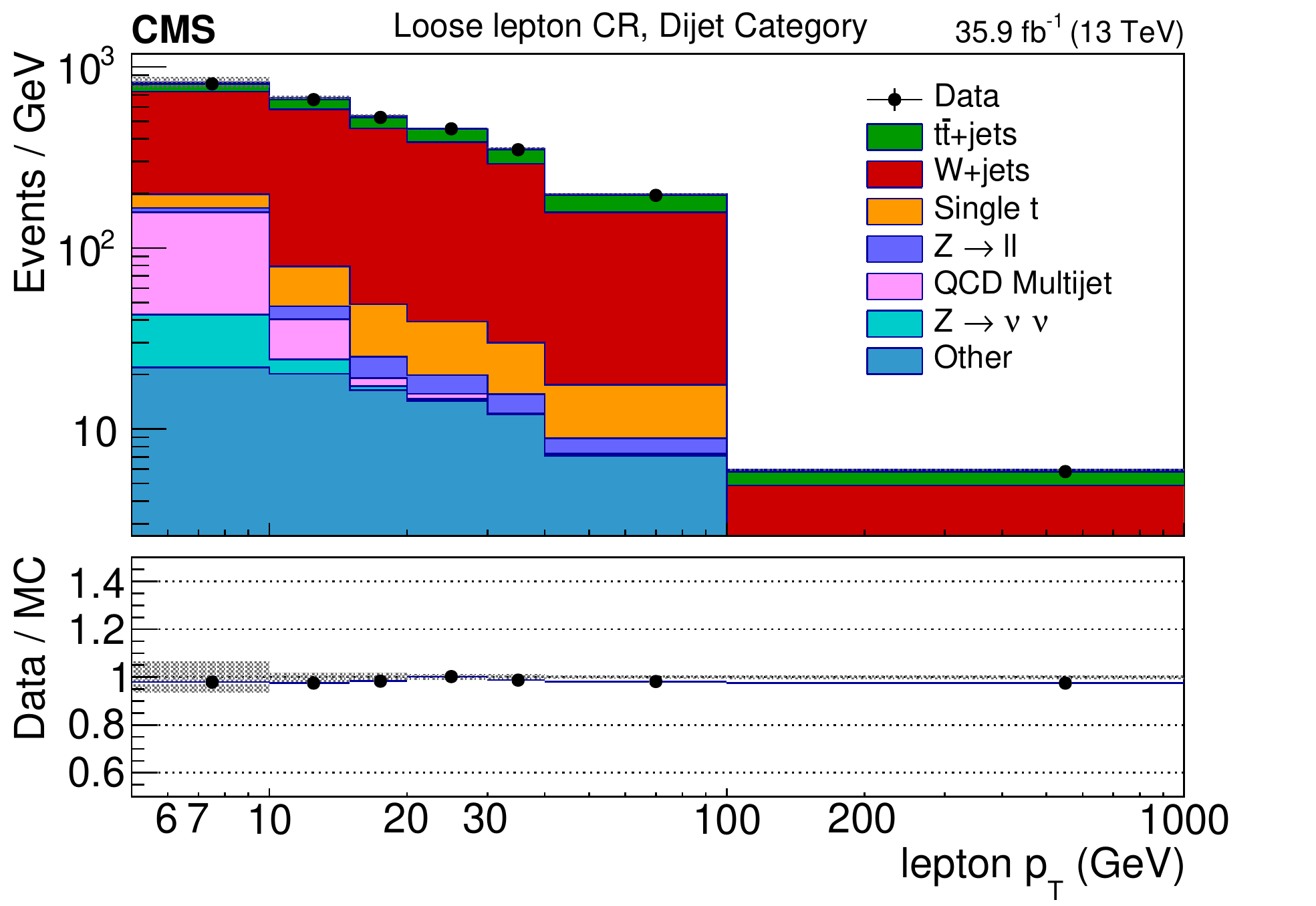}
\includegraphics[width=0.49\textwidth]{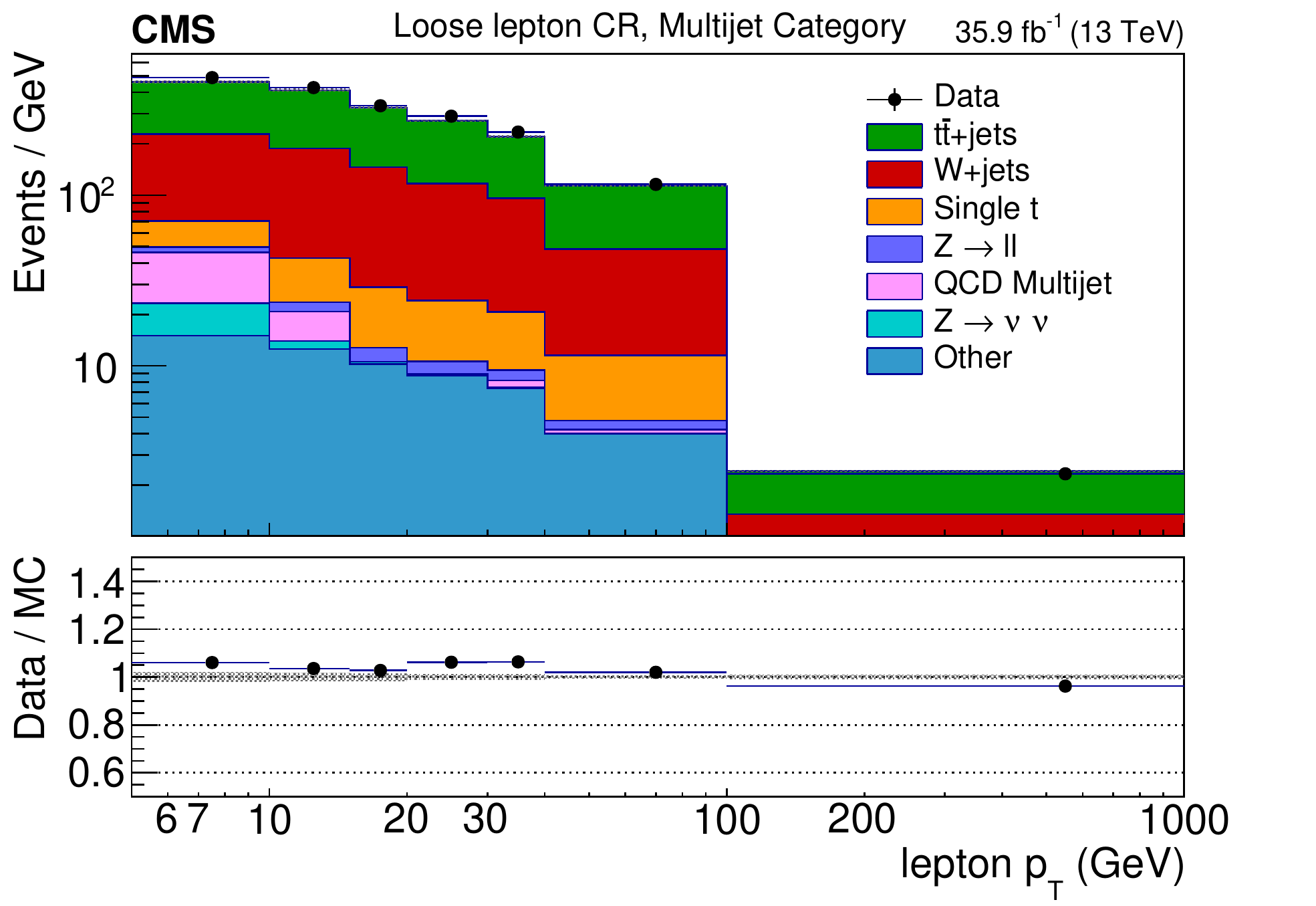}
\caption{ The $\MR$ distribution in the $\ttbar$ dilepton CR (upper row) and lepton $\pt$ distribution
in the loose lepton CR (lower row) are displayed
in the 2--3 (left) and 4--6 (right) jet categories along with the corresponding MC predictions.
The corrections derived from the $\ttbar$ and $\PW$+jets CR have been applied.
The ratio of data to the MC prediction is shown on the bottom panel, with
the statistical uncertainty expressed through the data point error bars and the systematic uncertainty in the
background prediction represented by the shaded region.
}
\label{fig:inclusive_ttbar2L}
\end{figure}

The $\ttbar$ background in the tight lepton CR is composed mostly of lepton+jets
$\ttbar$ events, where one top quark decayed fully hadronically and the other top quark decayed
leptonically.  In the leptonic analysis SRs, the $\mT$
requirement suppresses lepton+jets $\ttbar$ events, and the dominant remaining $\ttbar$ background
consists of $\ttbar$ events where both top quarks decayed leptonically, and one of the two leptons
is not identified.  It is therefore important
to validate that the corrections to the $\ttbar$ simulation derived in the tight lepton CR
also describe dileptonic $\ttbar$ events well.  We perform this check by selecting an event sample
enriched in dileptonic $\ttbar$ events, applying the corrections on the $\ttbar$ simulation prediction
derived in the tight lepton CR, and evaluating the consistency of the
data with the corrected prediction.  This check is performed separately for each jet multiplicity category
used in the analysis SRs.
The dilepton $\ttbar$-enriched sample consists of events with
two tight electrons or muons with $\pt > 30\GeV$ and invariant mass larger than $20\GeV$,
at least one \PQb-tagged jet with $\pt > 40\GeV$, and $\ptmiss > 40\GeV$.  Events with two same-flavor leptons
with invariant mass between $76$~and~$106\GeV$ are rejected to suppress Drell--Yan background.
The $\ptmiss$ and the $\mT$ variables are computed treating one of the leptons in each event
as visible and the other as invisible, and the requirement on the $\mT$ is subsequently applied.
A systematic uncertainty in the dilepton $\ttbar$ background
is assessed by comparing data with the MC prediction in the $\MR$
distribution for each jet multiplicity category. The $\MR$ distributions in the $\ttbar$ dilepton CR
for the two to three and four to six jet event categories are displayed in the upper row of
Fig.~\ref{fig:inclusive_ttbar2L}.

The MC prediction for the hadronic SRs can be affected by potential mismodeling of the
identification efficiency for electrons, muons, and $\tauh$ candidates.
The loose lepton and $\tauh$ CRs are defined
in order to assess the modeling of this efficiency in simulation.  Events in the loose lepton ($\tauh$)
CR are required to have at least one loose electron or muon ($\tauh$ candidate) and pass one of the
hadronic razor triggers.  These events must also have
$\mT$ between $30$~and~$100\GeV$, $\MR > 400\GeV$, $\Rtwo > 0.25$,
and at least two jets with $\pt>80\GeV$.  The data and MC prediction are compared in bins of
lepton $\pt$ and $\eta$ for each jet multiplicity category.
A systematic uncertainty of about $25\%$ is assigned to cover the difference between data and prediction
in the lepton $\pt$ spectrum. No further systematic mismodeling is observed
in the lepton $\eta$ distributions, and the size of the uncertainty
in each $\eta$ bin is propagated as an uncertainty in the analysis SR predictions.
The lepton $\pt$ distributions obtained in the loose lepton CR
for the categories with two to three and four to six jets are
displayed in the lower row of Fig.~\ref{fig:inclusive_ttbar2L}.

\subsection{The \texorpdfstring{$\cPZ\to\nu\overline{\nu}$}{Z to nu nubar} background}
\label{sec:ZInvCR}
The background prediction for the $\cPZ(\nu\overline{\nu})$+jets process is made
using the same methodology as for the $\ttbar$ and $\PW(\ell\nu)$ background processes.
We take advantage of the kinematic similarities between the $\cPZ\to\ell\ell$, $\PW(\ell\nu)$+jets, and
$\cPgg$+jets processes~\cite{Khachatryan:2015ira,Ellis:1985vn,Berends:1989cf}.
Corrections to the hadronic recoil and jet multiplicity spectra
are obtained in a control sample enriched in $\cPgg$+jets events, and the validity of these
corrections is checked in a second control sample enriched in $\PW(\ell\nu)$+jets events.
A third control sample, enriched in $\cPZ\to\ell\ell$ events, is used to normalize the
obtained correction factors and to provide an additional consistency check of the MC prediction.

The $\cPgg$+jets control sample consists of events having at least one selected photon
and passing a set of kinematic requirements.  Photons are required to have $\pt > 185\GeV$
and pass loose identification and isolation criteria.
The photon is treated as invisible---its $\pt$ is added vectorially to the \ptvecmiss,
and it is ignored in the calculation of $\MR$---in order to simulate
the invisible $\cPZ$~boson decay products in a $\cPZ\to\nu\overline{\nu}$+jets event.
Selected events must pass a single-photon
trigger, have two jets with $\pt > 80\GeV$, and have $\MR > 400\GeV$ and $\Rtwo > 0.25$.

The contribution of misidentified photons to the yield in this control sample is estimated
via a template fit to the distribution of the photon charged isolation, the $\pt$ sum
of all charged PF particles within a $\Delta$R cone of size $0.4$ centered on the
photon momentum axis.  The fit is performed
in bins of $\MR$ and $\Rtwo$ and yields an estimate of the purity of the photon sample
in each bin. Contributions from other background processes such as $\ttbar\cPgg$ are estimated using simulation
and account for about $1$--$2\%$.
Additionally, events in which the photon is produced within a jet are considered to be background.
Corrections to the hadronic recoil in simulation are derived in this CR by subtracting the
estimated background yields from the number of observed counts, and comparing the resulting
yield with the prediction from the $\cPgg$+jets simulation, in each bin of $\MR$ and $\Rtwo$.

As in the tight lepton CR described in Section~\ref{sec:TTBarWJetsCR}, an additional correction
is derived to account for possible mismodeling in simulation as a function of the jet multiplicity.
This correction is derived for events with two or three jets, with four to six jets, and
with seven or more jets.
After these corrections are applied, the data in the CR are compared with the MC
prediction in bins of the number of \PQb-tagged jets.
As in the tight lepton CR, the $\MR$ spectra in simulation are
corrected to match the data in each \PQb~tag category,
and a systematic uncertainty in the $\cPZ(\nu\overline{\nu})$+jets background
is assigned based on the size of the uncertainty in each bin of $\Rtwo$.

\begin{figure}[!tbp] \centering
\includegraphics[width=0.49\textwidth]{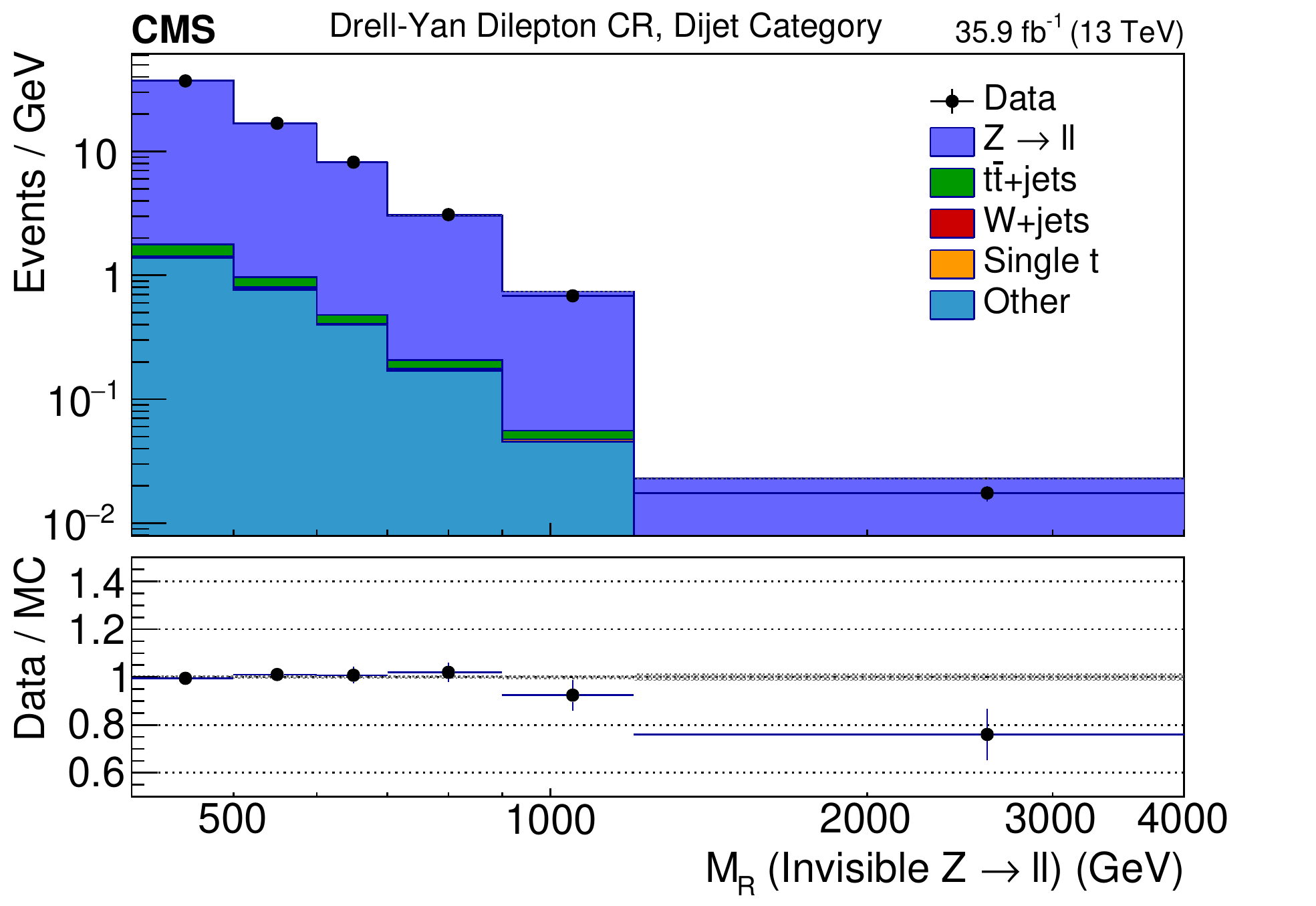}
\includegraphics[width=0.49\textwidth]{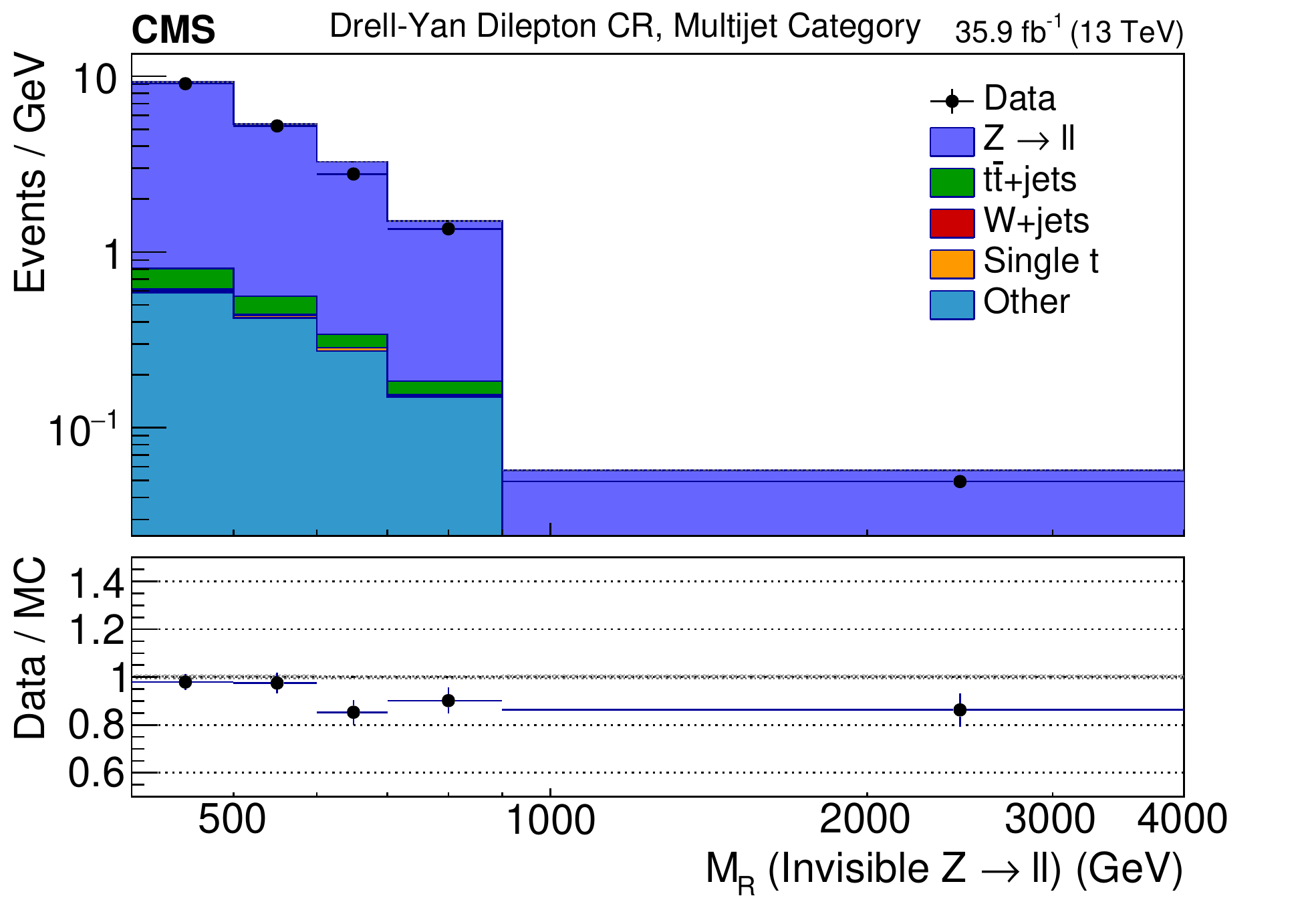}
\caption{ The $\MR$ distribution in the $\cPZ\to\ell\ell$+jets CR is displayed
in the 2--3 (left) and 4--6 (right) jet categories along with the corresponding MC predictions.
The corrections derived from the $\cPgg$+jets CR, as well as the
overall normalization correction, have been applied in this figure.
}
\label{fig:inclusive_dyjets}
\end{figure}

A check of the $\cPZ(\nu\overline{\nu})$+jets prediction is performed with a sample enriched
in $\cPZ\to\ell\ell$ decays.  Events in this sample are required to have two tight
electrons or two tight muons having an invariant mass consistent with the $\cPZ$ mass.
The two leptons are treated as invisible for the purpose of computing the razor variables.
Events must have no \PQb-tagged jets, two or more jets with $\pt > 80\GeV$, $\MR > 400\GeV$,
and $\Rtwo > 0.25$.
The correction factors obtained from the $\cPgg$+jets CR are normalized
so that the total MC prediction in the $\cPZ\to\ell^+\ell^-$+jets CR
matches the observed data yield.  This corrects for the difference between the true
$\cPgg$+jets cross section and the leading order cross section used to normalize the simulated samples.
The $\MR$ distributions in this CR for the two to three and four to six
jet categories are shown in Fig.~\ref{fig:inclusive_dyjets}.
The observed residual disagreements between data and simulation in the $\MR$ and $\Rtwo$ distributions
are propagated as systematic uncertainties in the $\cPZ(\nu\overline{\nu})$+jets prediction.

The MC corrections derived in the $\cPgg$+jets CR are checked against a second
set of corrections derived in a CR enriched in $\PW(\ell\nu)$+jets events.
This CR is identical to the $\PW(\ell\nu)$+jets sample described in Section~\ref{sec:TTBarWJetsCR},
except that the selected lepton is treated as invisible for the purpose of computing
$\MR$ and $\Rtwo$.  Correction factors are derived in the same way as in the $\PW(\ell\nu)$+jets
CR.  The full difference between these corrections and those obtained from the
$\cPgg$+jets CR is taken as a systematic uncertainty in the
$\cPZ(\nu\overline{\nu})$+jets prediction in the SR,
and is typically between $10$~and~$20\%$, depending on the bin.

\subsection{The QCD multijet background}
\label{sec:QCDCR}
Multijet events compose a nonnegligible fraction of the total event yield in the hadronic
SRs.  Such events are characterized by a significant undermeasurement
of the energy of a jet, and consequently a large amount of $\ptmiss$, usually pointing towards
the mismeasured jet.  A large fraction of QCD multijet events are
rejected by the requirement that the azimuthal angle $\dPhiR$ between the axes of the two razor
megajets is less than 2.8. We treat the events with $\dPhiR \ge 2.8$ as a
CR of QCD multijet events, while the events with $\dPhiR < 2.8$ define the SRs.

We estimate the number of QCD multijet events in this CR in bins of $\MR$ and $\Rtwo$ by
subtracting the predicted contribution of other processes from the total event yield
in each bin.
This is done for each jet multiplicity category.
We observe in simulation that the fraction of QCD multijet events at each
\PQb~tag multiplicity is independent of $\MR$, $\Rtwo$, and
$\dPhiR$. The event yields in the QCD CRs
are therefore measured inclusively in the number of \PQb~tags
and then scaled according to the fraction of QCD multijet events
at each multiplicity of \PQb-tagged jets.

We then predict the number of QCD multijet events in the SRs via the transfer factor
$\zeta$, defined as
\begin{equation}
\zeta = \frac{N(\abs{\dPhiR}<2.8)}{N(\abs{\dPhiR}>2.8)}.
\end{equation}
It is calculated using control regions in data and validated with simulation.
The QCD background prediction in each bin ($N^{\text{QCD}}_{\text{SR}\ \text{bin}\ \text{i}}$) is made as:
\begin{equation}
N^{\text{QCD}}_{\text{SR}\ \text{bin}\ \text{i}} = \zeta(N^{\text{data}}_{\text{CR}\ \text{bin}\ \text{i}} - N^{\text{bkg}}_{\text{CR}\ \text{bin}\ \text{i}}),
\end{equation}
where $N^{\text{data}}_{\text{CR}\ \text{bin}\ \text{i}}$ is the number of events observed in the data CR
and $N^{\text{bkg}}_{\text{CR}\ \text{bin}\ \text{i}}$ is the contribution from background processes other
than the QCD multijet process and is predicted from the corrected MC.

We observe in simulation that $\zeta$ changes slowly with $\MR$ and
increases roughly linearly with $\Rtwo$. In data we
therefore compute $\zeta$ in bins of $\MR$ and $\Rtwo$
in a low-$\Rtwo$ region defined by $0.20 < \Rtwo < 0.30$ and fit the computed values
with a linear function in $\MR$ and $\Rtwo$.
We then use the linear fit and its uncertainty to estimate
the value of $\zeta$ in the analysis SRs.
The fit is performed separately in each category of
jet multiplicity, but inclusively in the number of \PQb-tagged jets,
as $\zeta$ is observed in simulation not to depend on the \PQb~tag multiplicity.
For the category with seven or more jets, the fit function is
allowed to depend on $\Rtwo$ only, because of the low
number of events in the fit region.

The statistical uncertainty in the CR
event counts and the fitted uncertainty of the
transfer factor extrapolation are propagated as systematic
uncertainties of the QCD multijet background prediction.
Another systematic uncertainty of $30\%$ is propagated in order to
cover the dependence of the transfer factor on the number of
\PQb-tagged jets in different CRs. Furthermore,
we make an alternative extrapolation for the transfer factor
where we allow a dependence on $\MR$ and $\Rtwo$ for the
Seven-jet category, and a quadratic dependence on $\MR$ for
the Dijet and Multijet categories. The difference in the
QCD multijet background prediction between the default
and alternative transfer factor extrapolation is propagated
as an additional systematic uncertainty, whose size
ranges from $10\%$ for $\MR$ below $1\TeV$ to $70$--$90\%$
for $\MR$ above $1.6\TeV$.

\subsection{Background modeling in boosted event categories}
\label{sec:bgboost}

The dominant SM background processes in the boosted categories are the same as in
the non-boosted categories. An additional, but important source of background comes from
processes where one of the jets in the event is mistagged as a
boosted hadronic \PW~boson or top quark.

Requiring boosted objects in the selection results in a smaller number of events in the SRs or CRs.
As a general rule, in cases where no MC events exist in SR bins for a given
background process, MC counts in these bins are extrapolated from a looser version of the
signal selection obtained by relaxing the $N$-subjettiness criteria for $\PW$ or \cPqt~tagging.
For cases where there are no counts or very low statistical precision in the CR
bins, these depleted bins are temporarily merged to obtain coarser bins with increased event
count.  Background estimation is done in two steps, where first the yields are estimated
using the coarser bins, and next, the yields in coarse bins are distributed to the finer
bins proportional to the background MC counts in the finer bins.

\subsubsection{The $\ttbar$+jets and $\PW$+jets background estimation for the boosted categories }
\label{sec:Tregion}

The CRs for the $\ttbar$ and $\PW$+jets backgrounds are defined similar to
the CRs used for the non-boosted categories. We require
exactly one loose electron or muon. To suppress contamination from signal
processes, $\mT$ is required to be less than $100\GeV$.
To mimic the signal selection, the $\dPhiR < 2.8$
requirement is applied.  To estimate the top quark background for
the boosted $\PW$ 4--5 jet and boosted $\PW$ 6 jet SR
categories, we require events in the CR to have
at least one boosted $\PW$~boson and one \cPqb-tagged jet, while for the boosted top category, we require one
boosted top quark. To estimate the $\PW(\ell\nu)$+jets background
for the boosted $\PW$ 4--5 jet and boosted $\PW$ 6 jet SR categories, we require events in the CR to have no
loosely tagged \cPqb~jets, while for the boosted top category we require no \cPqb-tagged subjets.
To maintain consistency with SR kinematics, we require a jet
which is tagged only using the \PW~boson or top quark mass requirement, but without the
$N$-subjettiness requirement. The background estimate for each SR $i$ is then
extrapolated from the corresponding CR via transfer factors calculated in MC:
$\lambda_{\text{i}} = N_{\text{i}}^{\text{SR,MC}} / N_{\text{i}}^{\text{CR,MC}}$.

For certain bins, the MC prediction of the transfer factors can have large statistical
fluctuations from the limited number of MC events. To smooth out these fluctuations
we use a combination of bin-merging and extrapolations from a region with looser
requirements on the $N$-subjettiness variables. While the fluctuations in
the nominal background prediction are smoothed out, the statistical uncertainties
from the limited MC sample size are still propagated as a systematic uncertainty.

\begin{figure}[tbp]
\centering
\includegraphics[width=0.49\textwidth]{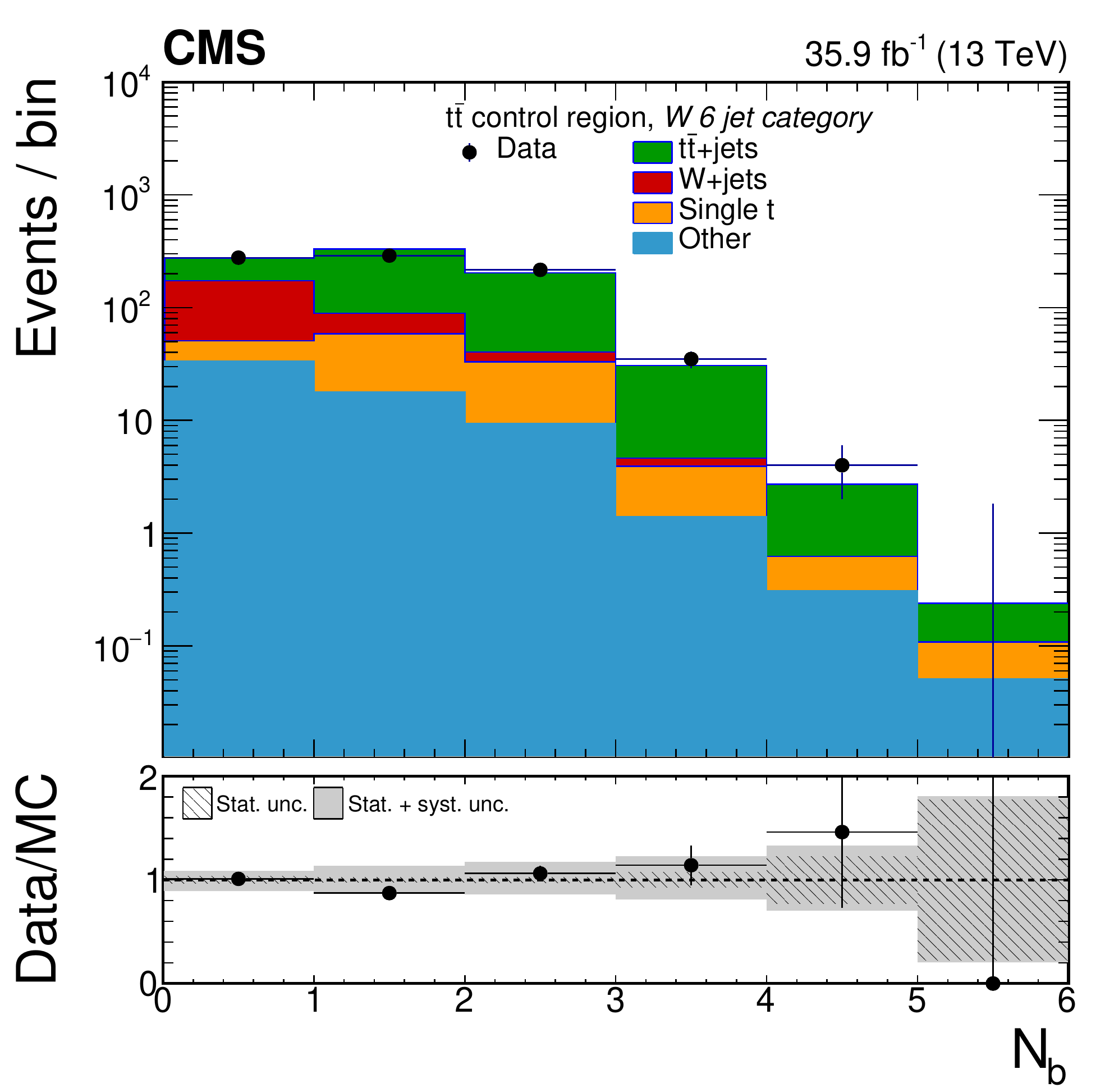}
\includegraphics[width=0.49\textwidth]{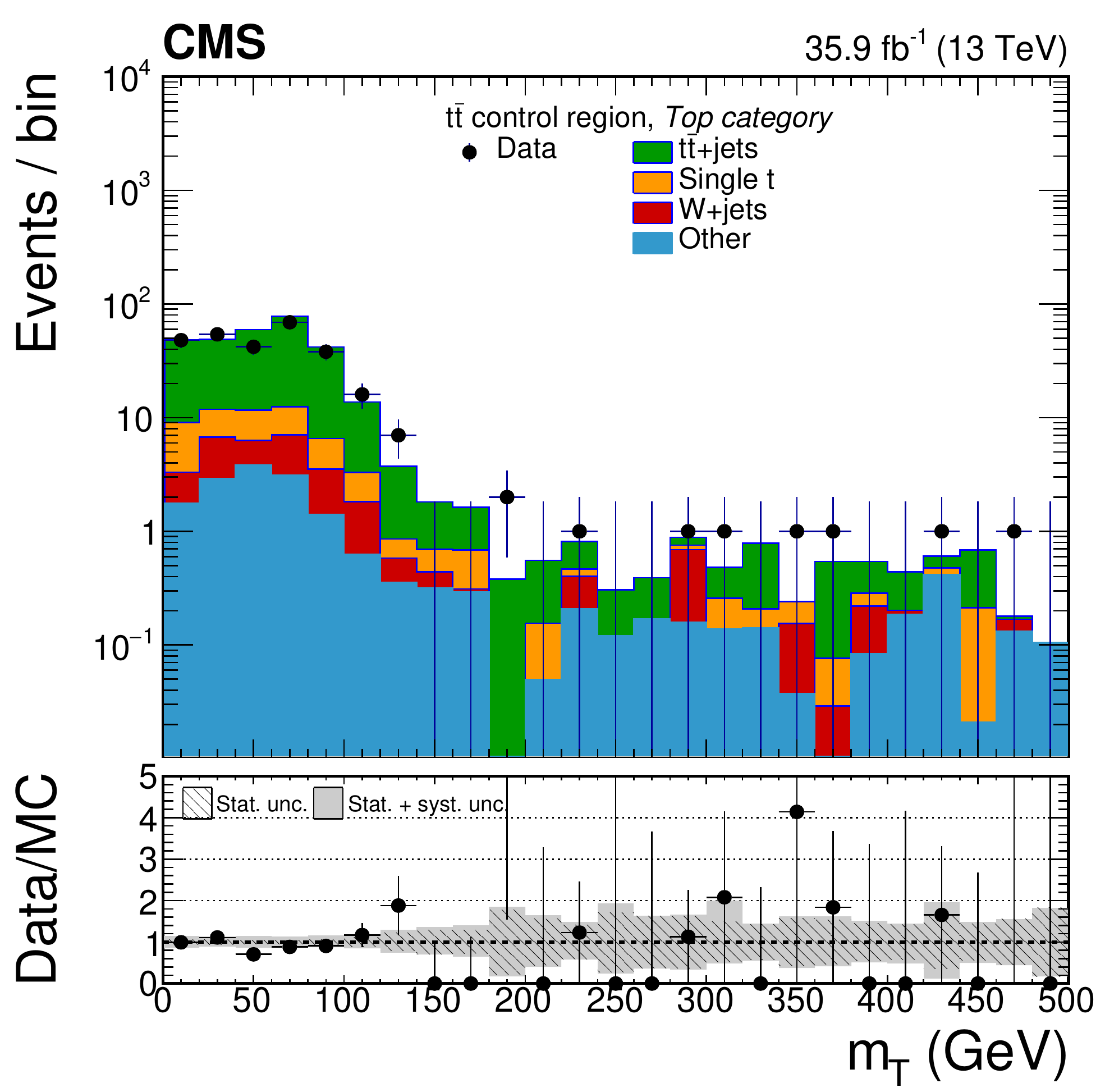}
\caption{The distribution of \cPqb-tagged jet multiplicity before applying the \cPqb~tagging selection
requirement in the $\ttbar$ CR of the boosted $\PW$ 6 jet category (left), and
the distribution in $\mT$ before applying the $\mT$ selection requirement
in the $\ttbar$ CR of the boosted top category (right) are shown.
The ratio of data over MC prediction is shown in the lower panels, where the gray band
is the total uncertainty and the dashed band is the statistical uncertainty in the MC prediction.
}
\label{fig:nminusone_T}
\end{figure}

\begin{figure}[tbp]
\centering
\includegraphics[width=0.49\textwidth]{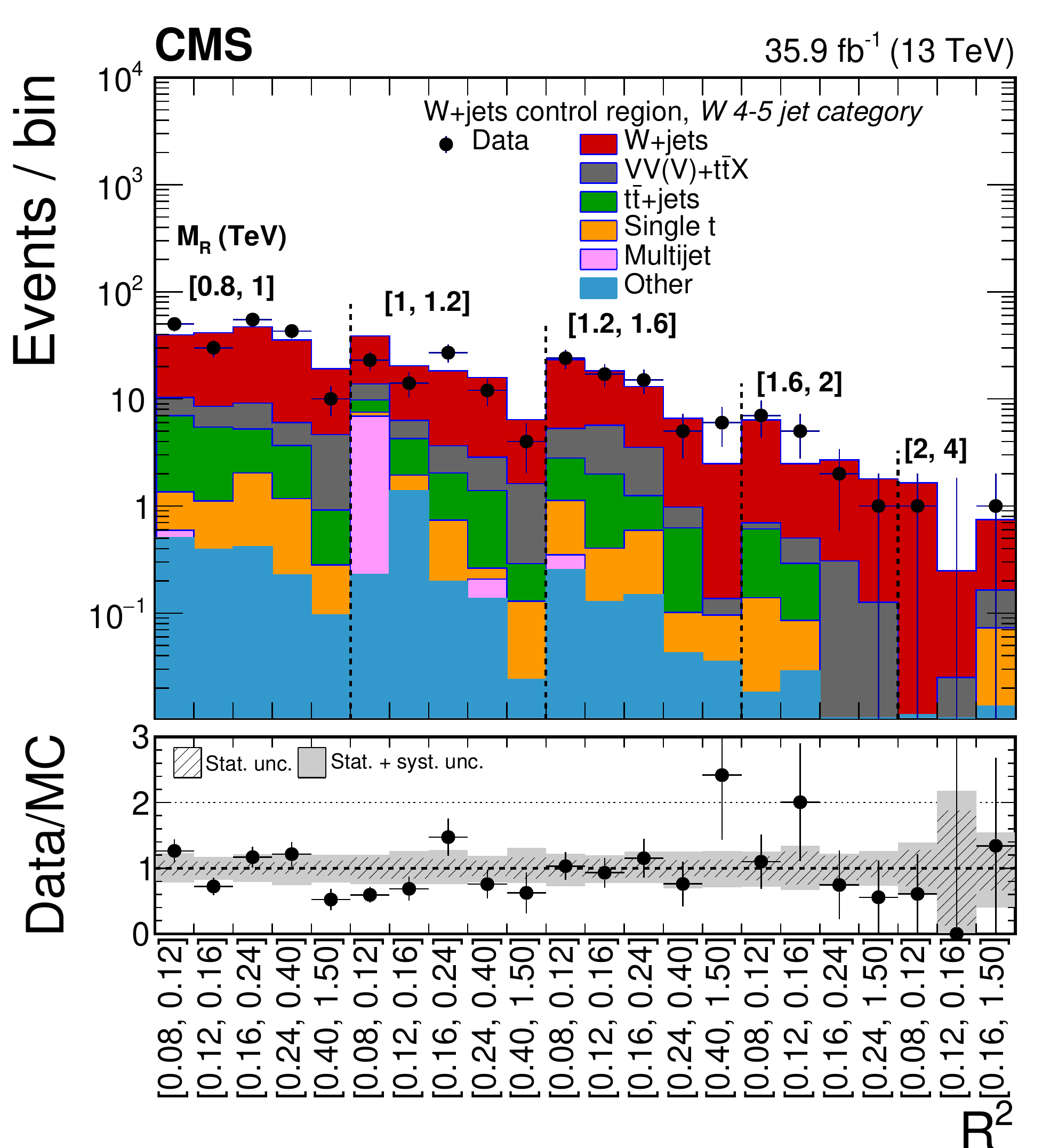}
\includegraphics[width=0.49\textwidth]{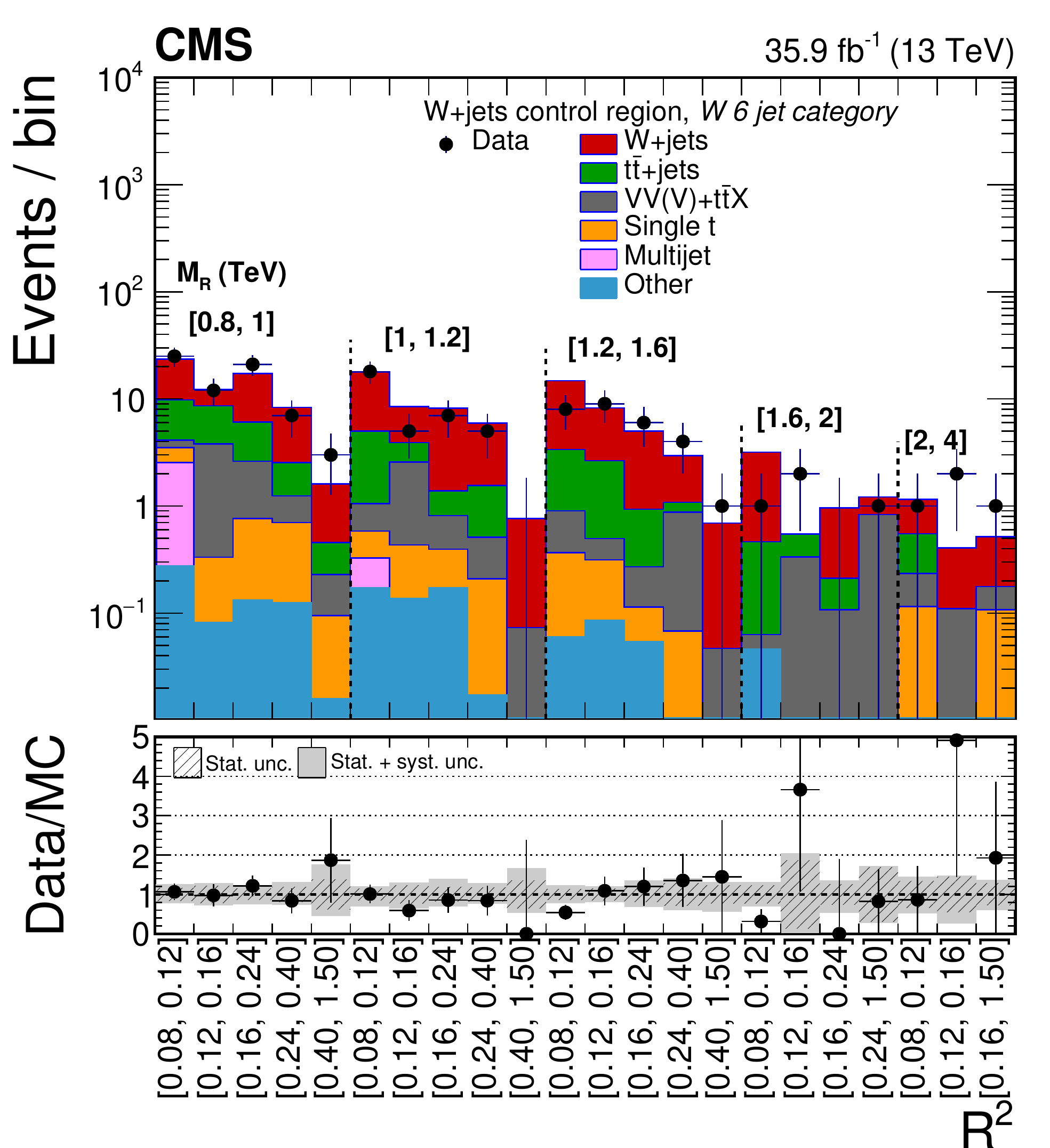}  \\
\includegraphics[width=0.49\textwidth]{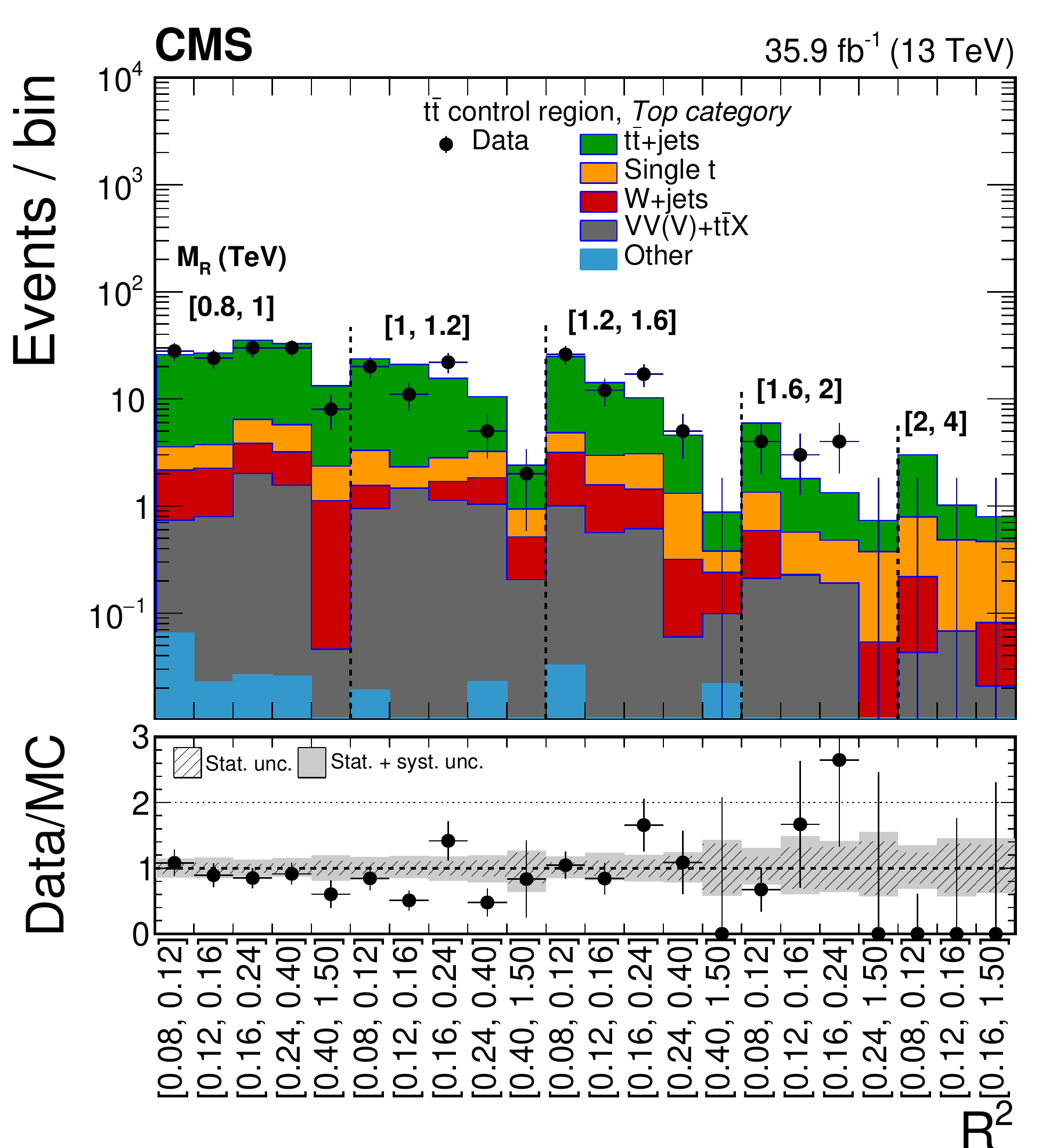}
\caption{$\MR$--$\Rtwo$ distributions in the $\PW$+jets CRs of the boosted
$\PW$ 4--5 jet (upper left) and boosted $\PW$ 6 jet (upper right) categories,
and the $\ttbar$ CR (lower) of the boosted top category.  The ratio of data over MC prediction is
shown in the lower panels, where the gray band is the total uncertainty and the dashed band is
the statistical uncertainty in the MC prediction.
\label{fig:Razor_WT}}
\end{figure}

Figure~\ref{fig:nminusone_T} shows the \cPqb-tagged jet multiplicity distribution,
identified with the medium \cPqb~jet tagger, for events in the boosted $\PW$ 6 jet category in
the $\ttbar$ CR before applying the \cPqb~tagging selection, and the $\mT$
distribution in the boosted top category in the $\ttbar$ CR before applying the $\mT$ selection.
Figure~\ref{fig:Razor_WT} shows the distribution in
$\MR$ and $\Rtwo$ bins for events in the boosted top category in the $\ttbar$ CR, and for
events in the boosted $\PW$ 4--5 jet and boosted $\PW$ 6 jet categories in the $\PW(\ell\nu)$+jets CR.
The purity of $\ttbar$+jets and single top events in
the $\ttbar$ CR is more than $80\%$, and the purity of the $\PW(\ell\nu)$+jets process in the
$\PW(\ell\nu)$+jets CR is also larger than $80\%$.

\subsubsection{The $\cPZ\to\nu\overline{\nu}$+jets background estimation for the boosted categories }
\label{sec:GZLregion}

\begin{figure}[tbp]
\centering
\includegraphics[width=0.49\textwidth]{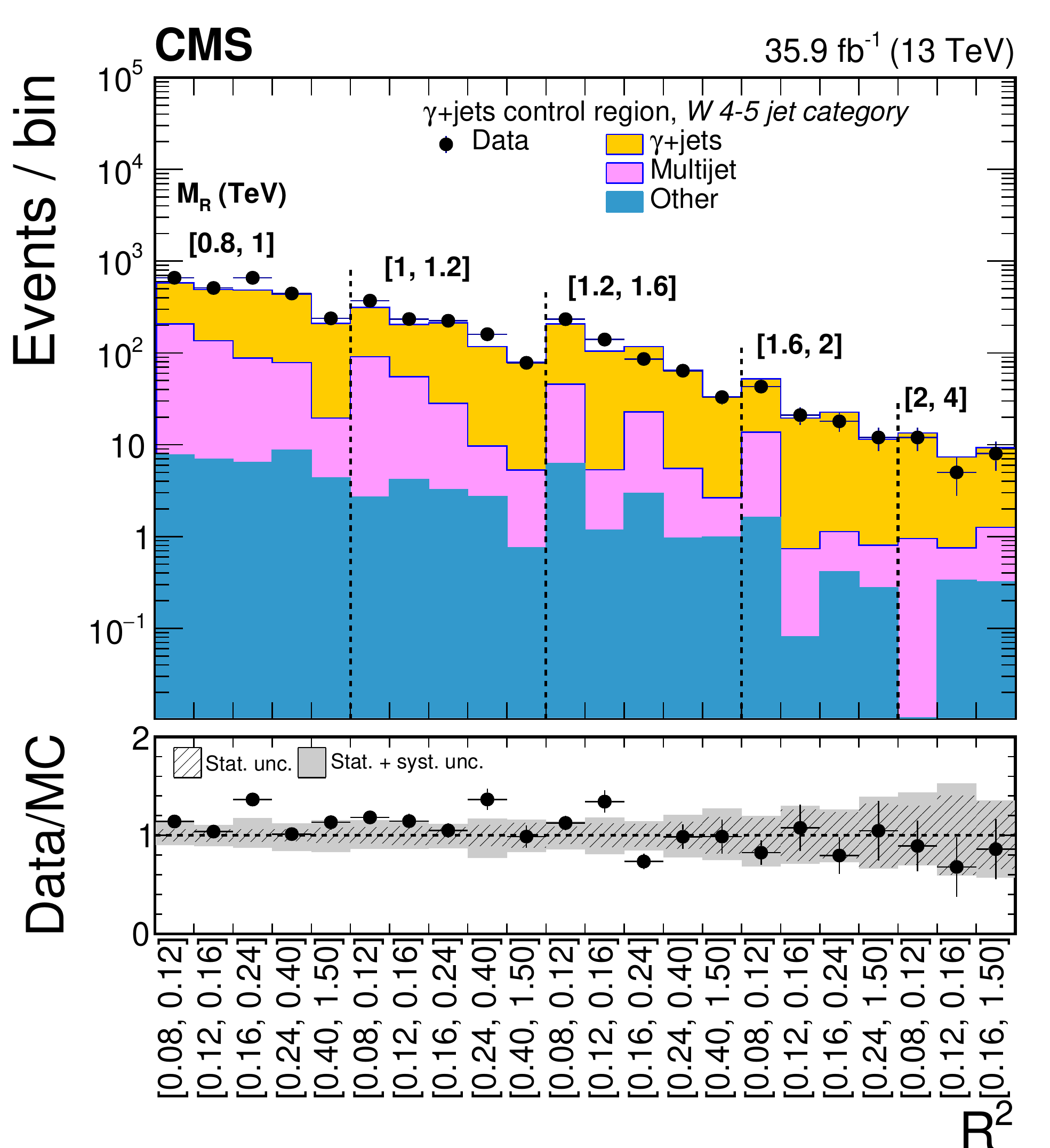}
\includegraphics[width=0.49\textwidth]{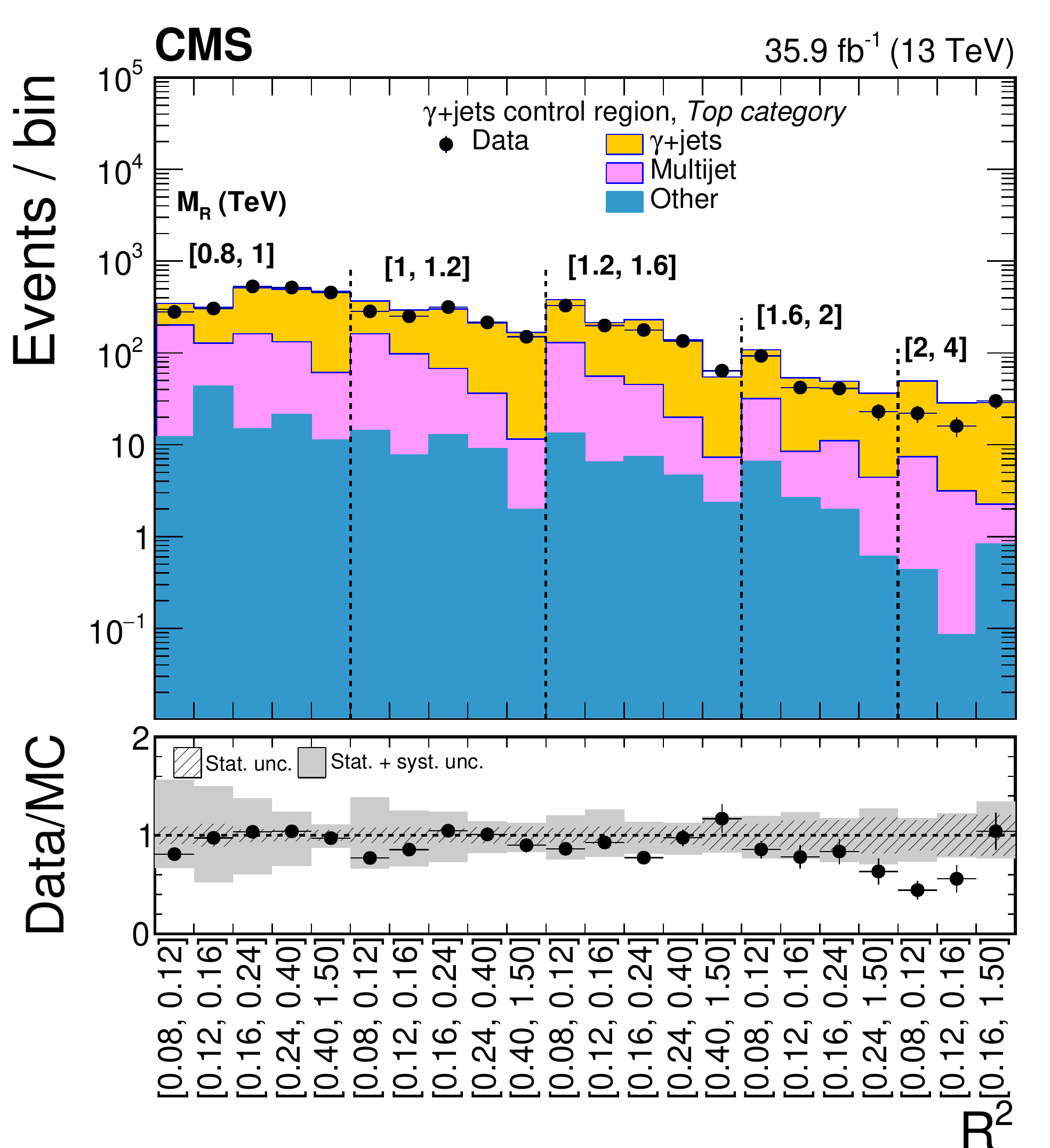}
\caption{$\MR$--$\Rtwo$ distributions for the $\cPgg$+jets CR of the boosted $\PW$ 4--5 jet (left)
and boosted top (right) category. The ratio of data over MC prediction is shown in the lower panel,
where the gray band is the total uncertainty and the
dashed band is the statistical uncertainty in the MC prediction.
\label{fig:Razor_G}}
\end{figure}

The background estimate for the $\cPZ\to\nu\overline{\nu}$+jets process is again similar to the method used
for the non-boosted categories. We make use of the similarity in the kinematics of the photon in $\cPgg$+jets
events and the $\cPZ$~boson in $\cPZ$+jets events to select a control sample of $\cPgg$+jets to mimic
the behavior of $\cPZ\to\nu\overline{\nu}$+jets events. The $\cPgg$+jets CR is selected by
requiring exactly one photon with $\pt > 80\GeV$ from data collected by jet and $\HT$ triggers.
The momentum of the photon is added to $\ptvecmiss$ to mimic the contribution of the neutrinos
from $\cPZ\to\nu\overline{\nu}$ decays.  We require that the events contain no loose leptons or $\tauh$ candidates,
and $\dPhiR$, computed after treating the photon as invisible, is required to be less than $2.8$.
One $\PW$-tagged or \cPqt-tagged jet is required for the boosted $\PW$ and top categories, respectively.
Figure~\ref{fig:Razor_G} shows the $\MR$--$\Rtwo$ distribution for the boosted top category.
The QCD multijet contribution to the $\cPgg$+jets CR is accounted for by
a template fit to the photon charged isolation variable in inclusive bins of $\MR$ and $\Rtwo$.
Other background processes in the $\cPgg$+jets CRs are small and predicted using MC.
Finally, the SR prediction for the $\cPZ\to\nu\overline{\nu}$+jets background is extrapolated
from the $\cPgg$+jets yields via the MC transfer factor
$\lambda_{\cPZ\to\nu\overline{\nu}} = N^{\mathrm{SR,MC}}_{\cPZ\to\nu\overline{\nu}} / N^{\mathrm{CR,MC}}_{\cPgg\text{+jets}}$.

We perform a cross check on the previous estimate using a CR enhanced in $\cPZ\to\ell\ell$ events.
The $\cPZ\to\ell\ell$ CR is defined by requiring
exactly two tight electrons or muons with $\pt > 10\GeV$ and dilepton mass satisfying
$\abs{m_{\ell\ell} - m_{\cPZ}} < 10\GeV$, where $m_{\cPZ}$ is the $\cPZ$~boson mass. All other requirements are the same as those
for the $\cPgg$+jets CR. The momentum of the dilepton system is added vectorially to
$\ptvecmiss$ to mimic an invisible decay of the $\cPZ$~boson. Similarly for the
non-boosted categories, the comparison between data and MC yields in the $\cPZ\to\ell\ell$ CR
are used to correct the MC transfer factor $\lambda$ to account for the impact of missing higher order
corrections on the total normalization predicted by the $\cPgg$+jets simulation.

\begin{figure}[tbp]
\centering
\includegraphics[width=0.49\textwidth]{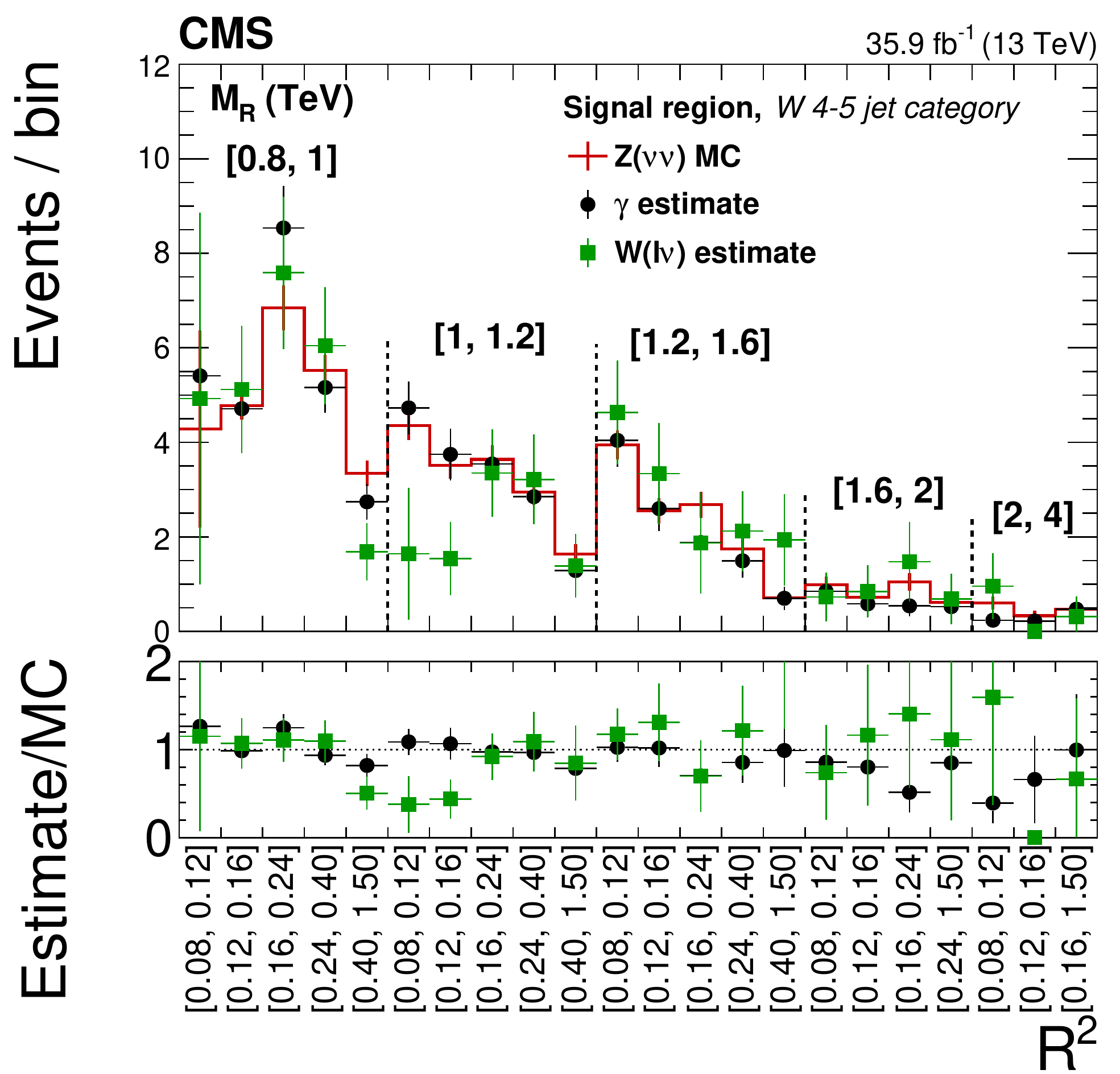}
\includegraphics[width=0.49\textwidth]{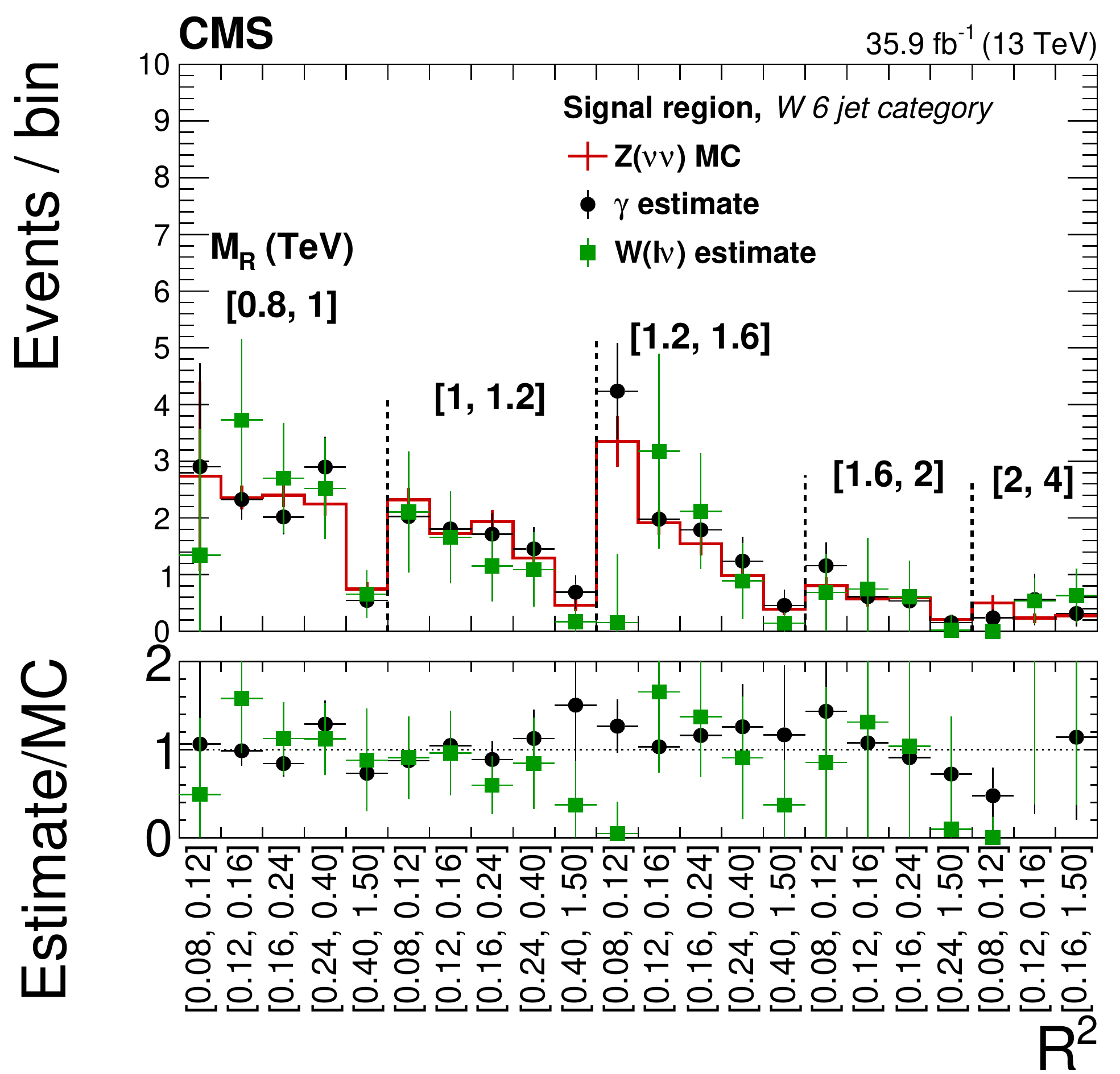} \\
\includegraphics[width=0.49\textwidth]{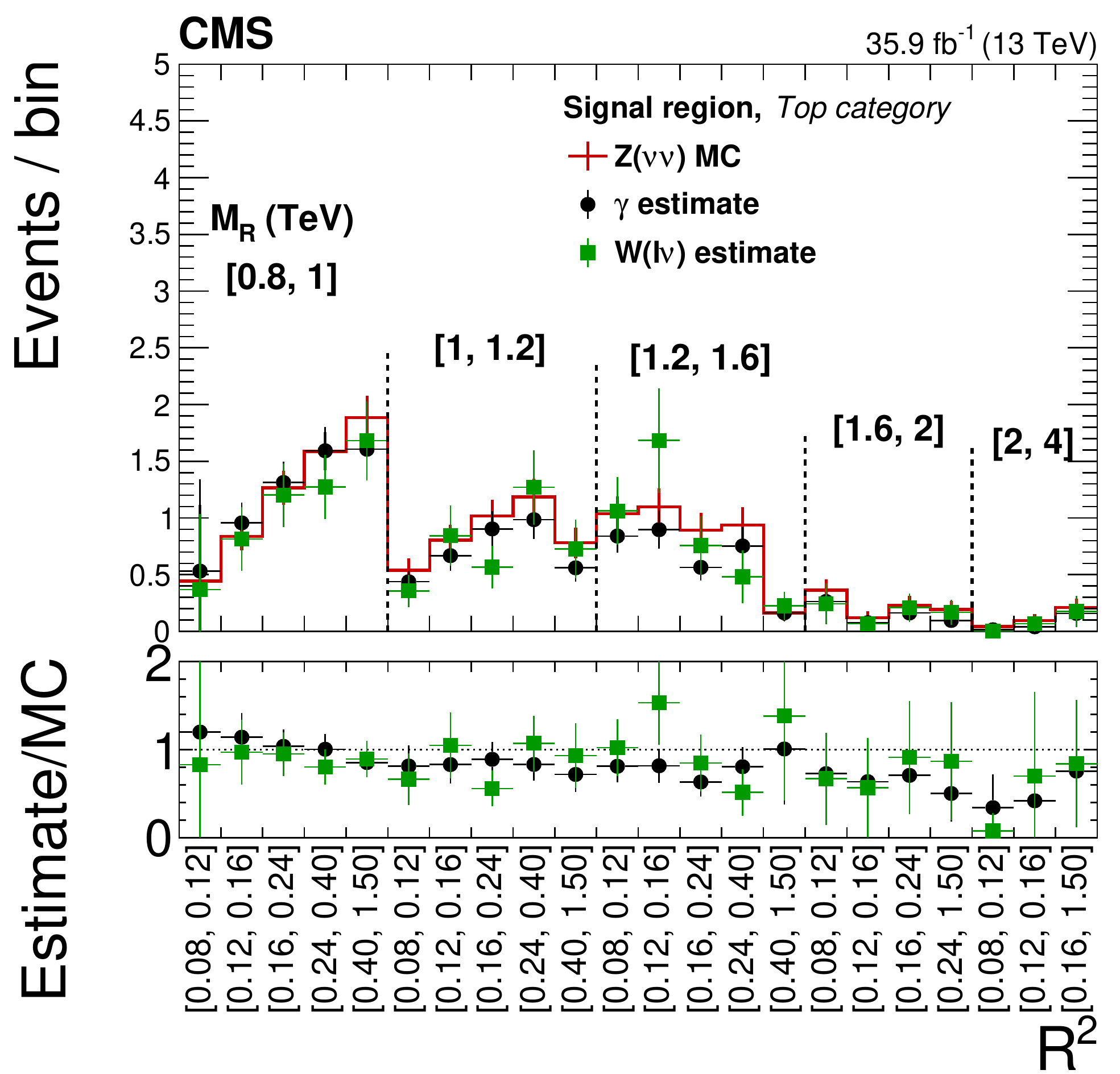}
\caption{Comparison of the estimate of the $Z(\to \nu\nu)$+jets background contribution in
the SR extrapolated from the $\cPgg$+jets CR with the estimate extrapolated from
the $W(\to\ell\nu)$+jets CR for the boosted $\PW$ 4--5 jet (upper left),
boosted $\PW$ 6 jet (upper right) and boosted top (lower) categories in bins of $\MR$ and $\Rtwo$.
The prediction from the uncorrected MC simulation is also shown.
The black labels indicate the range in $\MR$ that each set of bins correspond to.
\label{fig:Razor_zvv_result}}
\end{figure}

As for the inclusive categories, we obtain an alternative estimate from the $\PW(\to\ell\nu)$+jets-enriched
CR to validate the predictions from the $\cPgg$+jets CR. We
require the presence of exactly one tight electron or muon. $\mT$
is required to be between $30$~and~$100\GeV$.  The rest of the selection is the same as for the
$\cPgg$+jets CR.  The lepton momentum is added vectorially to $\ptvecmiss$ to mimic an
invisible decay. The $W(\to\ell\nu)$+jets CR yields are extrapolated
to the SR via transfer factors calculated from simulation to obtain the alternative $\cPZ\to\nu\overline{\nu}$+jets
background estimate. Figure~\ref{fig:Razor_zvv_result} compares the estimates from the $\cPgg$+jets CR,
the $W(\to\ell\nu)$+jets CR, and the MC simulation. The difference between the
two alternative estimates based on CRs in data is propagated as a systematic uncertainty.

\subsubsection{Multijet background estimation in the boosted categories
\label{sec:Qregion}}

\begin{figure}[tbp]
\centering
\includegraphics[width=0.49\textwidth]{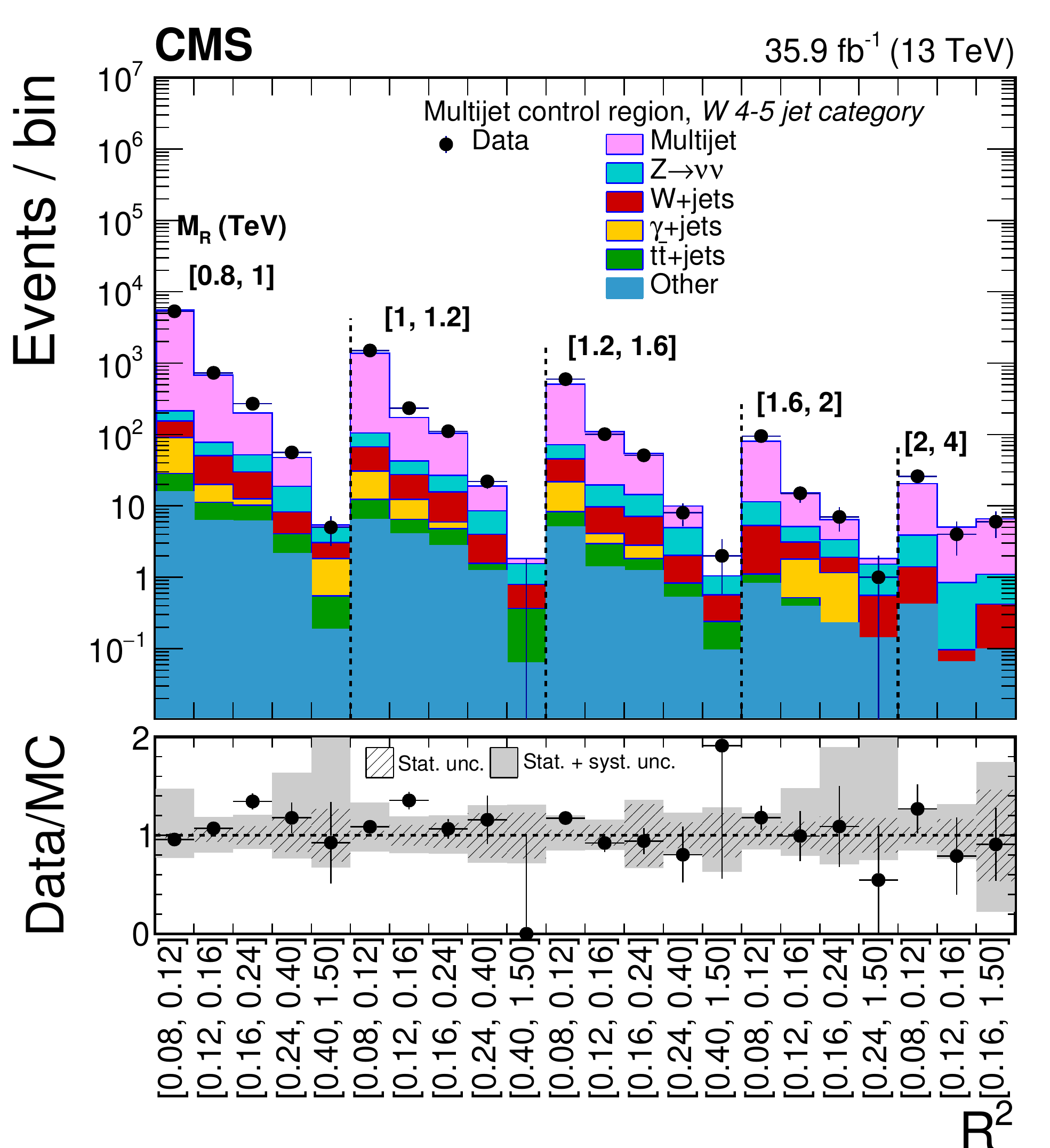}
\includegraphics[width=0.49\textwidth]{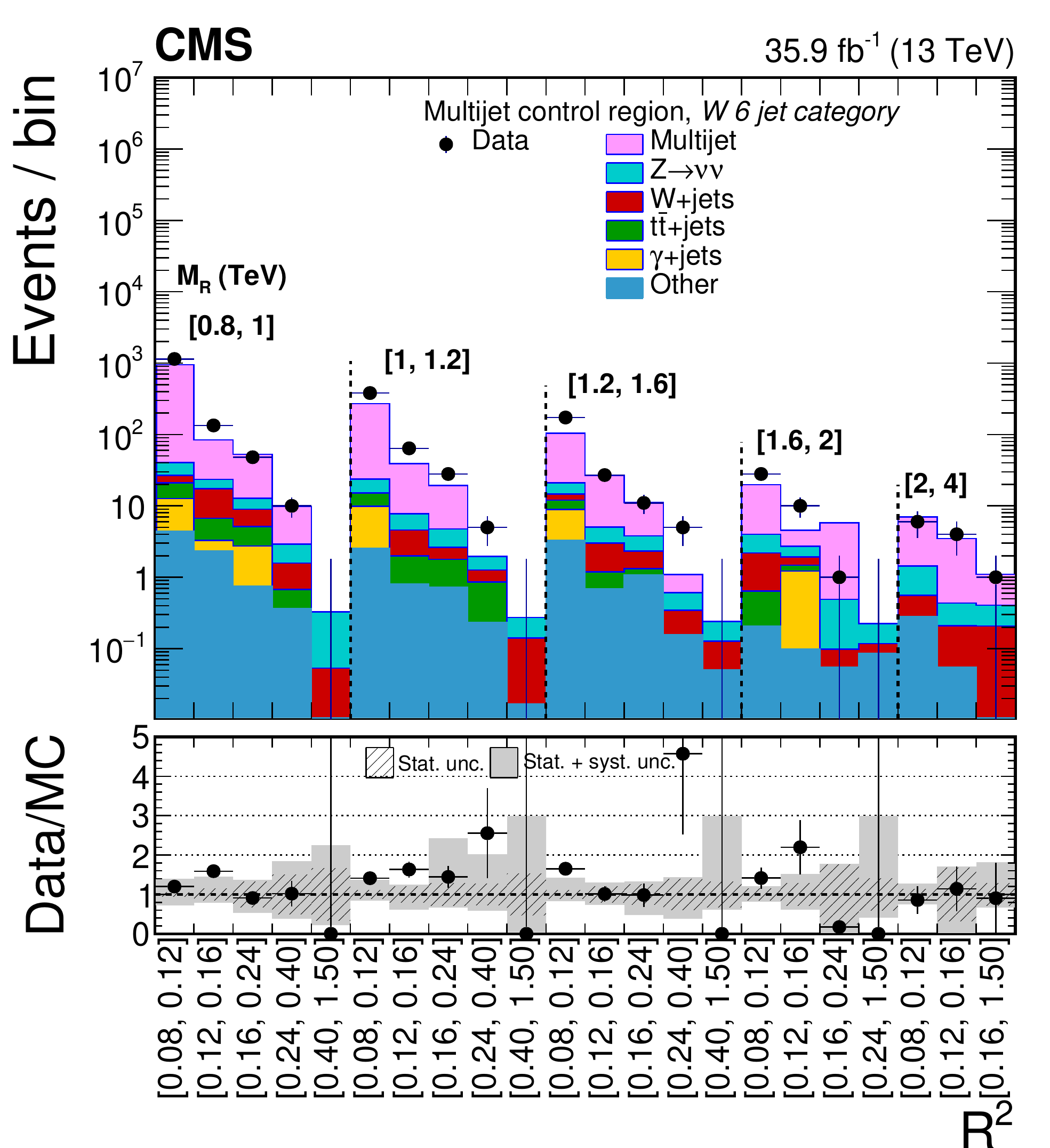} \\
\includegraphics[width=0.49\textwidth]{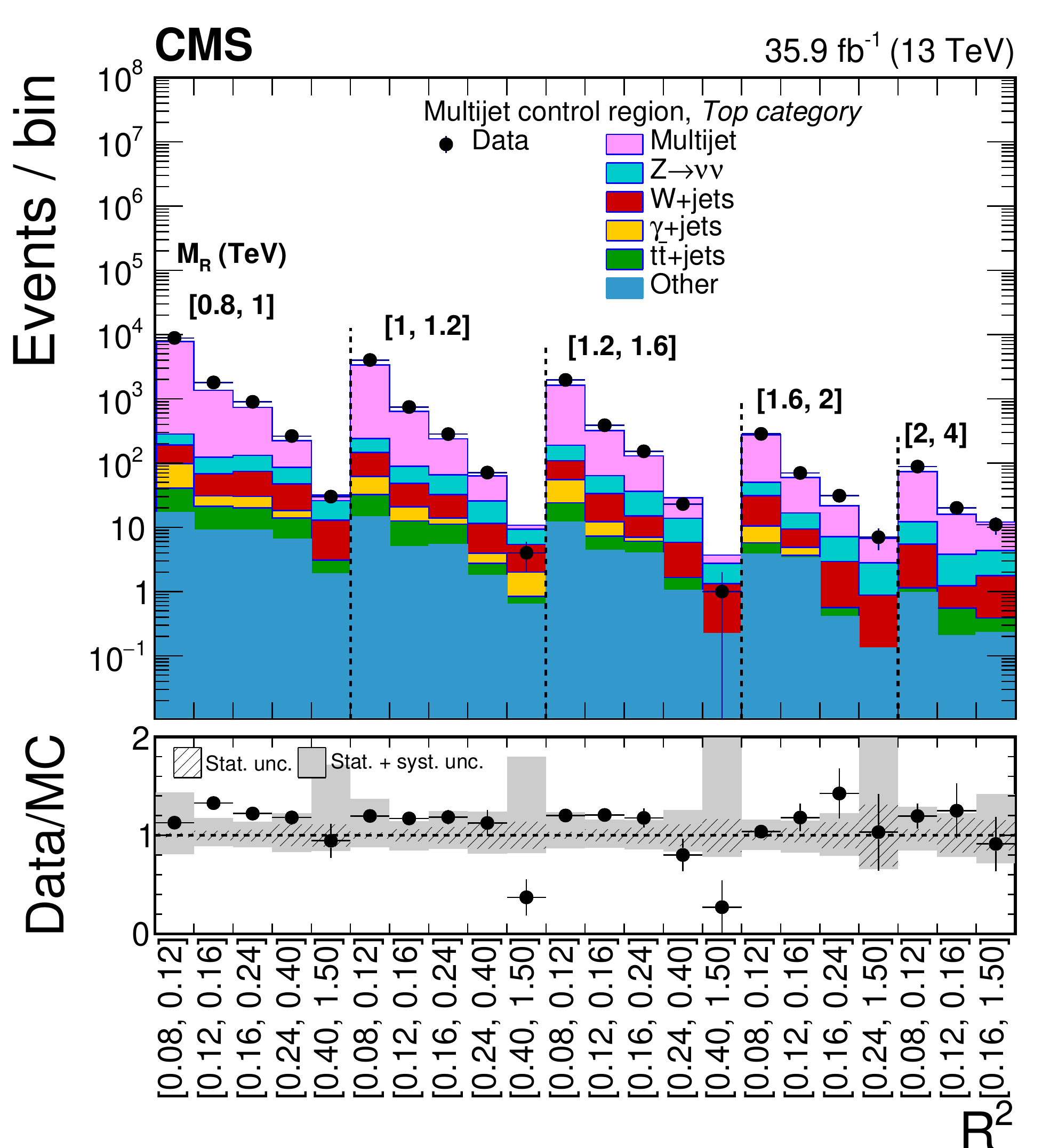}
\caption{The $\MR$--$\Rtwo$ distributions in the QCD multijet CRs of the boosted $\PW$ 4--5 jet (upper left),
boosted $\PW$ 6 jet (upper right), and boosted top (lower) categories.
The ratios of data over MC prediction is shown in the lower panels, where the gray band is the total uncertainty
and the dashed band is the statistical uncertainty in the MC prediction.
\label{fig:Razor_Q}}
\end{figure}

The CR enriched in QCD multijet background is defined by inverting
the $\dPhiR$ requirement, and requiring antitagged \PW~boson or top quark candidates
by inverting the $N$-subjettiness criteria and subjet \cPqb~tagging
for \cPqt-tagged jets. Figure~\ref{fig:Razor_Q} shows the distribution in the $\MR$
and $\Rtwo$ bins for the boosted $\PW$ 4--5 jet, boosted $\PW$ 6 jet and boosted top categories.
The purity achieved with the selection described above is about $90\%$.
The QCD multijet background is predicted by extrapolating the event
yields from this QCD multijet CR to the SRs via
transfer factors calculated from simulation.

The effects of inaccuracies in the modeling of the multijet background estimate are taken
into account by propagating a systematic uncertainty computed based on the level of disagreement
between data and simulation in the \cPqb~jet multiplicity, $N$-subjettiness  and $\dPhiR$ distributions
before applying these selections. The resulting overall systematic uncertainties are $13$ and $24\%$ for
boosted $\PW$ and top categories, respectively.

\subsubsection{Validating the background estimation with closure tests in boosted categories}

Two validations are performed in CRs similarly to that for the QCD multijet
CR but by inverting only one of the two requirements. These validations are
intended to verify the reliability of the background estimation method
for each requirement individually.

\begin{figure}[tpb]
\centering
\includegraphics[width=0.49\textwidth]{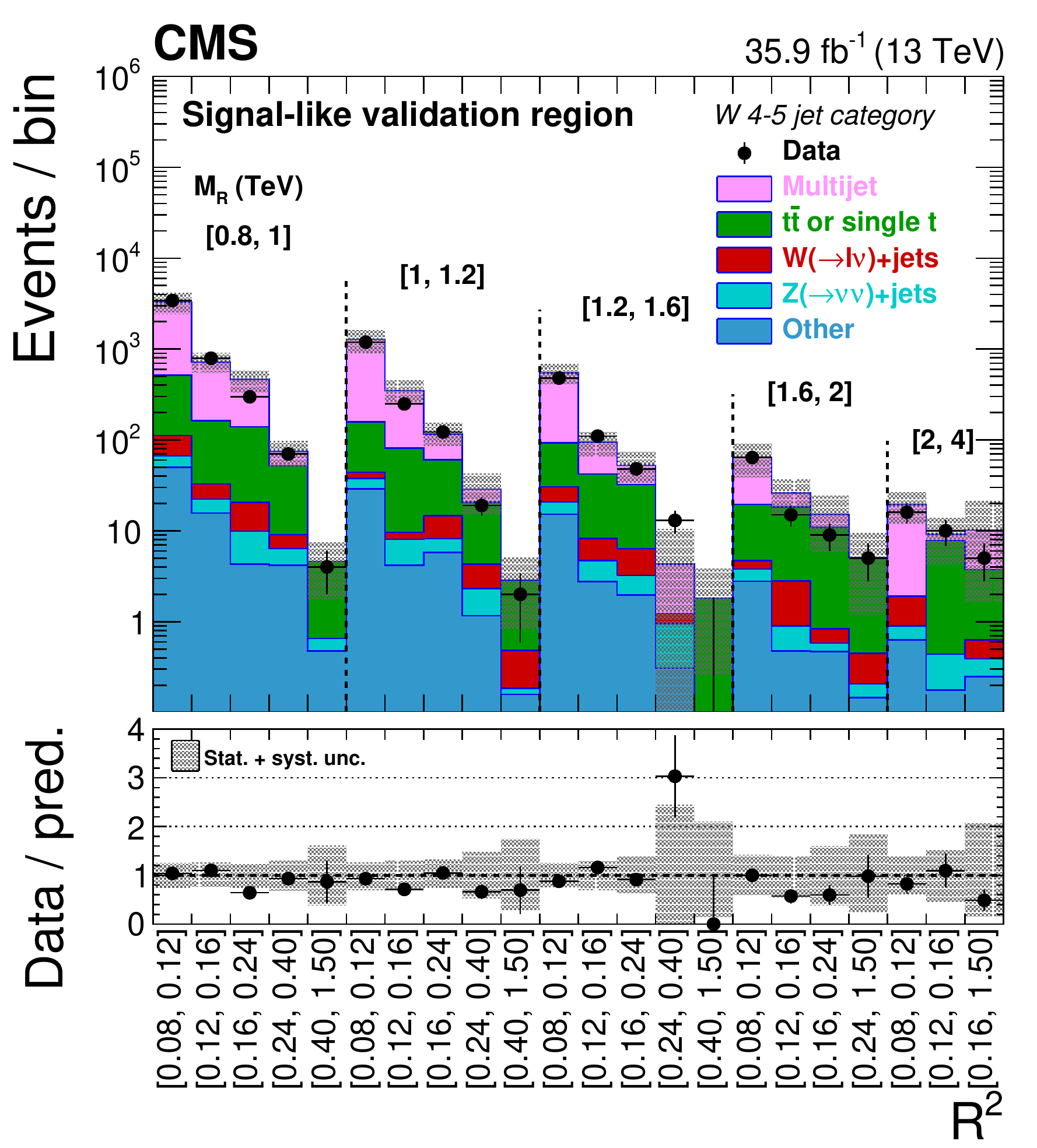}
\includegraphics[width=0.49\textwidth]{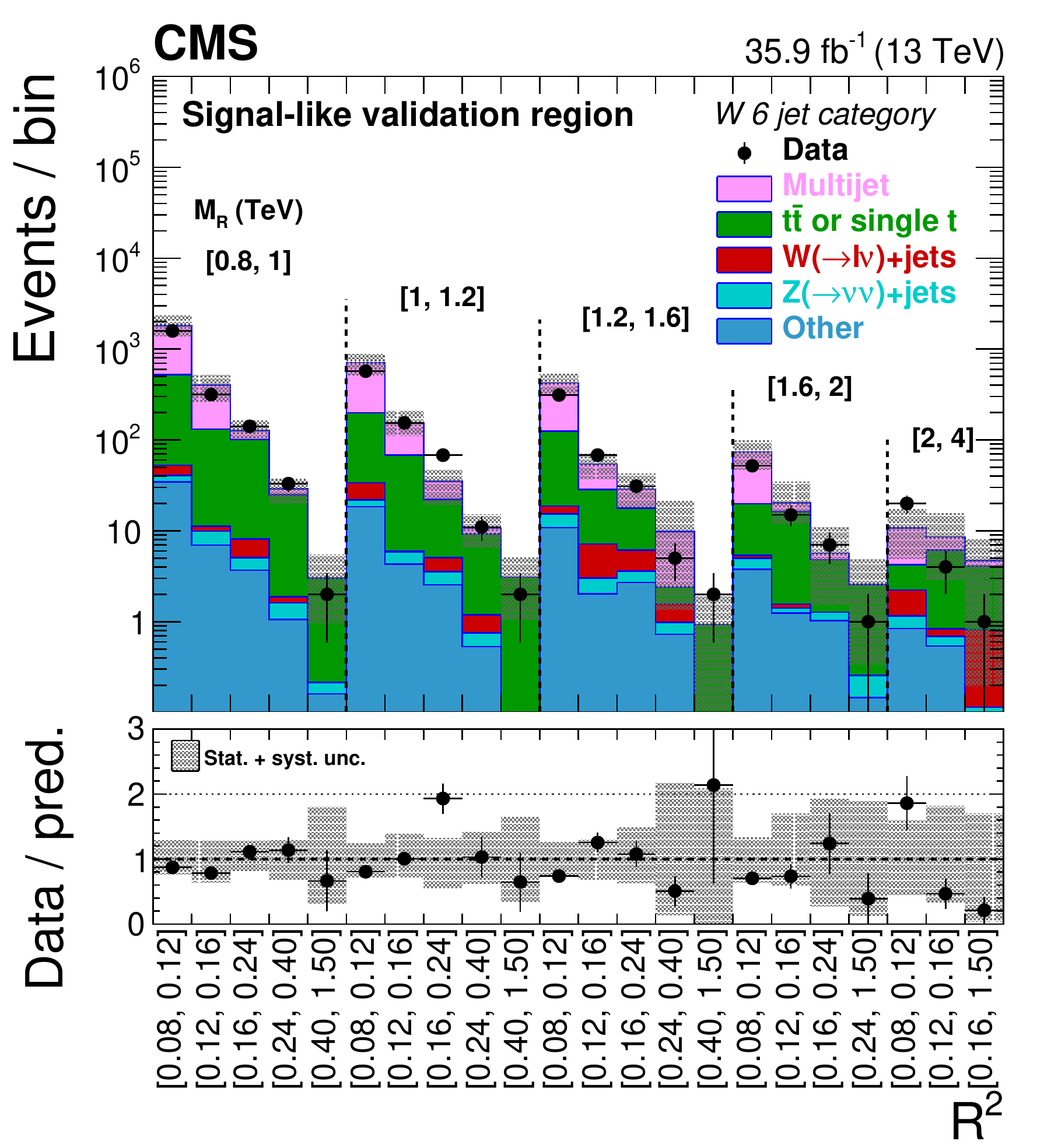} \\
\includegraphics[width=0.49\textwidth]{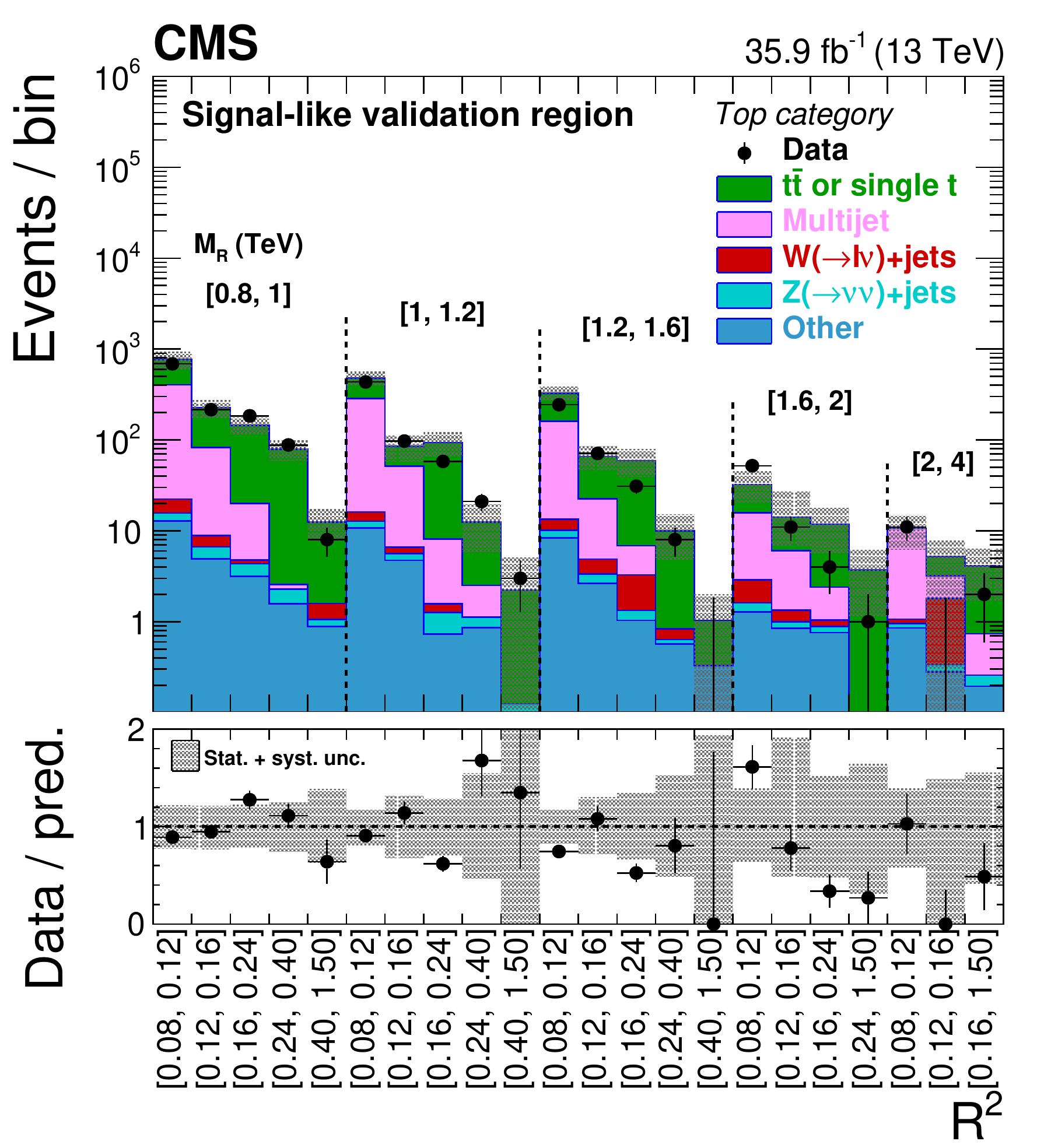}
\caption{Comparisons between data and the predicted background for the
inverted $\dPhiR$ validation region for the boosted $\PW$ 4--5 jet (upper left),
boosted $\PW$ 6 jet (upper right), and boosted top (lower) categories.
\label{fig:boostclosure1}}
\end{figure}

The first validation is performed in
a CR that is defined identically to the SR except that we
invert the $\dPhiR$ requirement. The comparison between data and predicted background
validates the MC modeling of \cPqb~tagging, the $\dPhiR$ shape, the extrapolation in the
lepton multiplicity, and the accuracy of the efficiency for $\PW$~boson and top quark tagging.
Figure~\ref{fig:boostclosure1} shows
the results for the boosted $\PW$~4--5~jet, boosted $\PW$~6~jet,
and boosted top categories. Overall, the estimation agrees with data within uncertainties.

\begin{figure}[tpb]
\centering
\includegraphics[width=0.49\textwidth]{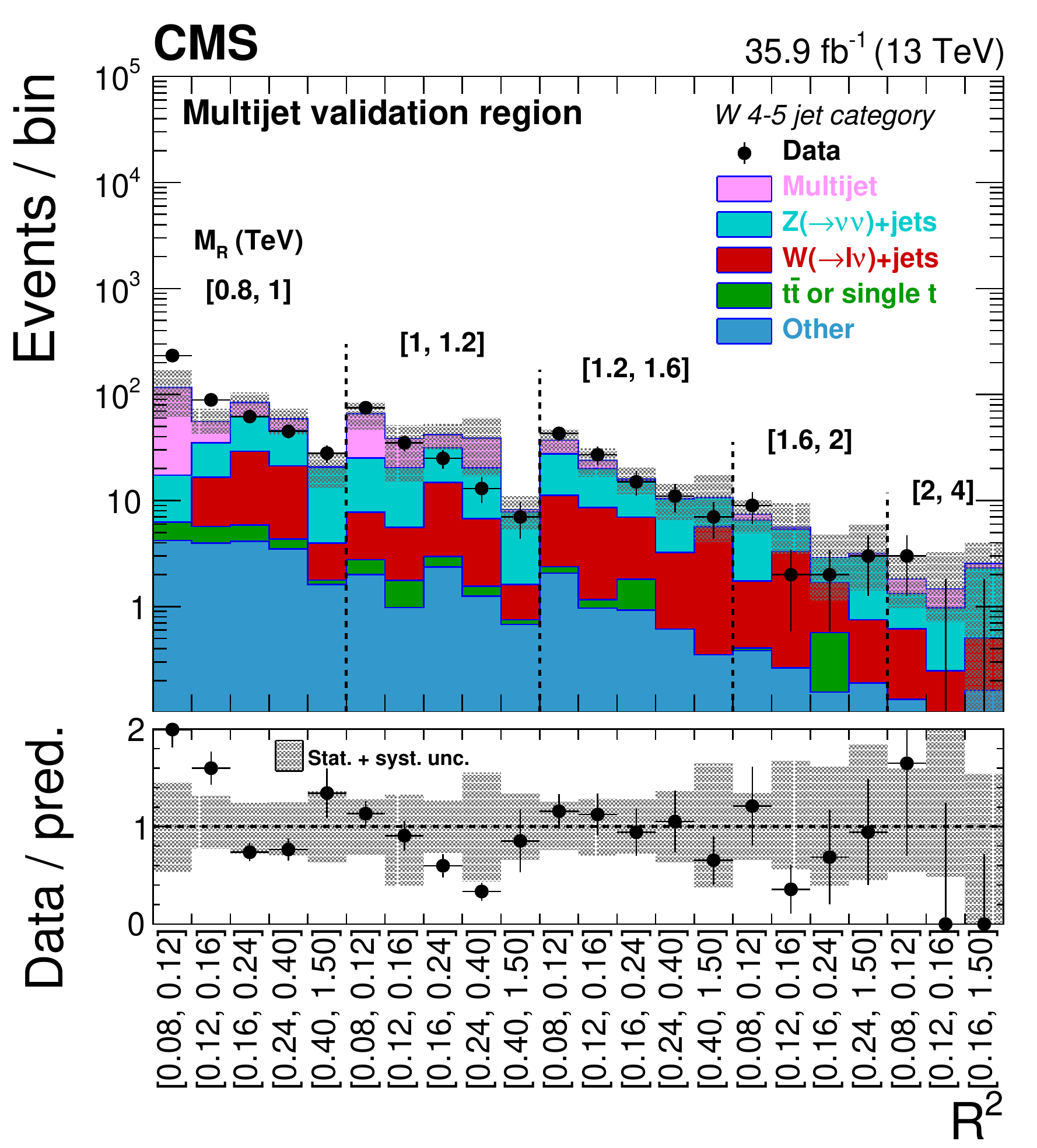}
\includegraphics[width=0.49\textwidth]{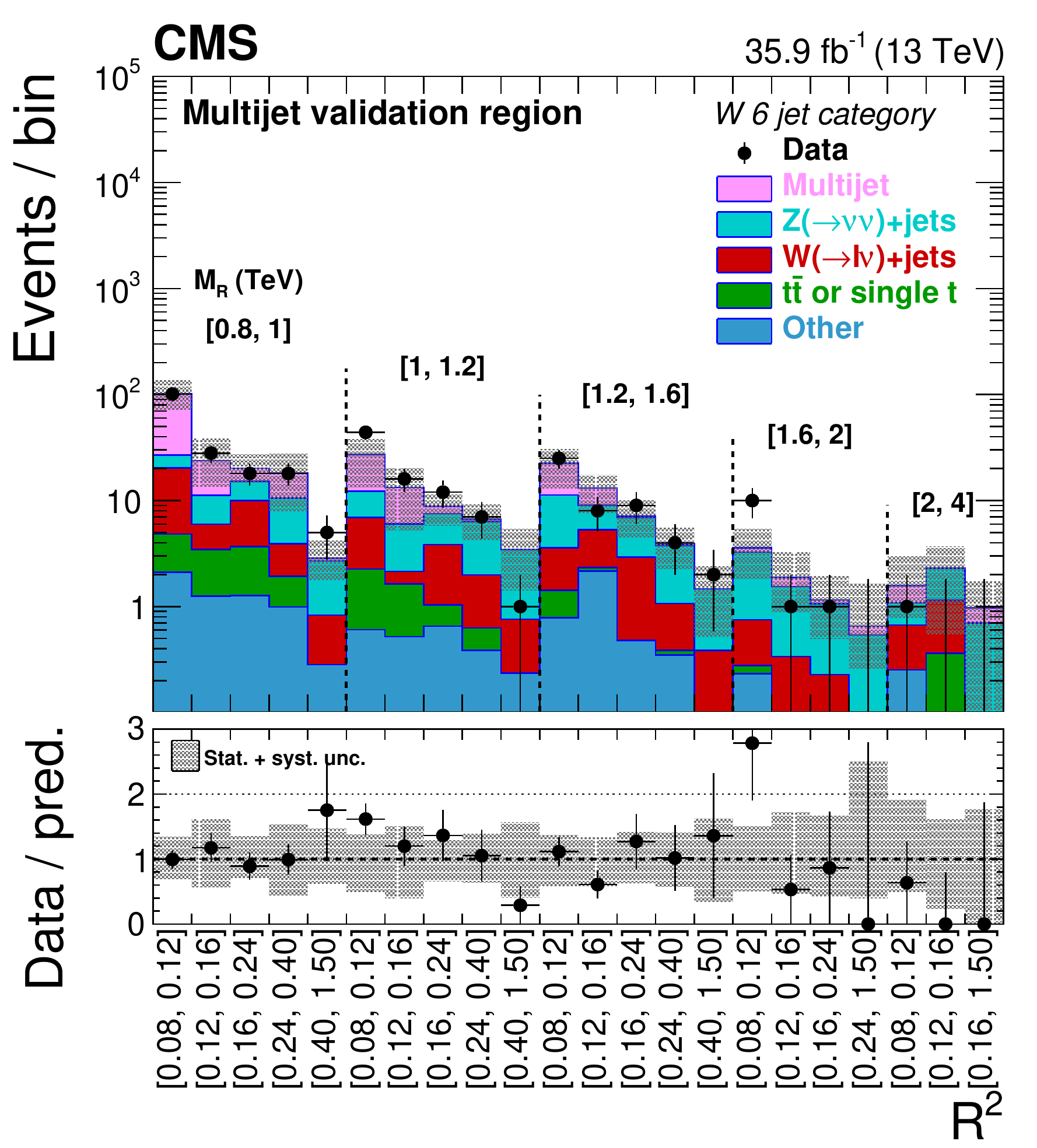} \\
\includegraphics[width=0.49\textwidth]{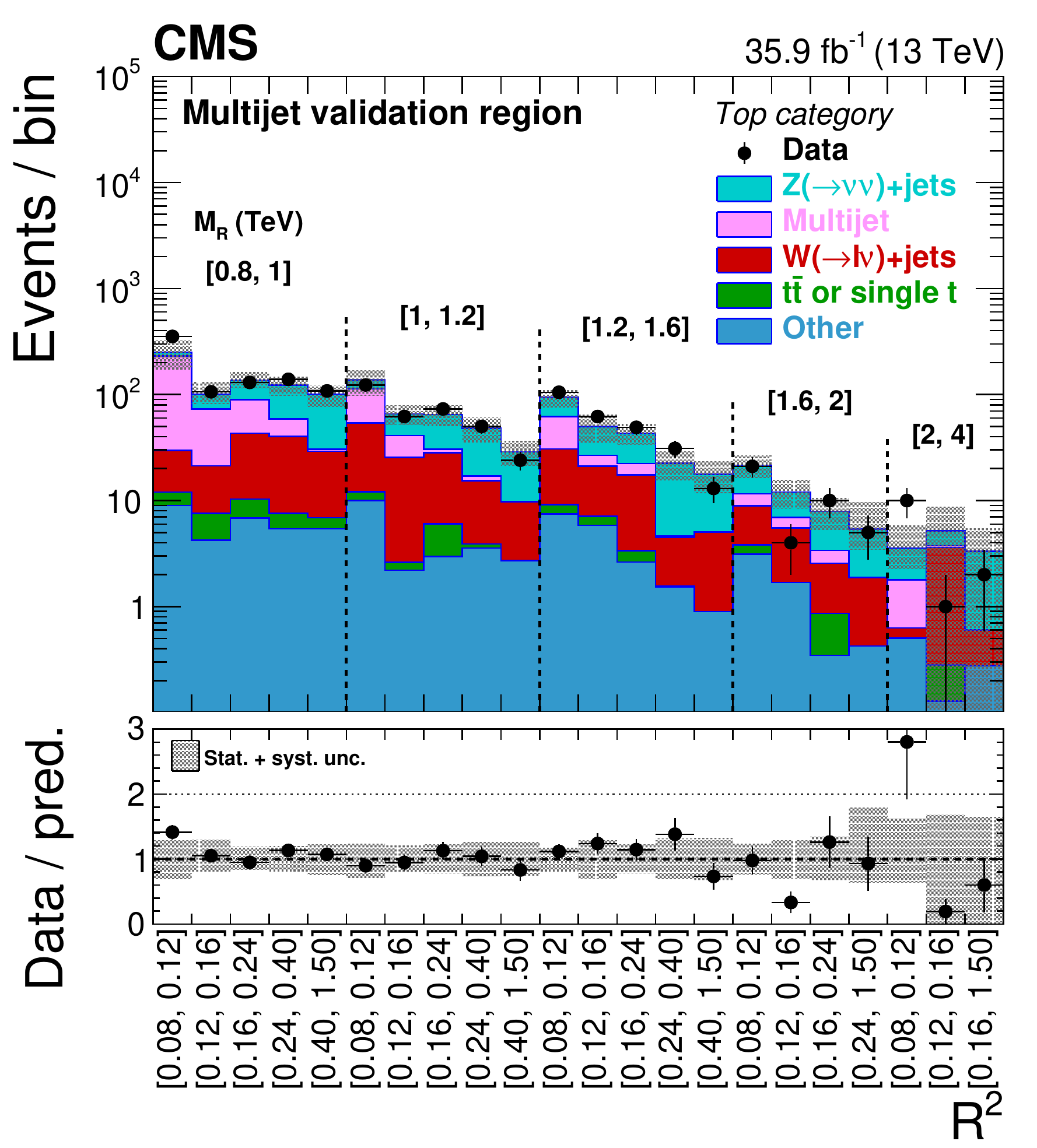}
\caption{Comparisons between data and the predicted background for the
validation region with antitagged \PW~boson or top quark candidates for the boosted $\PW$~4--5~jet (upper left),
boosted $\PW$~6~jet (upper right), and boosted top (lower) categories.
\label{fig:boostclosure2}}
\end{figure}

The second validation is performed in a CR defined identically to the SR
but requiring antitagged \PW~boson or top quark candidates. This validation is
designed to check the modeling of the $\dPhiR$ variable in the
QCD multijet and $\cPZ(\nu\overline{\nu})$+jets simulation. The plots in Fig.~\ref{fig:boostclosure2}
show the estimation results compared to data for the boosted $\PW$~4--5~jet, boosted $\PW$~6~jet,
and boosted top categories.  Overall, the estimation agrees with data within uncertainties.

\section{Systematic uncertainties}
\label{sec:Systematics}

Systematic uncertainties considered in this analysis can
be broadly categorized into three types: uncertainties from the limited accuracy of calibrations,
auxiliary measurements, and theoretical predictions; uncertainties
from the data-driven background prediction methodology; and uncertainties
specific to the fast simulation prediction of the signal.

Systematic uncertainties of the first type are propagated as shape
uncertainties in the signal and background predictions
in all event categories. Uncertainties in the trigger and lepton selection efficiency, and in the
integrated luminosity~\cite{CMS:2017sdi}, primarily affect the total normalization. Uncertainties in
the \cPqb~tagging efficiency affect the relative yields between different \cPqb~tag categories.
Systematic uncertainties in the modeling of the \PW~boson and top
quark tagging and mistagging efficiencies affect the yields of the
boosted categories. The uncertainties from missing higher-order corrections and the
uncertainties in the jet energy and lepton momentum scales affect the shapes of the $\MR$
and $\Rtwo$ distributions. In Table~\ref{tab:BackgroundSystematics} we summarize
these systematic uncertainties and their typical impact on the background and signal
predictions.

\begin{table}[!tbp]
\centering
\topcaption{Summary of the main instrumental and theoretical systematic uncertainties.}
\label{tab:BackgroundSystematics}
\begin{tabular}{ccc}
\hline
\multirow{2}{*}{Systematic uncertainty source}  & \multirow{2}{*}{On signal and/or bkg}   & Typical impact of     \\
                                                &                                         & uncertainty on yields ($\%$) \\
\hline
Jet energy scale                   & Both                   & $6$--$16$ \\
Lepton momentum scale              & Both                   & $1$ \\
Muon efficiency                    & Both                   & $1$ \\
Electron efficiency                & Both                   & $1$--$2$ \\
Trigger efficiency                 & Both                   & $1$ \\
\PQb-tagging  efficiency           & Both                   & $1$--$7$ \\
\PQb~mistagging  efficiency        & Both                   & $2$--$20$ \\
\PW$/$\cPqt-tagging  efficiency        & Both                   & $1$--$8$ \\
\PW$/$\cPqt-mistagging  efficiency     & Both                   & $1$--$3$ \\
Higher-order corrections           & Both                   & $10$--$25$ \\
Luminosity                         & Both                   & $2.6$ \\
Pileup                             & Both                   & $1$--$3$ \\
Monte Carlo event count            & Both                   & $1$--$50$ \\
Fast simulation corrections        & Signal only            & $1$--$5$ \\
Initial-state radiation            & Signal only            & $4 $--$ 25$ \\
\hline
\end{tabular}
\end{table}

The second type of systematic uncertainty is related to the
background prediction methodology.
Statistical uncertainties
of the CR data range from $1$--$20\%$ depending on the
$\MR$ and $\Rtwo$ bin. Systematic uncertainties of the
background processes that we are not targeting in each
CR contribute at the level of a few percent.
Systematic uncertainties related to the accuracy of
assumptions made by the background estimation method
are estimated through closure tests in different
CRs as discussed in
Section~\ref{sec:Background}. These systematic uncertainties capture the potential modeling
inadequacies of the simulation after applying the corrections
derived as part of the analysis procedure. They are summarized
in Table~\ref{tab:MultiJetMRClosureTestSystematics}.

For the closure tests performed in each $N_{\text{jets}}$ bin
in the $\ttbar$ dilepton and the $\cPZ(\nu\overline{\nu})$+jets
dilepton CRs, and the test of the
$\pt$ distributions in the loose lepton and $\tauh$ CRs,
the uncertainties are applied correlated across all bins.
For the checks of the $\Rtwo$ distributions in each \cPqb~tag category
in the tight lepton and photon CRs, and of
the lepton $\eta$ distributions in the loose lepton and
$\tauh$ CRs, the systematic uncertainties are assigned
based on the size of the statistical uncertainty in the
CRs and are assumed to be uncorrelated from bin to bin.

For the $\cPZ(\nu\overline{\nu})$+jets process, the difference in the correction factors
computed in the $\cPgg$+jets and tight lepton CRs are propagated
as a systematic uncertainty. This systematic uncertainty
estimates the potential differences in the MC mismodeling of the hadronic recoil
between the $\cPgg$+jets process and the $\cPZ(\nu\overline{\nu})$+jets process.
These systematic uncertainties range up to $20\%$.

\begin{table}[!ht]
\centering
\topcaption{Summary of systematic uncertainties from
the background estimation methodology expressed as relative or fractional uncertainties.}
\label{tab:MultiJetMRClosureTestSystematics}
\begin{tabular}{ccc}
\hline
Uncertainty source & Background process & Size ($\%$) \\
\hline
\multicolumn{3}{c}{Non-Boosted categories} \\
1-lepton CR, $\Rtwo$ closure test & $\ttbar$, $\PW$+jets & $1$--$95$ \\
$\ttbar$ 2-lepton closure test & $\ttbar$ & $1$--$12$ \\
Loose lepton $\pt$ closure test & $\ttbar$, $\PW$+jets & $4$--$50$ \\
Loose lepton $\eta$ closure test & $\ttbar$, $\PW$+jets & $5$--$40$ \\
$\tauh$ $\pt$ closure test & $\ttbar$, $\PW$+jets & $2$--$43$ \\
$\tauh$ $\eta$ closure test & $\ttbar$, $\PW$+jets & $2$--$28$ \\
$\cPgg$+jets CR, transfer factor uncertainty and $\Rtwo$ closure test & $\cPZ(\nu\overline{\nu})$+jets & $1$--$40$ \\
DY+jets 2-lepton closure test & $\cPZ(\nu\overline{\nu})$+jets & $1$--$25$ \\
QCD multijet transfer factor extrapolation & QCD multijet & $30$--$90$ \\[\cmsTabSkip]

\multicolumn{3}{c}{Boosted categories} \\
QCD multijet modeling     & QCD multijet & $  13$--$24$ \\
DY+jets modeling          & $\cPZ(\nu\overline{\nu})$+jets & $19$--$29$ \\
$\cPZ(\nu\overline{\nu})$+jets closure test & $\cPZ(\nu\overline{\nu})$+jets & $19$--$98$ \\
\hline

\hline
\end{tabular}
\end{table}

Finally, there are systematic uncertainties specific to the
fast simulation prediction of the signal.
These include systematic uncertainties because of
possible inaccuracies of the fast simulation in modeling the efficiencies for
lepton selection, \cPqb~tagging, and boosted \PW~boson and top quark tagging.
To account for possible mismodeling of the signal acceptance because of differences
in the data and signal MC pileup distributions, we employ a linear fit
that extrapolates the acceptance in each analysis bin to the range of
pileup values observed in data.  Uncertainty in this method is propagated
to the signal yield predictions. An additional uncertainty is applied to account
for known tendencies for the fast simulation to mismodel the $\ptmiss$
in some events. Finally, we propagate an uncertainty in the modeling of the ISR
for signal predictions, ranging from $4$--$25\%$ depending on the
number of jets from ISR.

\section{Results and interpretation}
\label{sec:Results}

\begin{figure}[!tbp] \centering
\includegraphics[width=0.75\textwidth]{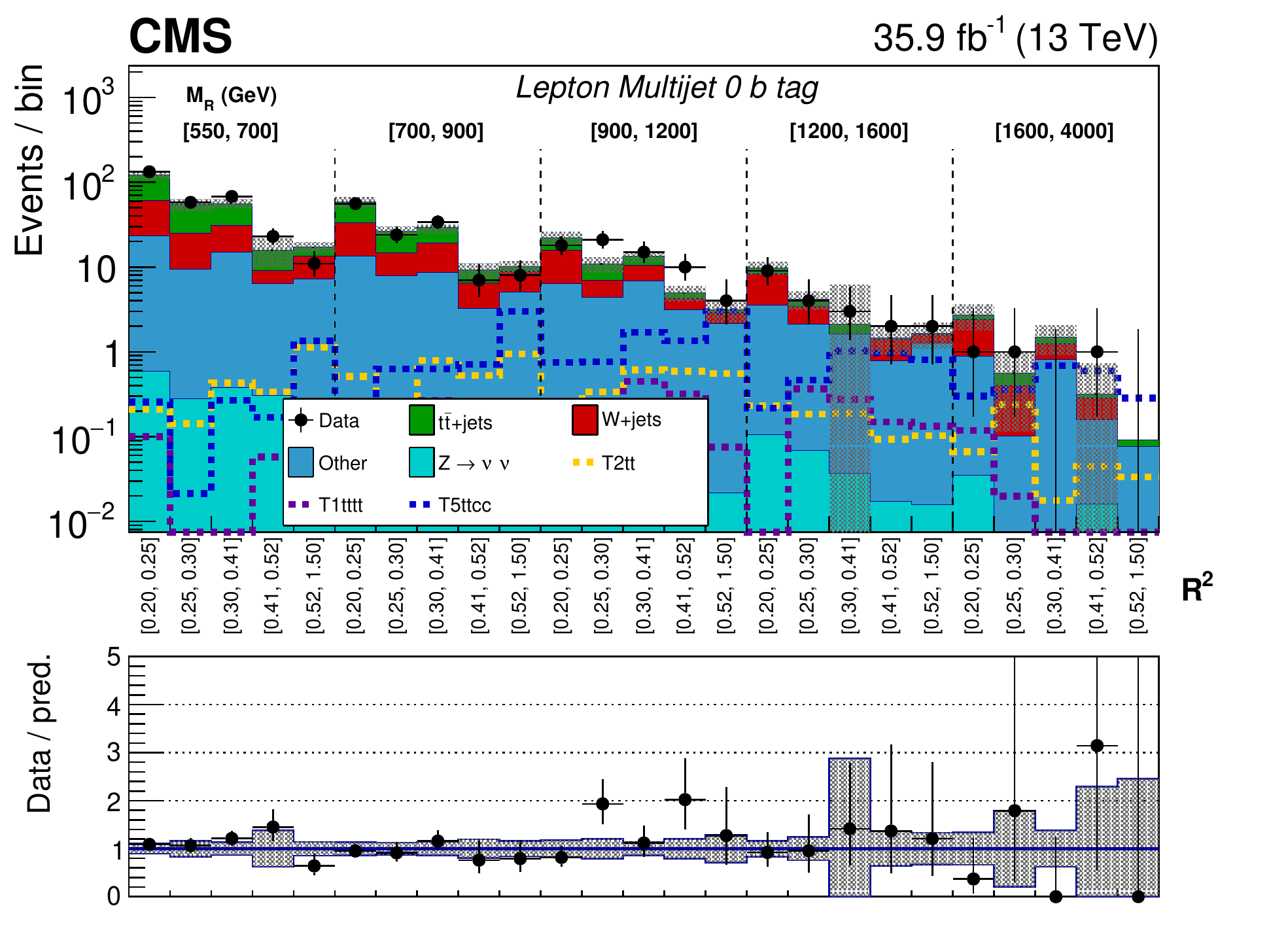}\\
\includegraphics[width=0.75\textwidth]{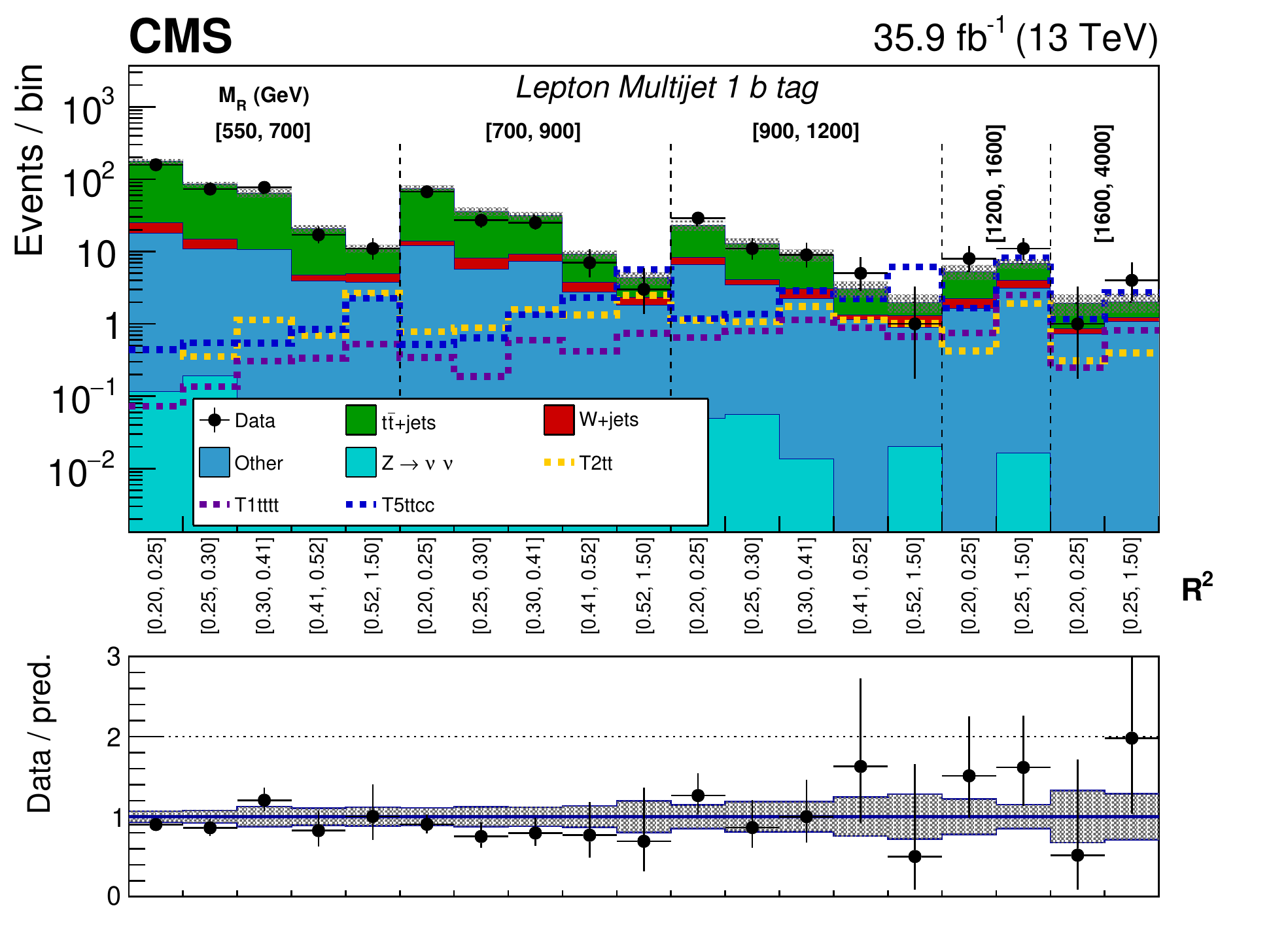}
\caption{ The $\MR$--$\Rtwo$ distribution observed in data is shown along with the pre-fit background prediction
obtained for the Lepton Multijet event category in the 0~\PQb~tag (upper) and 1~\PQb~tag (lower) bins. The two-dimensional $\MR$--$\Rtwo$ distribution is shown
in a one-dimensional representation, with each $\MR$ bin denoted by the dashed lines and labeled above,
and each $\Rtwo$ bin labeled below. The background labeled as ``Other'' includes single top quark
production, diboson production, associated production of a top quark pair and a \PW or \PZ~boson, and
triboson production. The ratio of data to the background prediction is shown on the bottom panel, with
the statistical uncertainty expressed through the data point error bars and the systematic uncertainty in the
background prediction represented by the shaded region. Signal benchmarks shown
are T5ttcc with $m_{\PSg} = 1.4\TeV$, $m_{\PSQt} = 320\GeV$ and $m_{\PSGczDo} = 300\GeV$;
T1tttt with $m_{\PSg} = 1.4\TeV$ and $m_{\PSGczDo} = 300\GeV$; and T2tt with
$m_{\PSQt} = 850\GeV$ and $m_{\PSGczDo} = 100\GeV$. The diagrams corresponding to these
signal models are shown in Fig.~\ref{fig:FeynmanDiagrams}.
}
\label{fig:ResultsLeptonMultiJet0b1b}
\end{figure}

\begin{figure}[!tbp] \centering
\includegraphics[width=0.8\textwidth]{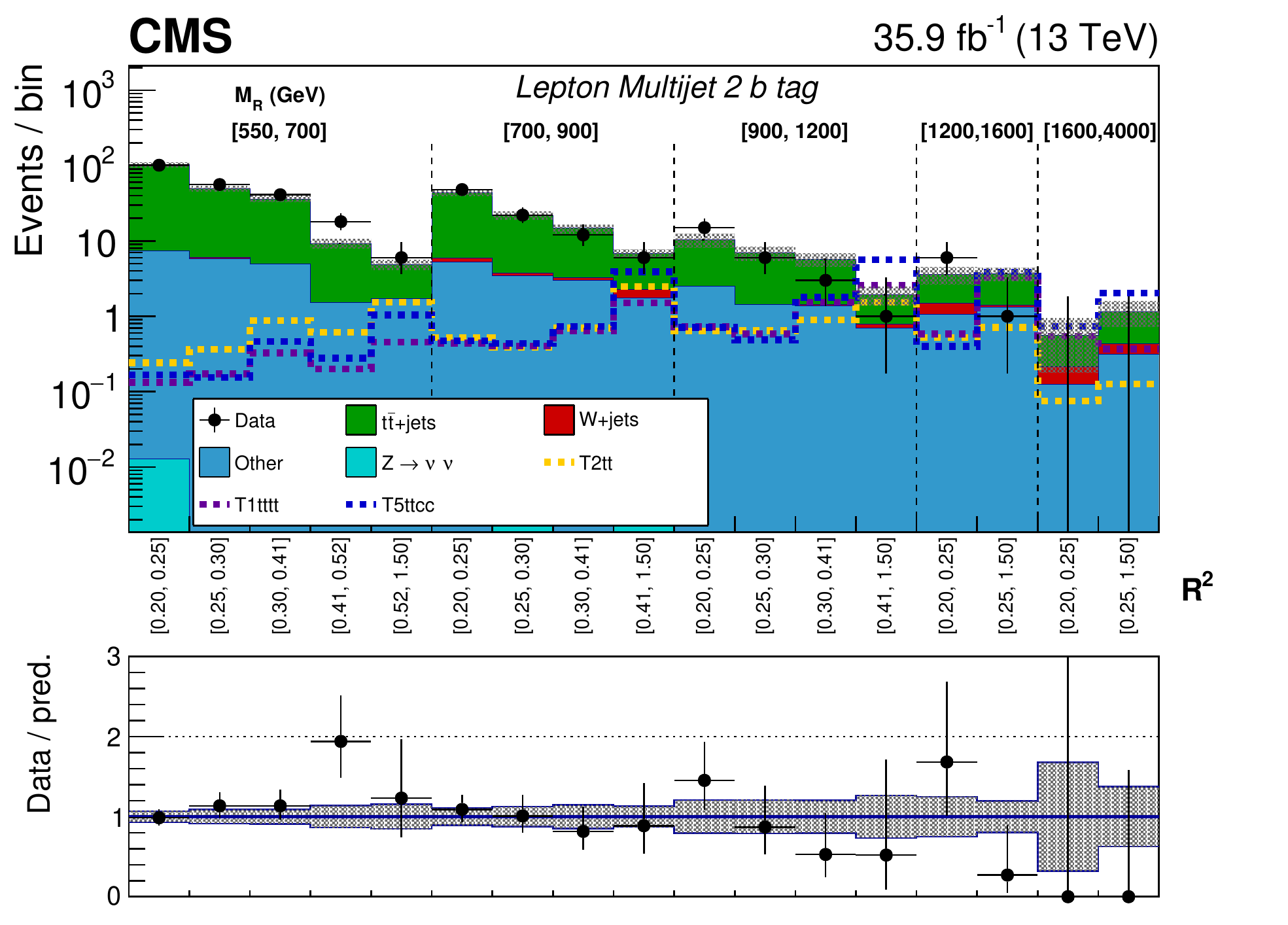}\\
\includegraphics[width=0.8\textwidth]{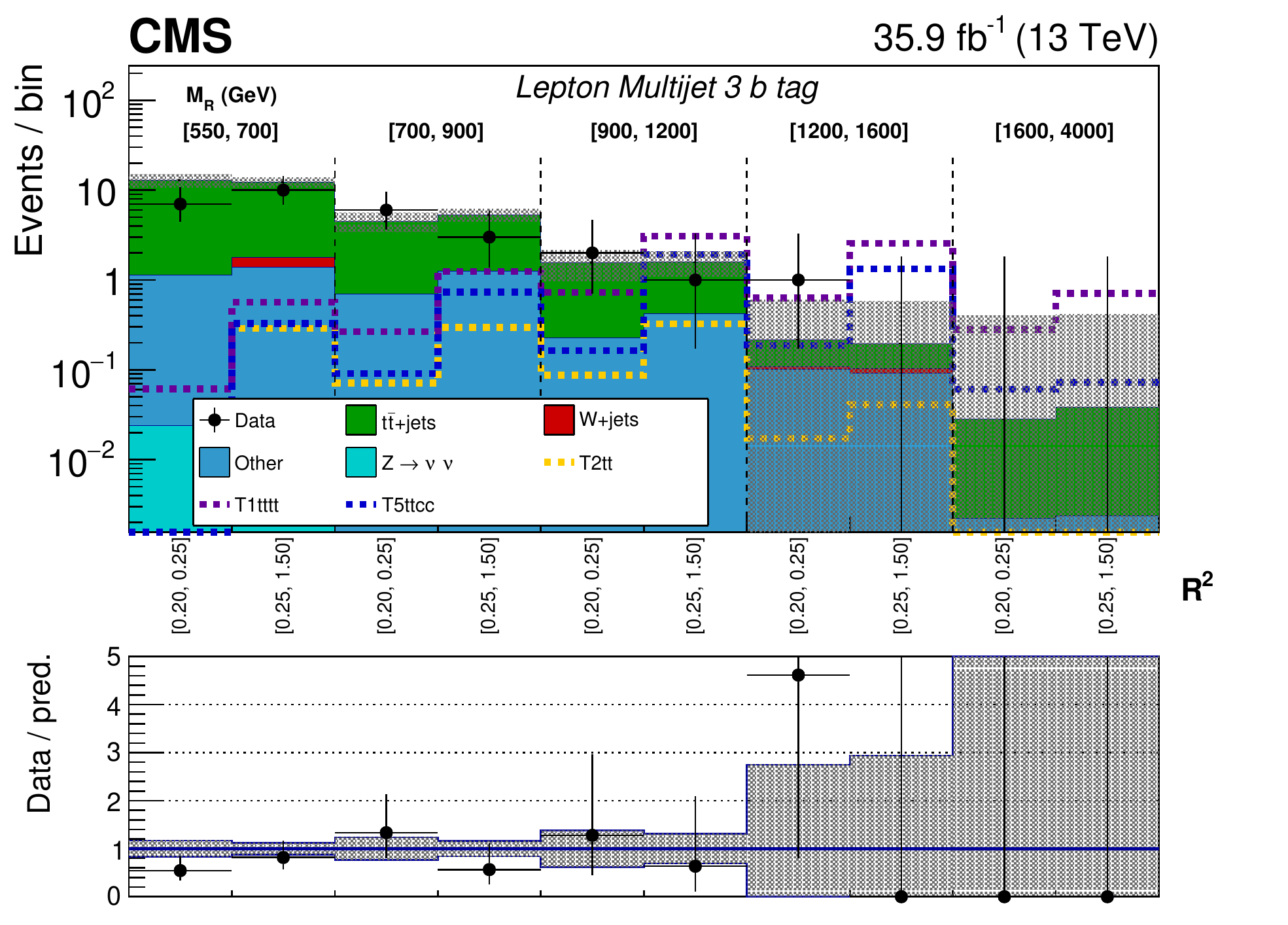}
\caption{ The $\MR$--$\Rtwo$ distribution observed in data is
shown along with the pre-fit background prediction
obtained for the Lepton Multijet event category in the 2
\PQb~tag (upper) and 3 or more \PQb~tag (lower) bins. Further details of the plots are explained in the caption of Fig.~\ref{fig:ResultsLeptonMultiJet0b1b}.
}
\label{fig:ResultsLeptonMultiJet2b3b}
\end{figure}

\begin{figure}[!tbp] \centering
\includegraphics[width=0.8\textwidth]{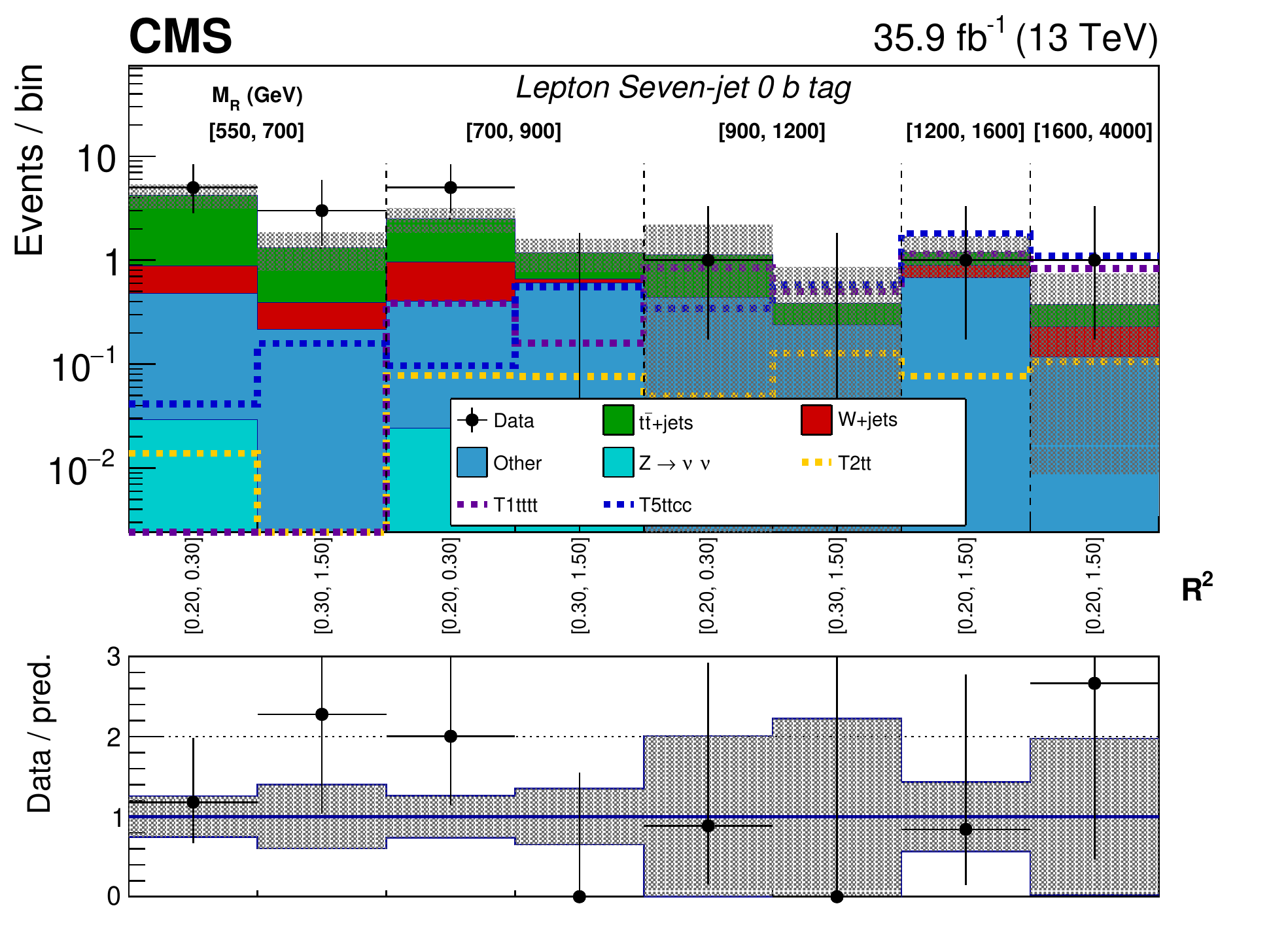}\\
\includegraphics[width=0.8\textwidth]{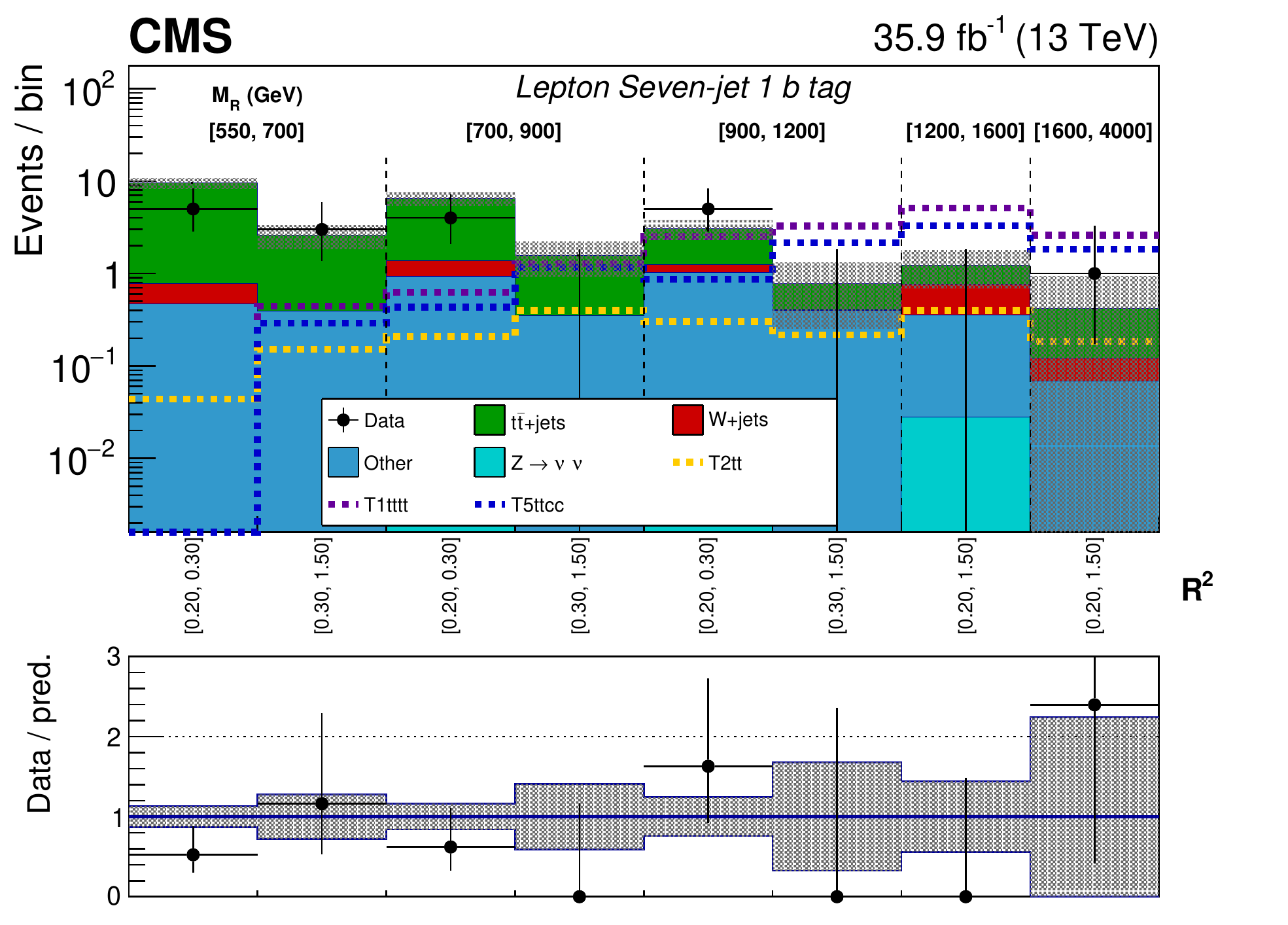}
\caption{ The $\MR$--$\Rtwo$ distribution observed in data is shown along with the pre-fit background prediction
obtained for the Lepton Seven-jet event category in the 0
\PQb~tag (upper) and 1 \PQb~tag (lower) bins. Further details of the plots are explained in the caption of Fig.~\ref{fig:ResultsLeptonMultiJet0b1b}.
}
\label{fig:ResultsLeptonSevenJet0b1b}
\end{figure}

\begin{figure}[!tbp] \centering
\includegraphics[width=0.8\textwidth]{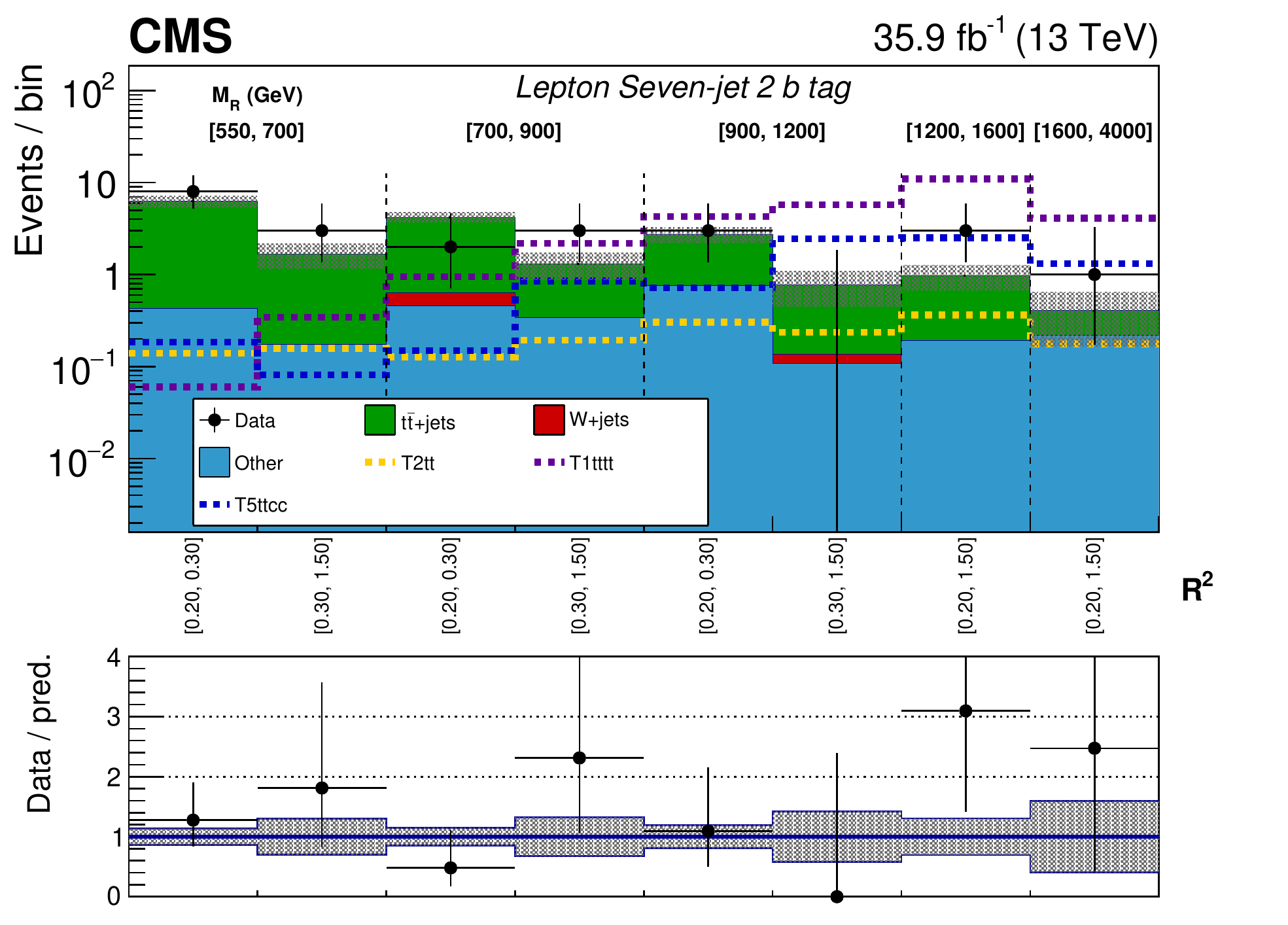}\\
\includegraphics[width=0.8\textwidth]{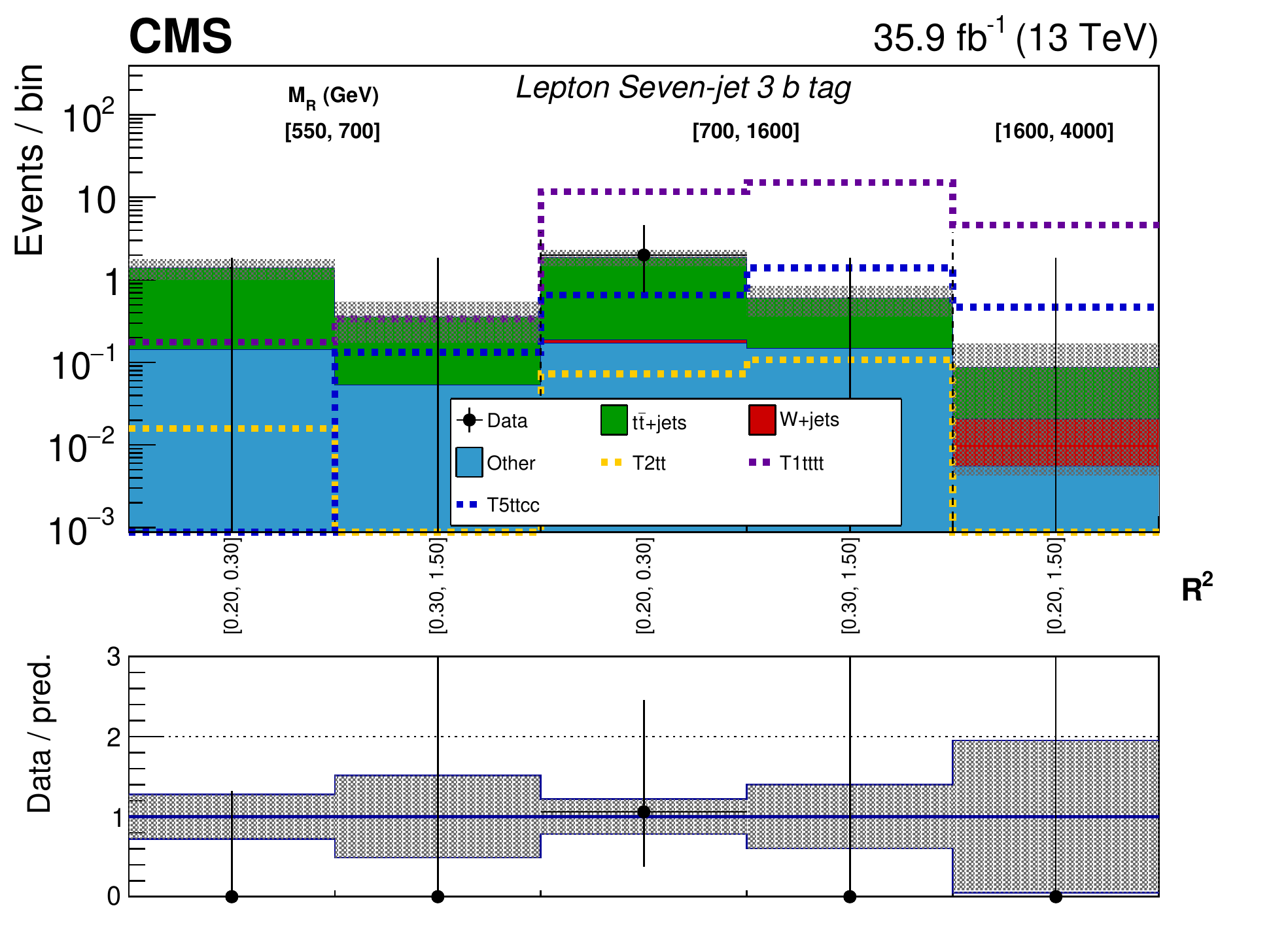}
\caption{ The $\MR$--$\Rtwo$ distribution observed in data is shown along with the pre-fit background prediction
obtained for the Lepton Seven-jet event category in the 2
\PQb~tag (upper) and 3 or more  \PQb~tag (lower) bins. Further details of the plots are explained in the caption of Fig.~\ref{fig:ResultsLeptonMultiJet0b1b}.
}
\label{fig:ResultsLeptonSevenJet2b3b}
\end{figure}

The observed data yields in the SRs are compatible
with the background prediction from SM processes. The results are summarized
in the distributions of the $\MR$ and $\Rtwo$ bins of the SRs.
The results for the one-lepton categories are shown in
Figs.~\ref{fig:ResultsLeptonMultiJet0b1b}--\ref{fig:ResultsLeptonSevenJet2b3b}.
The main backgrounds are $\PW$+jets and $\ttbar$ production, with $\ttbar$
becoming more dominant with increasing number of \cPqb-tagged jets.
The three signal scenarios used to interpret the results are also shown.

\begin{figure}[!tbp] \centering
\centering
\includegraphics[width=0.49\textwidth]{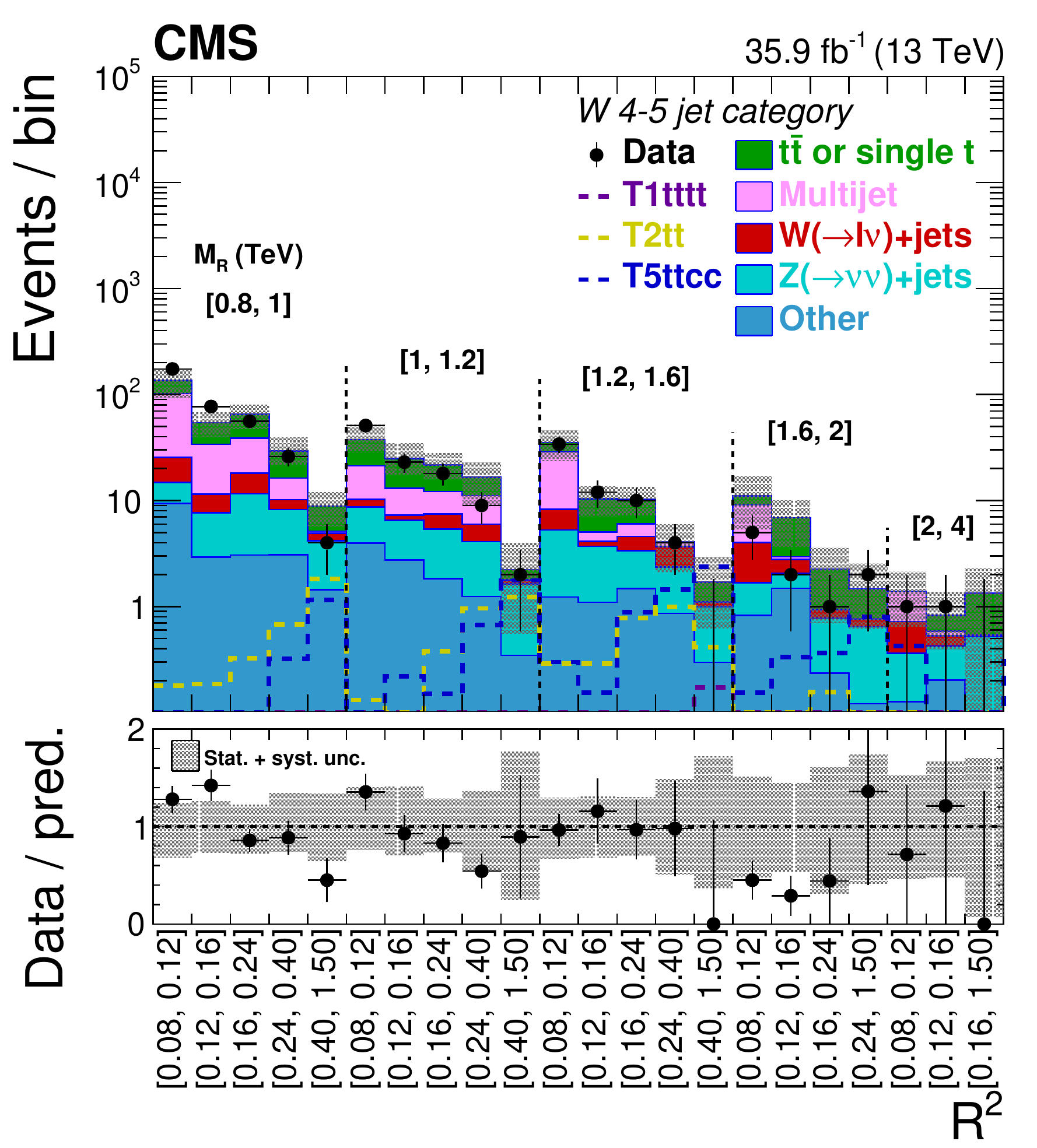}
\includegraphics[width=0.49\textwidth]{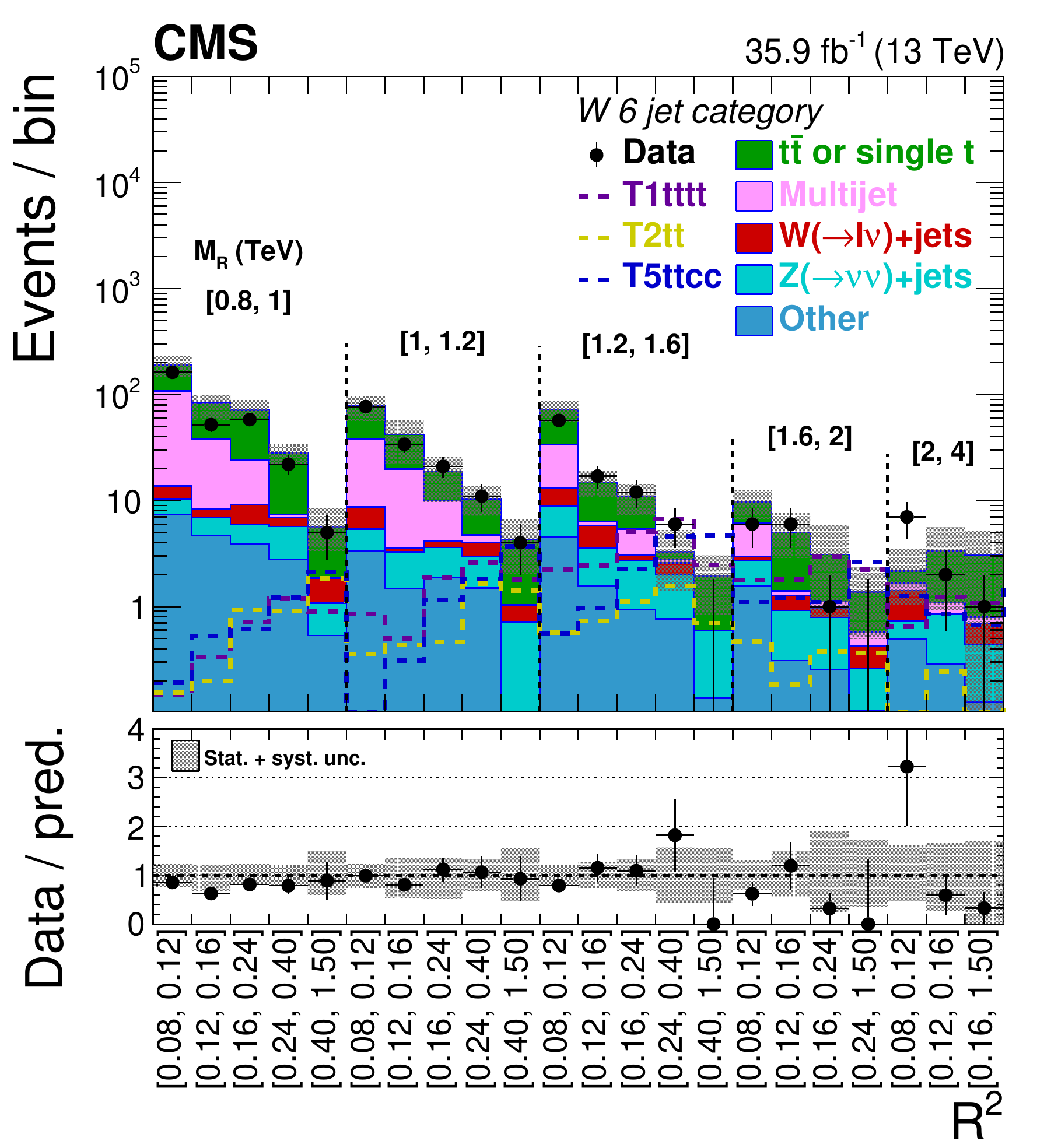} \\
\includegraphics[width=0.49\textwidth]{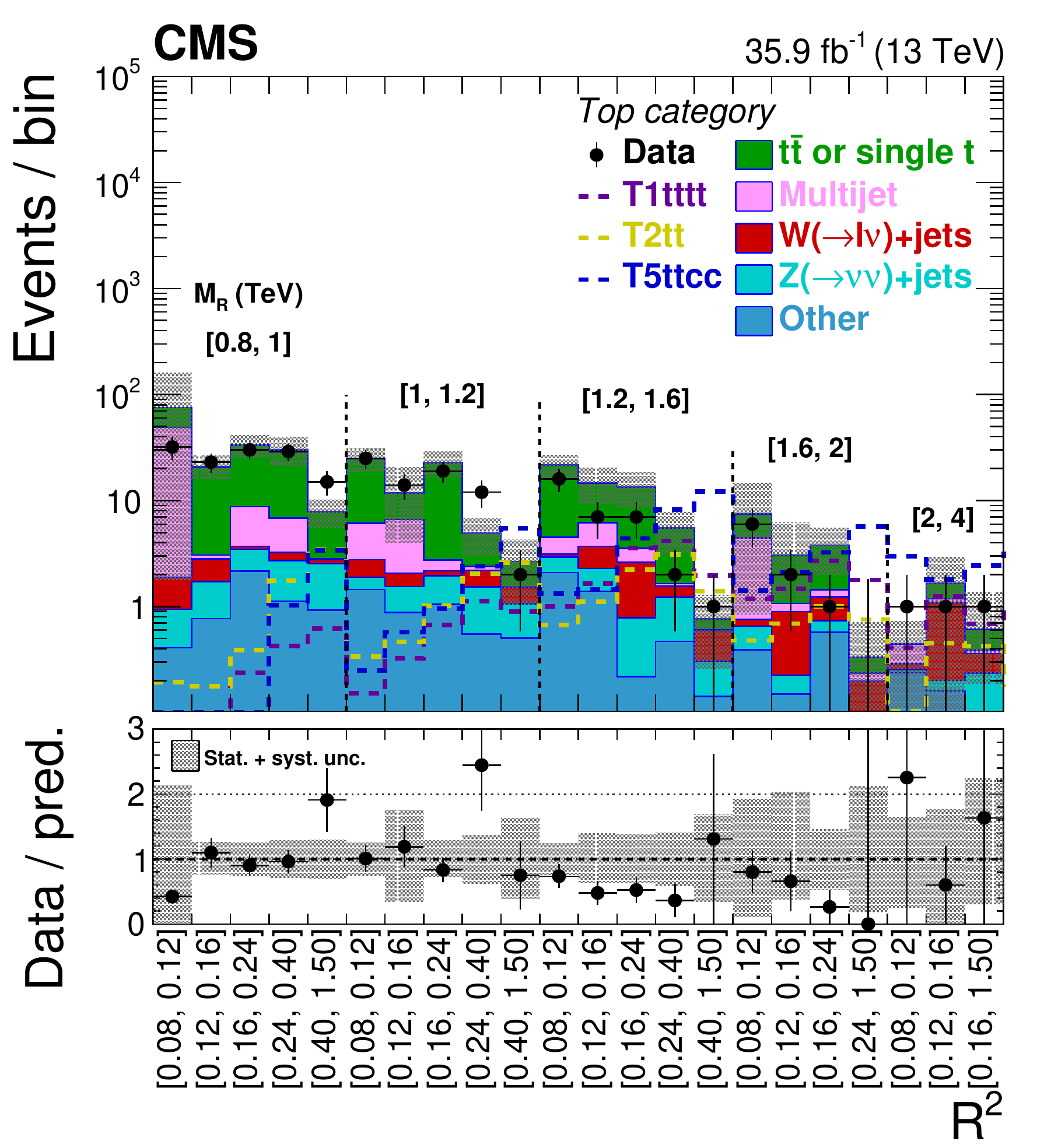}
\caption{ The $\MR$--$\Rtwo$ distribution observed in data is shown along with the pre-fit background prediction
obtained for the boosted $\PW$ 4--5 jet (upper left), boosted $\PW$ 6 jet (upper right),
and Top (lower) categories. Further details of the plots are explained in the
caption of Fig.~\ref{fig:ResultsLeptonMultiJet0b1b}.
}
\label{fig:ResultsBoost}
\end{figure}

The results for the zero-lepton boosted categories are shown in
Fig.~\ref{fig:ResultsBoost}, where $\ttbar$ is the dominant
background process in all subcategories.

\begin{figure}[!tbp] \centering
\includegraphics[width=0.8\textwidth]{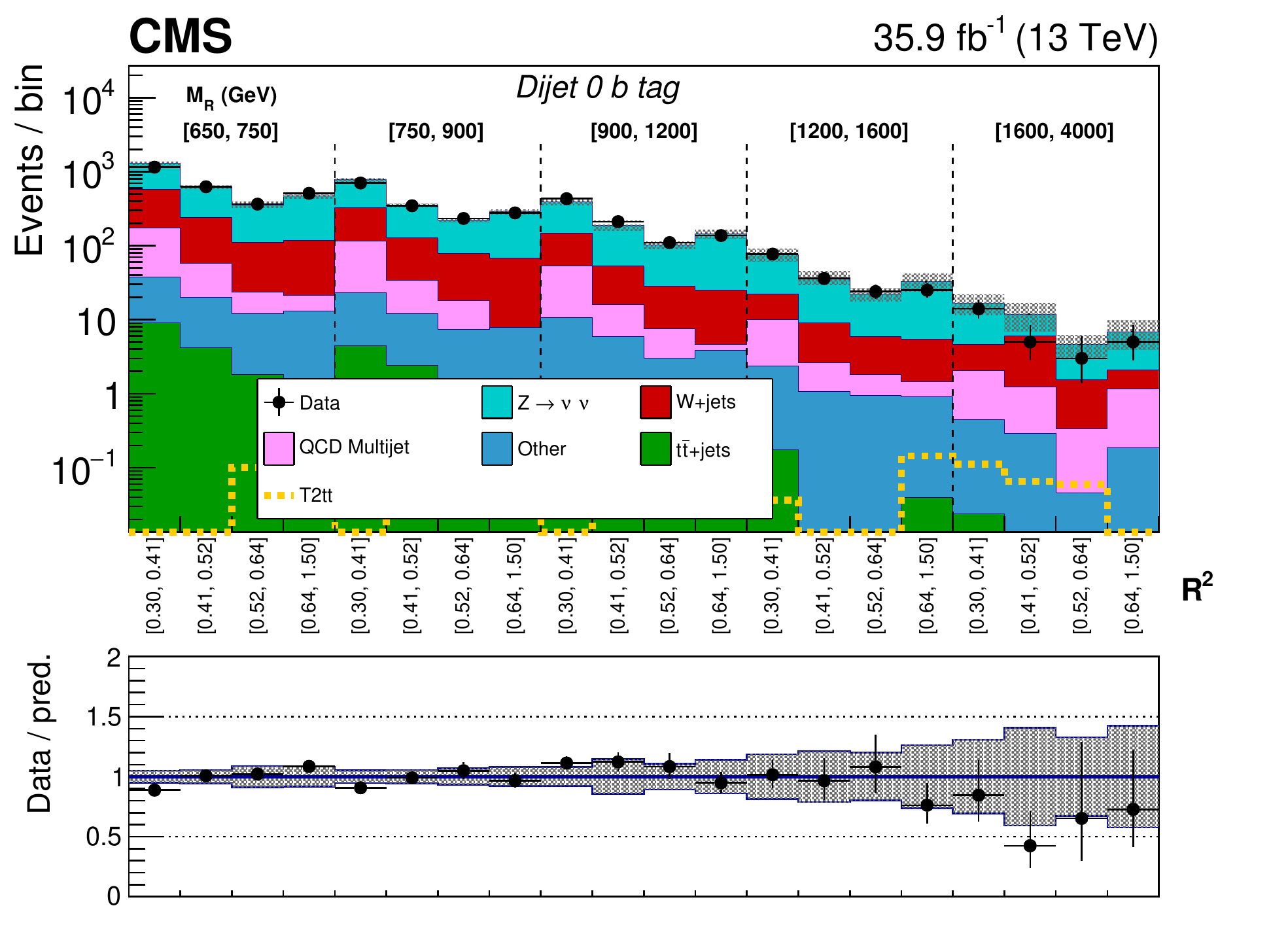}\\
\includegraphics[width=0.8\textwidth]{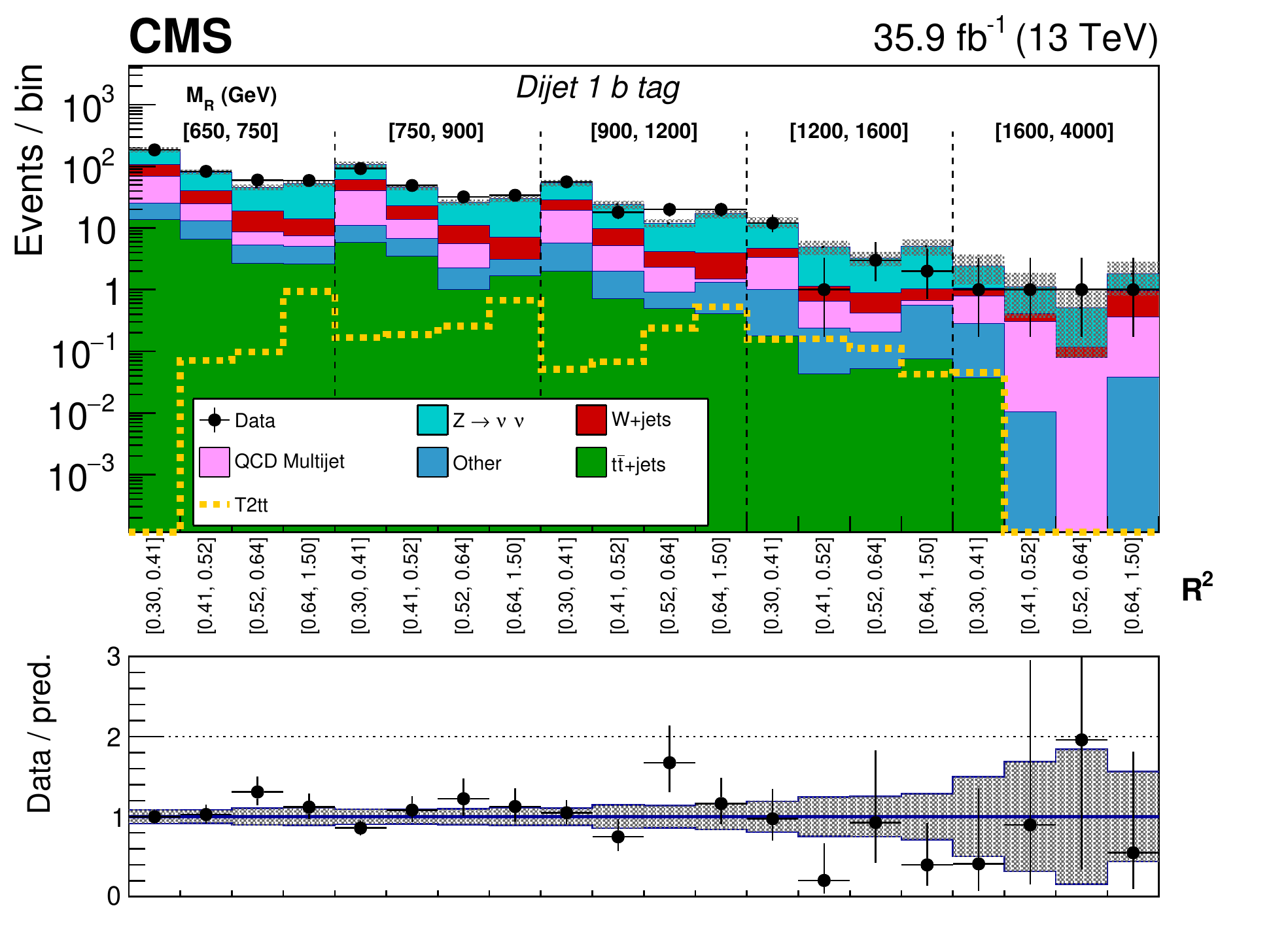}
\caption{ The $\MR$--$\Rtwo$ distribution observed in data is shown along with the pre-fit background prediction
obtained for the Dijet event category in the 0
\PQb~tag (upper) and 1 \PQb~tag (lower) bins.
Further details of the plots are explained in the caption of Fig.~\ref{fig:ResultsLeptonMultiJet0b1b}.
}
\label{fig:ResultsDiJet0b1b}
\end{figure}

\begin{figure}[!tbp] \centering
\includegraphics[width=0.8\textwidth]{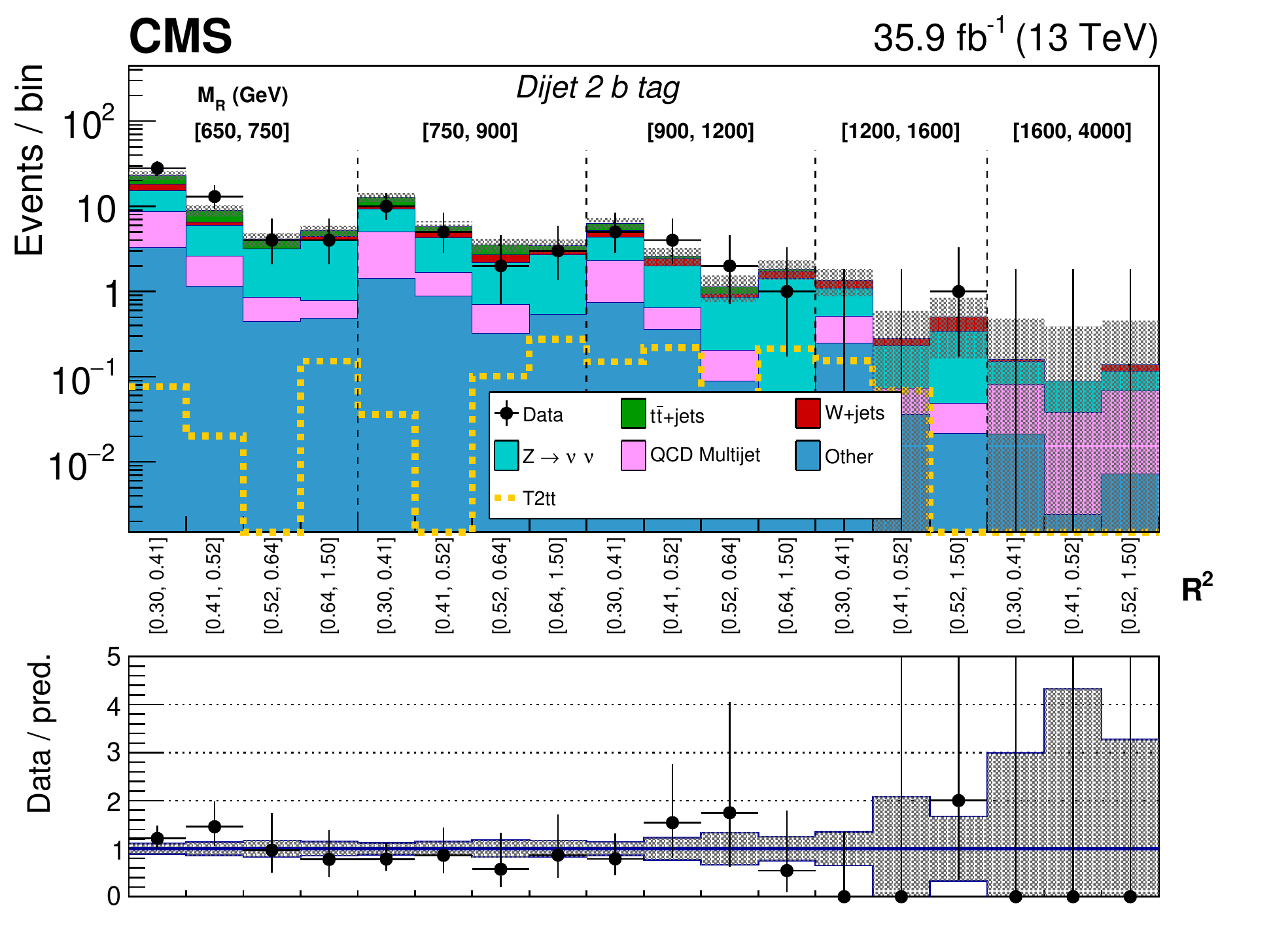}
\caption{ The $\MR$--$\Rtwo$ distribution observed in data is shown along with the pre-fit background prediction
obtained for the Dijet event category in the 2 or more \PQb~tag bin.
Further details of the plots are explained in the caption of Fig.~\ref{fig:ResultsLeptonMultiJet0b1b}.
}
\label{fig:ResultsDiJet2b}
\end{figure}

\begin{figure}[!tbp] \centering
\includegraphics[width=0.8\textwidth]{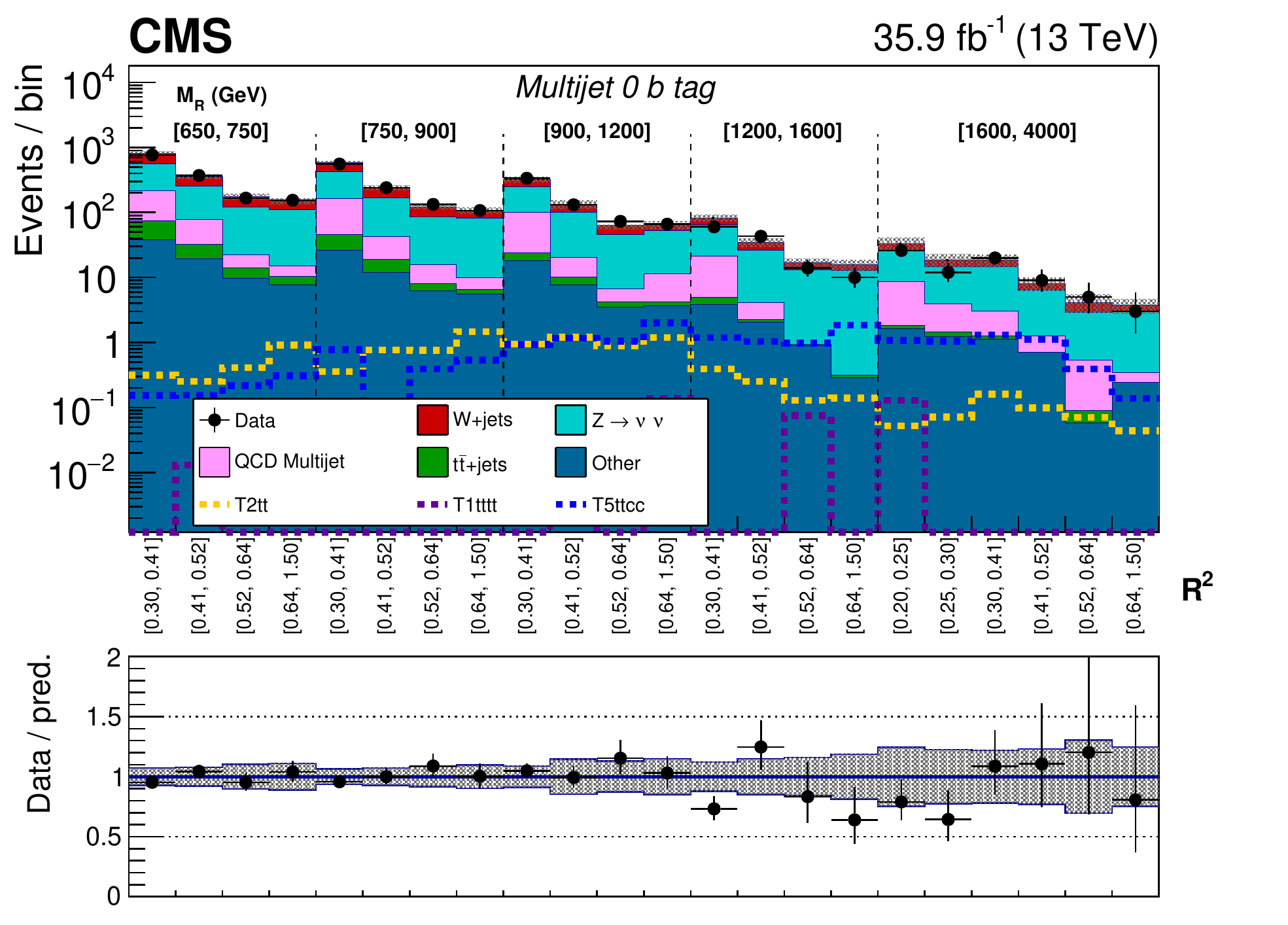}\\
\includegraphics[width=0.8\textwidth]{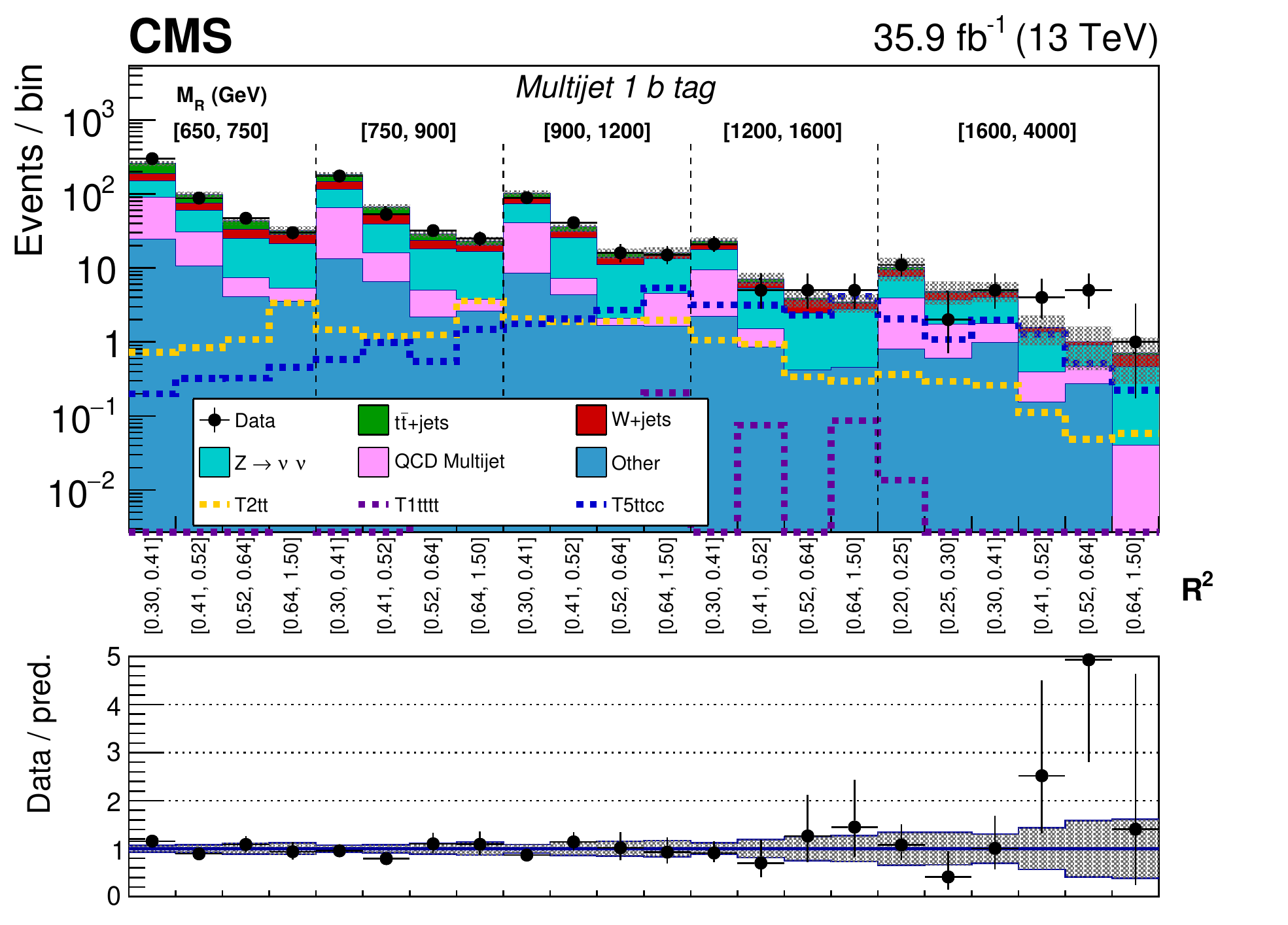}
\caption{ The $\MR$--$\Rtwo$ distribution observed in data is shown along with the pre-fit background prediction
obtained for the Multijet event category in the 0
\PQb~tag (upper) and 1 \PQb~tag (lower) bins. Further details of the plots are explained in the caption of Fig.~\ref{fig:ResultsLeptonMultiJet0b1b}.
}
\label{fig:ResultsMultiJet0b1b}
\end{figure}

\begin{figure}[!tbp] \centering
\includegraphics[width=0.8\textwidth]{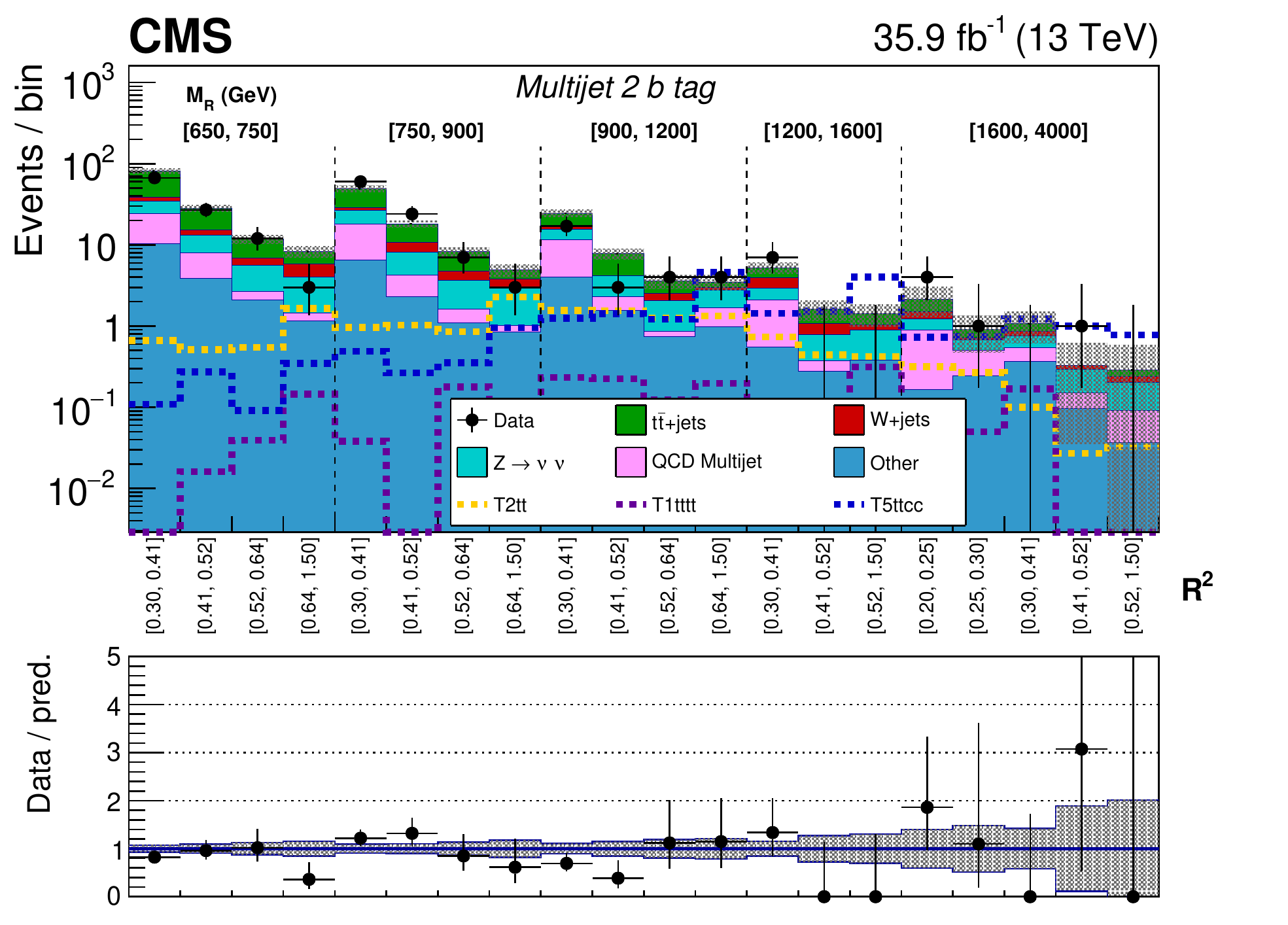}\\
\includegraphics[width=0.8\textwidth]{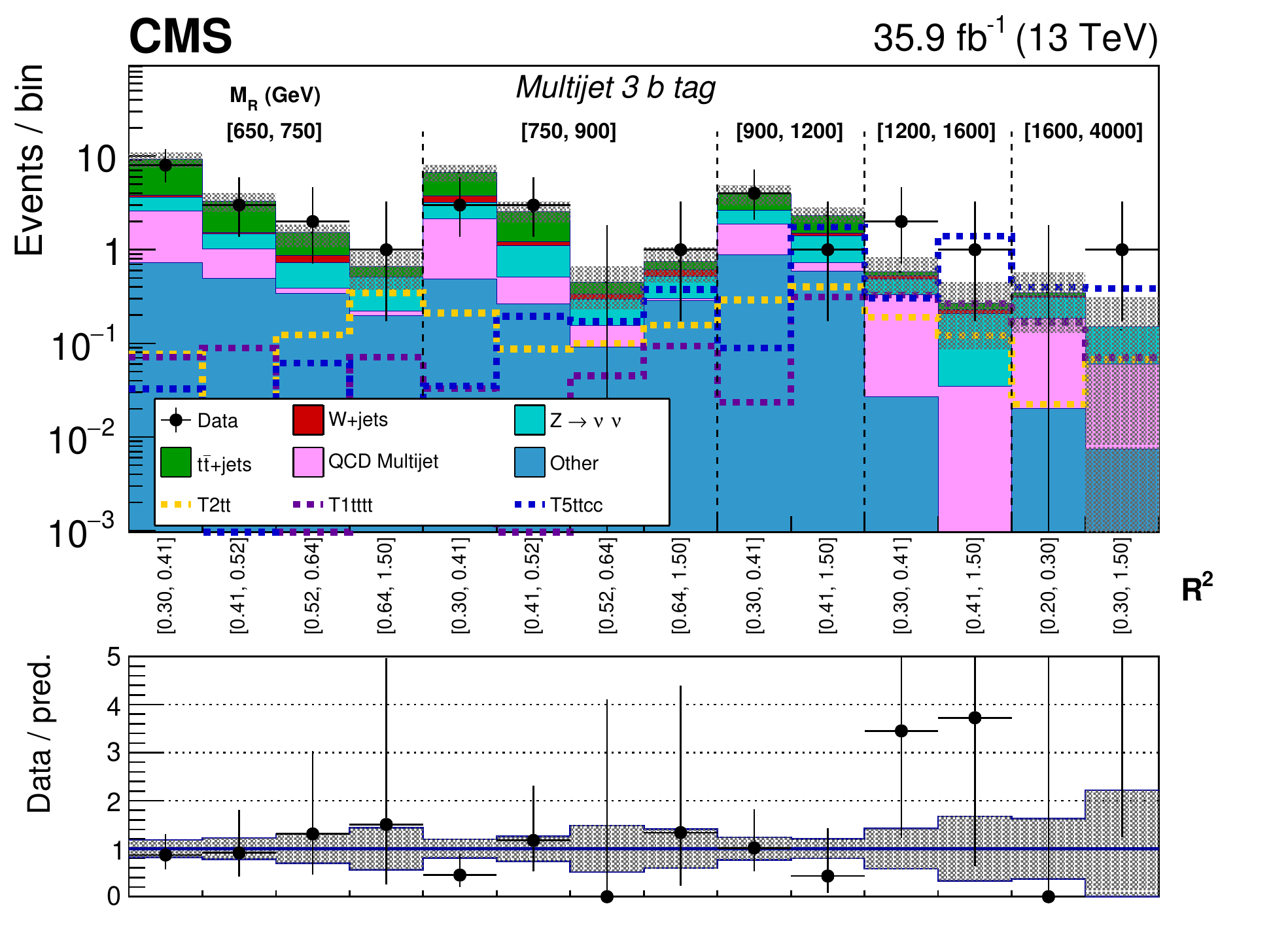}
\caption{ The $\MR$--$\Rtwo$ distribution observed in data is shown along with the pre-fit background prediction
obtained for the Multijet event category in the 2
\PQb~tag (upper) and 3 or more \PQb~tag (lower) bins. Further details of the plots are explained in the caption of Fig.~\ref{fig:ResultsLeptonMultiJet0b1b}.
}
\label{fig:ResultsMultiJet2b3b}
\end{figure}

\begin{figure}[!tbp] \centering
\includegraphics[width=0.8\textwidth]{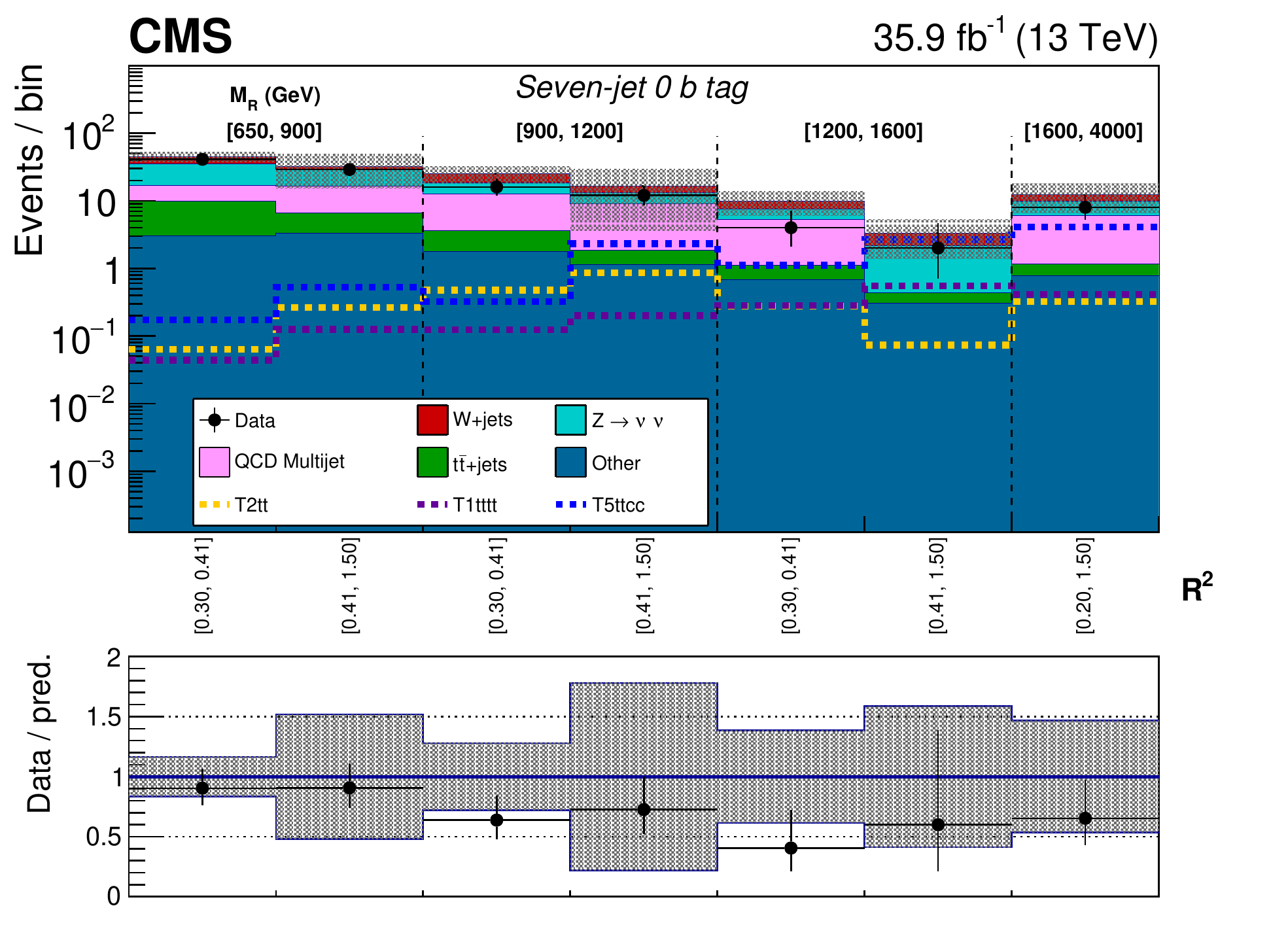}\\
\includegraphics[width=0.8\textwidth]{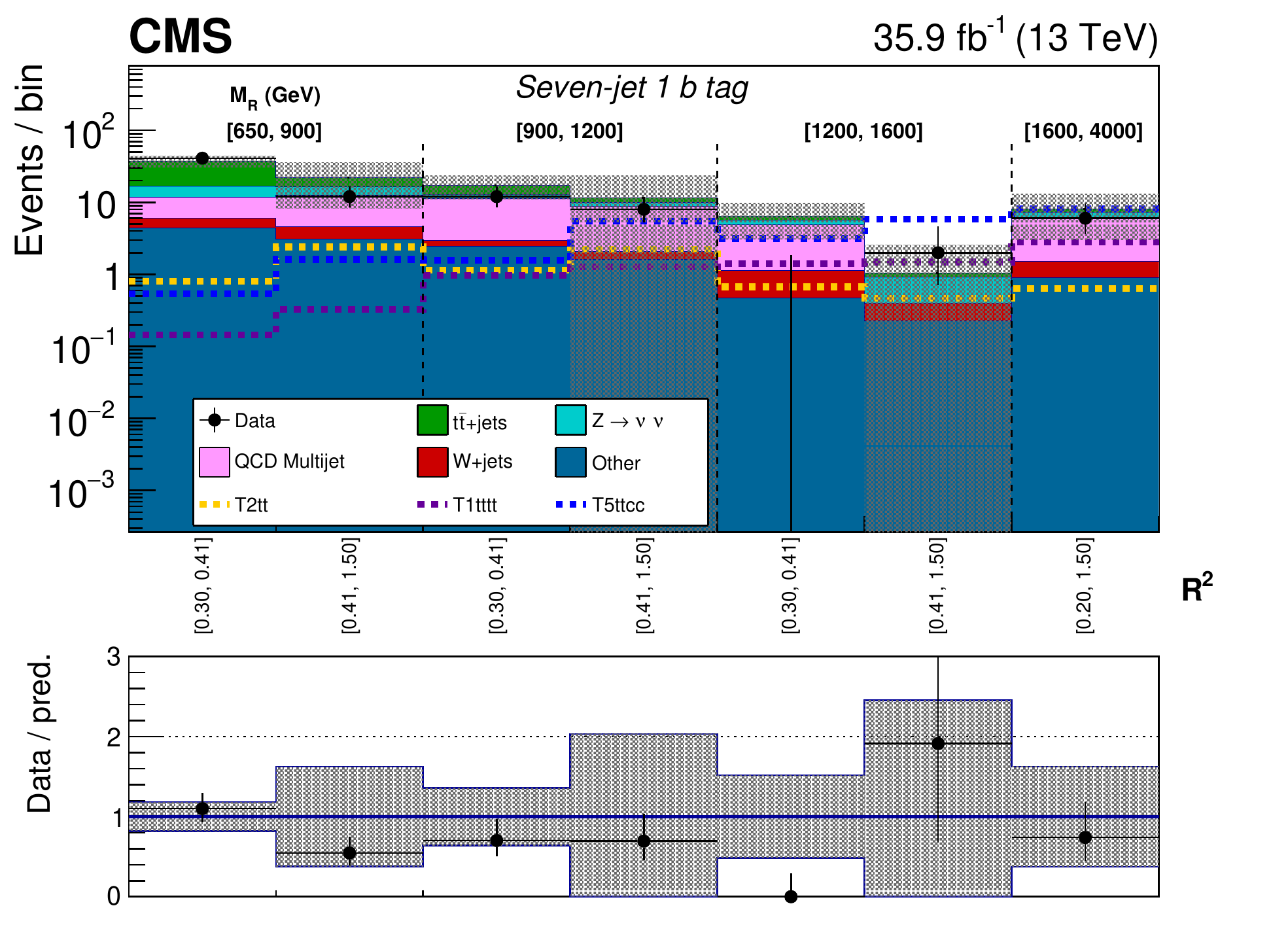}
\caption{ The $\MR$--$\Rtwo$ distribution observed in data is shown along with the pre-fit background prediction
obtained for the Seven-jet event category in the 0
\PQb~tag (upper) and 1 \PQb~tag (lower) bins. Further details of the plots are explained in the caption of Fig.~\ref{fig:ResultsLeptonMultiJet0b1b}.
}
\label{fig:ResultsSevenJet0b1b}
\end{figure}

\begin{figure}[!tbp] \centering
\includegraphics[width=0.8\textwidth]{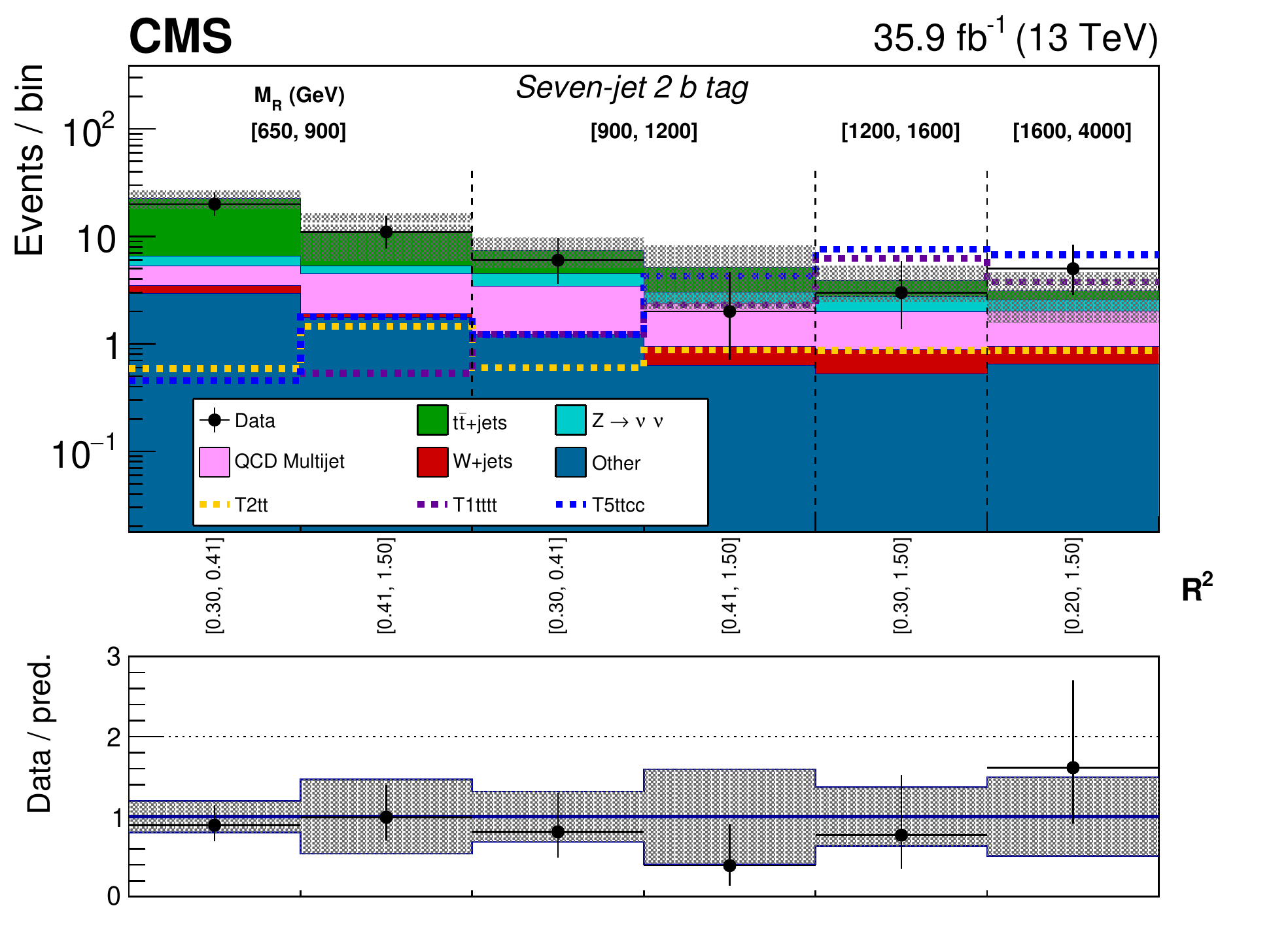}\\
\includegraphics[width=0.8\textwidth]{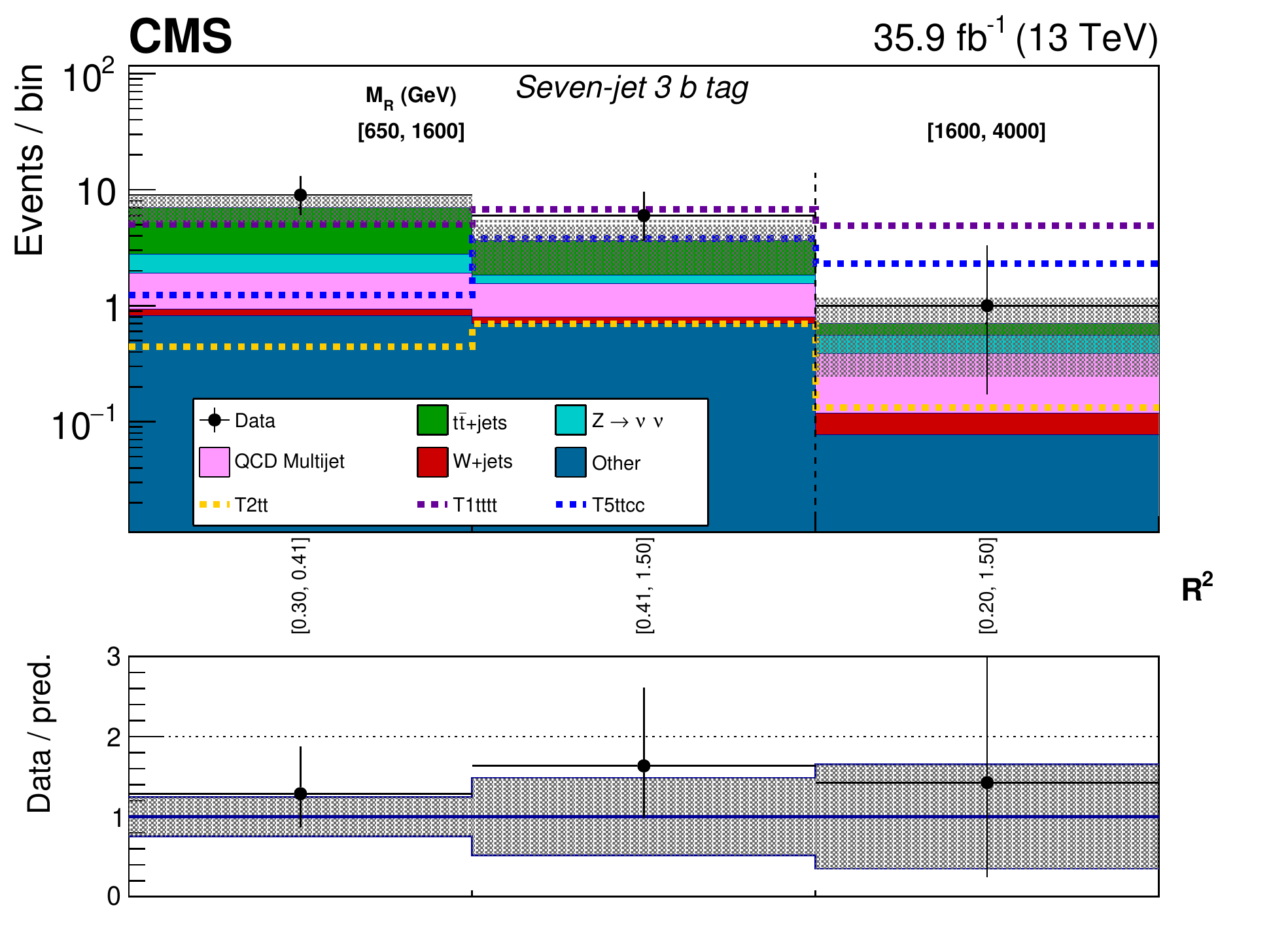}
\caption{ The $\MR$--$\Rtwo$ distribution observed in data is shown along with the pre-fit background prediction
obtained for the Seven-jet event category in the 2
\PQb~tag (upper) and 3 or more \PQb~tag (lower) bins. Further details of the plots are explained in the caption of Fig.~\ref{fig:ResultsLeptonMultiJet0b1b}.
}
\label{fig:ResultsSevenJet2b3b}
\end{figure}

Finally, the results for the zero-lepton
non-boosted categories are shown in
Figs.~\ref{fig:ResultsDiJet0b1b}--\ref{fig:ResultsSevenJet2b3b}.
The $\cPZ(\nu\overline{\nu})$+jets background is dominant for subcategories with
fewer jets and \cPqb-tagged jets, while the $\ttbar$ background
is dominant for subcategories with more jets and \cPqb-tagged jets.

We set upper limits
on the production cross sections of various SUSY simplified models.
We follow the LHC \CLs procedure~\cite{CLS1,CLS2,CMS-NOTE-2011-005} by using
the profile likelihood ratio test statistic and the asymptotic
formula to evaluate the 95\% confidence level (\CL) observed and expected limits on the
production cross section. Systematic uncertainties are propagated by
incorporating nuisance parameters that represent different sources of
systematic uncertainty, which are profiled in the maximum likelihood fit~\cite{CMS-NOTE-2011-005}.

Generally, the best signal sensitivity comes from the Lepton Multijet and Multijet
categories, and are dominated by bins with large $\MR$ when the mass splitting between
the gluino (or squark) and the LSP is large, and by bins with large $\Rtwo$
when the mass splitting is small. For signal models that produce many jets,
such as gluino pair production with gluinos decaying to two top quarks and the LSP, the Lepton Seven-jet
and Seven-jet categories dominate the sensitivity. For signal models with
boosted top quarks, such as top squark pair production, the boosted categories
contribute significantly to the sensitivity.

\begin{figure}[!tbp] \centering
\includegraphics[width=0.80\textwidth]{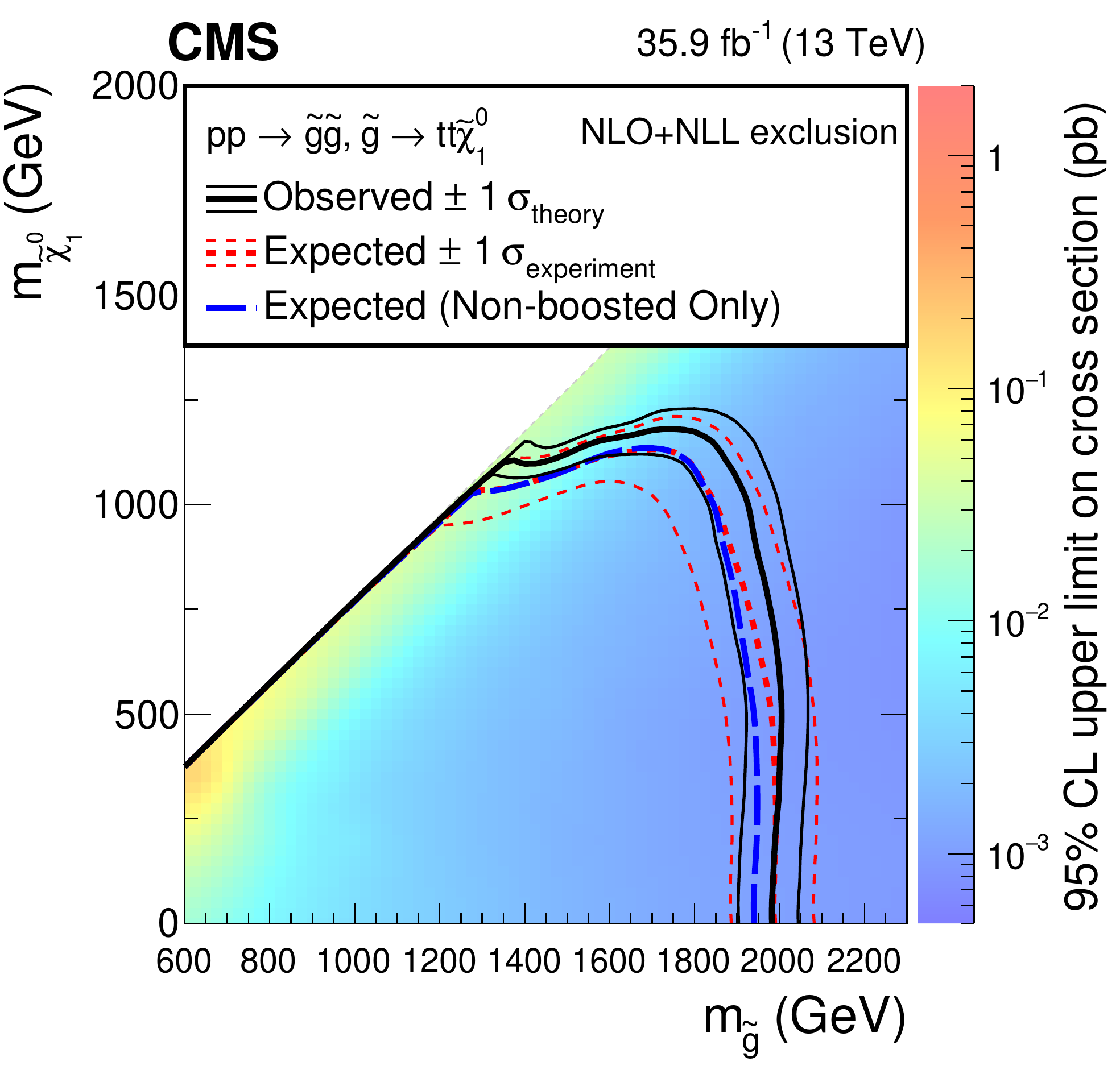} \\
\caption{ Expected and observed 95\% \CL limits on the production cross section
    for pair-produced gluinos each decaying to the LSP and top quarks. The blue dashed contour represents
    the expected 95\% \CL upper limit using data in the non-boosted categories only.
}
\label{fig:GluinoLimits}
\end{figure}

\begin{figure}[!tbp] \centering
\includegraphics[width=0.80\textwidth]{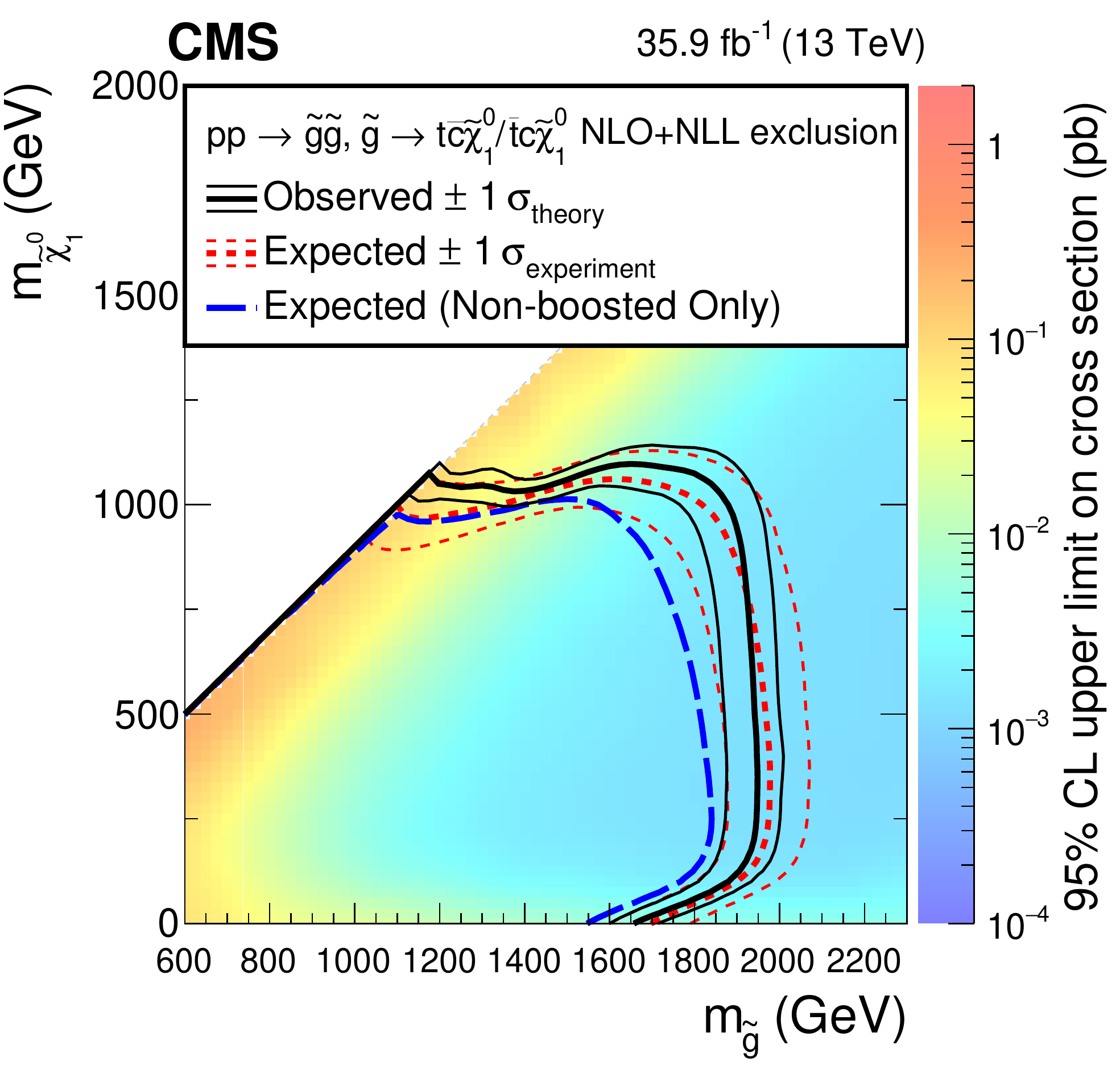}
\caption{ Expected and observed 95\% \CL limits on the production cross section
    for pair-produced gluinos each decaying to a top quark and a low mass top squark that
    subsequently decays to a charm quark and the LSP.
    The mass splitting ($m_{\PSQt}-m_{\chiz_1}$) is fixed to be $20\GeV$.
    The blue dashed contour represents
    the expected 95\% \CL upper limit using data in the non-boosted categories only.
}
\label{fig:GluinoLimits2}
\end{figure}

\begin{figure}[!tbp] \centering
\includegraphics[width=0.80\textwidth]{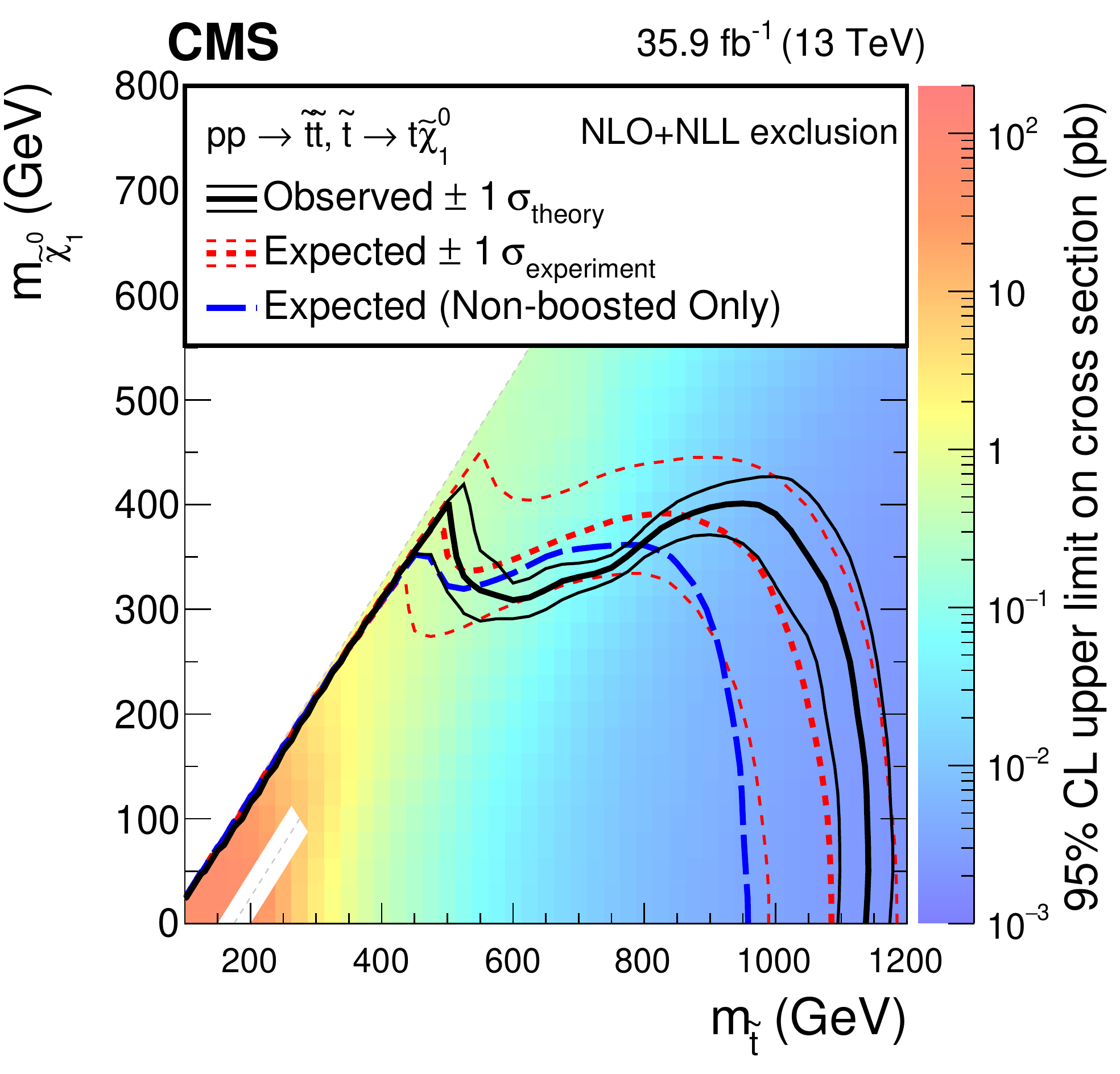}
\caption{ Expected and observed 95\% \CL limits on the production cross section
    for pair-produced squarks each decaying to a top quark and the LSP. The blue dashed contour represents
    the expected 95\% \CL upper limit using data in the non-boosted categories only.
    The white diagonal band corresponds to the region
    $\abs{m_{\PSQt}-m_{\PQt}-m_{\chiz_1}} < 25 \GeV$, where the
    mass difference between the $\PSQt$ and the $\chiz_1$ is very
    close to the top quark mass. In this region the signal
    acceptance depends strongly on the $\chiz_1$ mass and
    is therefore difficult to model.
}
\label{fig:SquarkLimits}
\end{figure}

First, we consider the scenario of pair produced gluinos
decaying to two top quarks and the LSP. The expected and observed limits for such gluino
decays are shown as a function of gluino and LSP masses in Fig.~\ref{fig:GluinoLimits}.
In this simplified model, we exclude gluino masses up to $2.0\TeV$ for LSP
mass below $700\GeV$. The limits for gluinos decaying to a top quark and a low mass top squark
that subsequently decays to a charm quark and the LSP,
is shown in Fig.~\ref{fig:GluinoLimits2}. For this simplified model, we exclude gluino masses
up to $1.9\TeV$ for LSP mass above $150$ and below $950\GeV$, extending the previous best limits~\cite{Sirunyan:2017pjw}
from the CMS experiment by about $100\GeV$ in the gluino mass. Finally, we consider pair produced top squarks decaying to the top quark
and the LSP. The expected and observed limits are shown in Fig.~\ref{fig:SquarkLimits}, and
we exclude top squark masses up to $1.14\TeV$ for LSP mass below $200\GeV$, extending the
previous best limits~\cite{Sirunyan:2017xse} from the CMS experiment by about $20\GeV$.
The dashed blue contour in each exclusion limit plot represents the expected limit
obtained using data from the non-boosted categories only. By comparing the expected limits obtained using only
the non-boosted categories with the expected limits using all categories, we observe clearly that the boosted categories
make an important contribution to the sensitivity for the signal models presented here.

\section{Summary}
\label{sec:Summary}

We have presented an inclusive search for supersymmetry (SUSY) in events
with no more than one lepton, a large multiplicity of energetic jets, and
evidence of invisible particles using the razor kinematic variables.
To enhance sensitivity to a broad range of signal models, the events
are categorized according to the number of leptons, the presence of
jets consistent with hadronically decaying \PW~bosons or top quarks,
and the number of jets and \PQb-tagged jets.
The analysis uses $\sqrt{s}=13\TeV$ proton-proton collision data collected by the CMS experiment
in 2016 and corresponding to an integrated luminosity of $35.9\fbinv$.
Standard model backgrounds were estimated using control regions in data and Monte Carlo simulation
yields in signal and control regions.
Background estimation procedures were verified using validation regions with kinematics
resembling that of the signal regions and closure tests.  Data are observed to be
consistent with the standard model expectation.

The results were interpreted in the context of simplified models of pair-produced gluinos
and direct top squark pair production.  Limits on the gluino mass extend to $2.0\TeV$, while limits on top squark masses
reach $1.14\TeV$. The combination of a large variety of final states enables this analysis to improve the
sensitivity in various signal scenarios. The analysis extended the exclusion limit of the gluino mass from the CMS experiment
by ${\approx}100\GeV$ in decays to a low-mass top squark and a top quark, and the exclusion limit of the top squark mass
by ${\approx}20\GeV$ in direct top squark pair production.

\clearpage

\begin{acknowledgments}
\hyphenation{Bundes-ministerium Forschungs-gemeinschaft Forschungs-zentren Rachada-pisek} We congratulate our colleagues in the CERN accelerator departments for the excellent performance of the LHC and thank the technical and administrative staffs at CERN and at other CMS institutes for their contributions to the success of the CMS effort. In addition, we gratefully acknowledge the computing centres and personnel of the Worldwide LHC Computing Grid for delivering so effectively the computing infrastructure essential to our analyses. Finally, we acknowledge the enduring support for the construction and operation of the LHC and the CMS detector provided by the following funding agencies: the Austrian Federal Ministry of Education, Science and Research and the Austrian Science Fund; the Belgian Fonds de la Recherche Scientifique, and Fonds voor Wetenschappelijk Onderzoek; the Brazilian Funding Agencies (CNPq, CAPES, FAPERJ, FAPERGS, and FAPESP); the Bulgarian Ministry of Education and Science; CERN; the Chinese Academy of Sciences, Ministry of Science and Technology, and National Natural Science Foundation of China; the Colombian Funding Agency (COLCIENCIAS); the Croatian Ministry of Science, Education and Sport, and the Croatian Science Foundation; the Research Promotion Foundation, Cyprus; the Secretariat for Higher Education, Science, Technology and Innovation, Ecuador; the Ministry of Education and Research, Estonian Research Council via IUT23-4 and IUT23-6 and European Regional Development Fund, Estonia; the Academy of Finland, Finnish Ministry of Education and Culture, and Helsinki Institute of Physics; the Institut National de Physique Nucl\'eaire et de Physique des Particules~/~CNRS, and Commissariat \`a l'\'Energie Atomique et aux \'Energies Alternatives~/~CEA, France; the Bundesministerium f\"ur Bildung und Forschung, Deutsche Forschungsgemeinschaft, and Helmholtz-Gemeinschaft Deutscher Forschungszentren, Germany; the General Secretariat for Research and Technology, Greece; the National Research, Development and Innovation Fund, Hungary; the Department of Atomic Energy and the Department of Science and Technology, India; the Institute for Studies in Theoretical Physics and Mathematics, Iran; the Science Foundation, Ireland; the Istituto Nazionale di Fisica Nucleare, Italy; the Ministry of Science, ICT and Future Planning, and National Research Foundation (NRF), Republic of Korea; the Ministry of Education and Science of the Republic of Latvia; the Lithuanian Academy of Sciences; the Ministry of Education, and University of Malaya (Malaysia); the Ministry of Science of Montenegro; the Mexican Funding Agencies (BUAP, CINVESTAV, CONACYT, LNS, SEP, and UASLP-FAI); the Ministry of Business, Innovation and Employment, New Zealand; the Pakistan Atomic Energy Commission; the Ministry of Science and Higher Education and the National Science Centre, Poland; the Funda\c{c}\~ao para a Ci\^encia e a Tecnologia, Portugal; JINR, Dubna; the Ministry of Education and Science of the Russian Federation, the Federal Agency of Atomic Energy of the Russian Federation, Russian Academy of Sciences, the Russian Foundation for Basic Research, and the National Research Center ``Kurchatov Institute"; the Ministry of Education, Science and Technological Development of Serbia; the Secretar\'{\i}a de Estado de Investigaci\'on, Desarrollo e Innovaci\'on, Programa Consolider-Ingenio 2010, Plan Estatal de Investigaci\'on Cient\'{\i}fica y T\'ecnica y de Innovaci\'on 2013-2016, Plan de Ciencia, Tecnolog\'{i}a e Innovaci\'on 2013-2017 del Principado de Asturias, and Fondo Europeo de Desarrollo Regional, Spain; the Ministry of Science, Technology and Research, Sri Lanka; the Swiss Funding Agencies (ETH Board, ETH Zurich, PSI, SNF, UniZH, Canton Zurich, and SER); the Ministry of Science and Technology, Taipei; the Thailand Center of Excellence in Physics, the Institute for the Promotion of Teaching Science and Technology of Thailand, Special Task Force for Activating Research and the National Science and Technology Development Agency of Thailand; the Scientific and Technical Research Council of Turkey, and Turkish Atomic Energy Authority; the National Academy of Sciences of Ukraine, and State Fund for Fundamental Researches, Ukraine; the Science and Technology Facilities Council, UK; the US Department of Energy, and the US National Science Foundation.
Individuals have received support from the Marie-Curie programme and the European Research Council and Horizon 2020 Grant, contract No. 675440 (European Union); the Leventis Foundation; the A. P. Sloan Foundation; the Alexander von Humboldt Foundation; the Belgian Federal Science Policy Office; the Fonds pour la Formation \`a la Recherche dans l'Industrie et dans l'Agriculture (FRIA-Belgium); the Agentschap voor Innovatie door Wetenschap en Technologie (IWT-Belgium); the F.R.S.-FNRS and FWO (Belgium) under the ``Excellence of Science - EOS" - be.h project n. 30820817; the Ministry of Education, Youth and Sports (MEYS) of the Czech Republic; the Lend\"ulet (``Momentum") Programme and the J\'anos Bolyai Research Scholarship of the Hungarian Academy of Sciences, the New National Excellence Program \'UNKP, the NKFIA research grants 123842, 123959, 124845, 124850 and 125105 (Hungary); the Council of Scientific and Industrial Research, India; the HOMING PLUS programme of the Foundation for Polish Science, cofinanced from European Union, Regional Development Fund, the Mobility Plus programme of the Ministry of Science and Higher Education, the National Science Center (Poland), contracts Harmonia 2014/14/M/ST2/00428, Opus 2014/13/B/ST2/02543, 2014/15/B/ST2/03998, and 2015/19/B/ST2/02861, Sonata-bis 2012/07/E/ST2/01406; the National Priorities Research Program by Qatar National Research Fund; the Programa de Excelencia Mar\'{i}a de Maeztu, and the Programa Severo Ochoa del Principado de Asturias; the Thalis and Aristeia programmes cofinanced by EU-ESF, and the Greek NSRF; the Rachadapisek Sompot Fund for Postdoctoral Fellowship, Chulalongkorn University, and the Chulalongkorn Academic into Its 2nd Century Project Advancement Project (Thailand); the Welch Foundation, contract C-1845; and the Weston Havens Foundation (USA).
\end{acknowledgments}

\bibliography{auto_generated}

\providecommand{\href}[2]{#2}\begingroup\raggedright\begin{thebibliography}{10}%
\makeatletter
\providecommand{\hrefCMSnoop }[0]{\@secondoftwo}%
\makeatother
\providecommand{\doi}{\texttt{doi:}\begingroup \urlstyle{tt}\Url}

\bibitem{Rogan:2010kb}
\hrefCMSnoop {}{C.~Rogan, ``{Kinematical variables towards new dynamics at the
  LHC}'',} (2010).
\href{http://www.arXiv.org/abs/1006.2727}{\texttt{arXiv:1006.2727}}.

\bibitem{razor2015}
\hrefCMSnoop {}{{CMS Collaboration}, ``{Inclusive search for supersymmetry
  using razor variables in pp collisions at $\sqrt s=$ 13 TeV}'',} \textit{
  Phys. Rev. D} \textbf{ 95} (2017) 012003,
  \href{http://dx.doi.org/10.1103/PhysRevD.95.012003}{\doi{10.1103/PhysRevD.95.012003}},
\href{http://www.arXiv.org/abs/1609.07658}{\texttt{arXiv:1609.07658}}.

\bibitem{razorboost}
\hrefCMSnoop {}{{CMS Collaboration}, ``{Search for supersymmetry in pp
  collisions at $\sqrt{s} = 8$ TeV in final states with boosted W bosons and b
  jets using razor variables}'',} \textit{ Phys. Rev. D} \textbf{ 93} (2016)
  092009,
  \href{http://dx.doi.org/10.1103/PhysRevD.93.092009}{\doi{10.1103/PhysRevD.93.092009}},
\href{http://www.arXiv.org/abs/1602.02917}{\texttt{arXiv:1602.02917}}.

\bibitem{Wess}
\hrefCMSnoop {}{J.~Wess and B.~Zumino, ``{Supergauge transformations in four
  dimensions}'',} \textit{ Nucl. Phys. B} \textbf{ 70} (1974) 39,
  \href{http://dx.doi.org/10.1016/0550-3213(74)90355-1}{\doi{10.1016/0550-3213(74)90355-1}}.

\bibitem{Golfand}
\href {http://www.jetpletters.ac.ru/ps/1584/article_24309.pdf}{Y.~A. Gol'fand
  and E.~P. Likhtman, ``Extension of the algebra of {Poincar\'e} group
  generators and violation of {P} invariance'',} \textit{ JETP Lett.} \textbf{
  13} (1971) 323.

\bibitem{Volkov}
\href {http://www.jetpletters.ac.ru/ps/1766/article_26864.pdf}{D.~V. Volkov and
  V.~P. Akulov, ``{Possible universal neutrino interaction}'',} \textit{ JETP
  Lett.} \textbf{ 16} (1972) 438.

\bibitem{Chamseddine}
\hrefCMSnoop {}{A.~H. Chamseddine, R.~L. Arnowitt, and P.~Nath, ``{Locally
  supersymmetric grand unification}'',} \textit{ Phys. Rev. Lett.} \textbf{ 49}
  (1982) 970,
  \href{http://dx.doi.org/10.1103/PhysRevLett.49.970}{\doi{10.1103/PhysRevLett.49.970}}.

\bibitem{Kane}
\hrefCMSnoop {}{G.~L. Kane, C.~F. Kolda, L.~Roszkowski, and J.~D. Wells,
  ``{Study of constrained minimal supersymmetry}'',} \textit{ Phys. Rev. D}
  \textbf{ 49} (1994) 6173,
  \href{http://dx.doi.org/10.1103/PhysRevD.49.6173}{\doi{10.1103/PhysRevD.49.6173}},
  \href{http://www.arXiv.org/abs/hep-ph/9312272}{\texttt{arXiv:hep-ph/9312272}}.

\bibitem{Fayet}
\hrefCMSnoop {}{P.~Fayet, ``{Supergauge invariant extension of the Higgs
  mechanism and a model for the electron and its neutrino}'',} \textit{ Nucl.
  Phys. B} \textbf{ 90} (1975) 104,
  \href{http://dx.doi.org/10.1016/0550-3213(75)90636-7}{\doi{10.1016/0550-3213(75)90636-7}}.

\bibitem{Barbieri}
\hrefCMSnoop {}{R.~Barbieri, S.~Ferrara, and C.~A. Savoy, ``{Gauge models with
  spontaneously broken local supersymmetry}'',} \textit{ Phys. Lett. B}
  \textbf{ 119} (1982) 343,
  \href{http://dx.doi.org/10.1016/0370-2693(82)90685-2}{\doi{10.1016/0370-2693(82)90685-2}}.

\bibitem{Hall}
\hrefCMSnoop {}{L.~J. Hall, J.~D. Lykken, and S.~Weinberg, ``{Supergravity as
  the messenger of supersymmetry breaking}'',} \textit{ Phys. Rev. D} \textbf{
  27} (1983) 2359,
  \href{http://dx.doi.org/10.1103/PhysRevD.27.2359}{\doi{10.1103/PhysRevD.27.2359}}.

\bibitem{Ramond}
\hrefCMSnoop {}{P.~Ramond, ``{Dual theory for free fermions}'',} \textit{ Phys.
  Rev. D} \textbf{ 3} (1971) 2415,
  \href{http://dx.doi.org/10.1103/PhysRevD.3.2415}{\doi{10.1103/PhysRevD.3.2415}}.

\bibitem{Witten:1981nf}
\hrefCMSnoop {}{E.~Witten, ``{Dynamical breaking of supersymmetry}'',} \textit{
  Nucl. Phys. B} \textbf{ 188} (1981) 513,
\href{http://dx.doi.org/10.1016/0550-3213(81)90006-7}{\doi{10.1016/0550-3213(81)90006-7}}.

\bibitem{Dimopoulos:1981zb}
\hrefCMSnoop {}{S.~Dimopoulos and H.~Georgi, ``{Softly broken supersymmetry and
  SU(5)}'',} \textit{ Nucl. Phys. B} \textbf{ 193} (1981) 150,
\href{http://dx.doi.org/10.1016/0550-3213(81)90522-8}{\doi{10.1016/0550-3213(81)90522-8}}.

\bibitem{Dine:1981za}
\hrefCMSnoop {}{M.~Dine, W.~Fischler, and M.~Srednicki, ``{Supersymmetric
  technicolor}'',} \textit{ Nucl. Phys. B} \textbf{ 189} (1981) 575,
\href{http://dx.doi.org/10.1016/0550-3213(81)90582-4}{\doi{10.1016/0550-3213(81)90582-4}}.

\bibitem{Dimopoulos:1981au}
\hrefCMSnoop {}{S.~Dimopoulos and S.~Raby, ``{Supercolor}'',} \textit{ Nucl.
  Phys. B} \textbf{ 192} (1981) 353,
\href{http://dx.doi.org/10.1016/0550-3213(81)90430-2}{\doi{10.1016/0550-3213(81)90430-2}}.

\bibitem{Sakai:1981gr}
\hrefCMSnoop {}{N.~Sakai, ``{Naturalness in supersymmetric GUTs}'',} \textit{
  Z. Phys. C} \textbf{ 11} (1981) 153,
\href{http://dx.doi.org/10.1007/BF01573998}{\doi{10.1007/BF01573998}}.

\bibitem{Kaul:1981hi}
\hrefCMSnoop {}{R.~K. Kaul and P.~Majumdar, ``{Cancellation of quadratically
  divergent mass corrections in globally supersymmetric spontaneously broken
  gauge theories}'',} \textit{ Nucl. Phys. B} \textbf{ 199} (1982) 36,
\href{http://dx.doi.org/10.1016/0550-3213(82)90565-X}{\doi{10.1016/0550-3213(82)90565-X}}.

\bibitem{Dimopoulos:1981yj}
\hrefCMSnoop {}{S.~Dimopoulos, S.~Raby, and F.~Wilczek, ``{Supersymmetry and
  the scale of unification}'',} \textit{ Phys. Rev. D} \textbf{ 24} (1981)
  1681,
\href{http://dx.doi.org/10.1103/PhysRevD.24.1681}{\doi{10.1103/PhysRevD.24.1681}}.

\bibitem{Marciano:1981un}
\hrefCMSnoop {}{W.~J. Marciano and G.~Senjanovic, ``{Predictions of
  supersymmetric grand unified theories}'',} \textit{ Phys. Rev. D} \textbf{
  25} (1982) 3092,
\href{http://dx.doi.org/10.1103/PhysRevD.25.3092}{\doi{10.1103/PhysRevD.25.3092}}.

\bibitem{Einhorn:1981sx}
\hrefCMSnoop {}{M.~B. Einhorn and D.~R.~T. Jones, ``{The weak mixing angle and
  unification mass in supersymmetric SU(5)}'',} \textit{ Nucl. Phys. B}
  \textbf{ 196} (1982) 475,
\href{http://dx.doi.org/10.1016/0550-3213(82)90502-8}{\doi{10.1016/0550-3213(82)90502-8}}.

\bibitem{Ibanez:1981yh}
\hrefCMSnoop {}{L.~E. Ibanez and G.~G. Ross, ``{Low-energy predictions in
  supersymmetric grand unified theories}'',} \textit{ Phys. Lett. B} \textbf{
  105} (1981) 439,
\href{http://dx.doi.org/10.1016/0370-2693(81)91200-4}{\doi{10.1016/0370-2693(81)91200-4}}.

\bibitem{Amaldi:1991cn}
\hrefCMSnoop {}{U.~Amaldi, W.~de~Boer, and H.~Furstenau, ``{Comparison of grand
  unified theories with electroweak and strong coupling constants measured at
  LEP}'',} \textit{ Phys. Lett. B} \textbf{ 260} (1991) 447,
\href{http://dx.doi.org/10.1016/0370-2693(91)91641-8}{\doi{10.1016/0370-2693(91)91641-8}}.

\bibitem{Langacker:1995fk}
\hrefCMSnoop {}{P.~Langacker and N.~Polonsky, ``{The strong coupling,
  unification, and recent data}'',} \textit{ Phys. Rev. D} \textbf{ 52} (1995)
  3081,
  \href{http://dx.doi.org/10.1103/PhysRevD.52.3081}{\doi{10.1103/PhysRevD.52.3081}},
\href{http://www.arXiv.org/abs/hep-ph/9503214}{\texttt{arXiv:hep-ph/9503214}}.

\bibitem{Ellis:1983ew}
J.~R. Ellis\hrefCMSnoop {}{ {et~al.}, ``{Supersymmetric relics from the Big
  Bang}'',} \textit{ Nucl. Phys. B} \textbf{ 238} (1984) 453,
\href{http://dx.doi.org/10.1016/0550-3213(84)90461-9}{\doi{10.1016/0550-3213(84)90461-9}}.

\bibitem{Jungman:1995df}
\hrefCMSnoop {}{G.~Jungman, M.~Kamionkowski, and K.~Griest, ``{Supersymmetric
  dark matter}'',} \textit{ Phys. Rept.} \textbf{ 267} (1996) 195,
  \href{http://dx.doi.org/10.1016/0370-1573(95)00058-5}{\doi{10.1016/0370-1573(95)00058-5}},
\href{http://www.arXiv.org/abs/hep-ph/9506380}{\texttt{arXiv:hep-ph/9506380}}.

\bibitem{Farrar:1978xj}
\hrefCMSnoop {}{G.~R. Farrar and P.~Fayet, ``{Phenomenology of the production,
  decay, and detection of new hadronic states associated with
  supersymmetry}'',} \textit{ Phys. Lett. B} \textbf{ 76} (1978) 575,
\href{http://dx.doi.org/10.1016/0370-2693(78)90858-4}{\doi{10.1016/0370-2693(78)90858-4}}.

\bibitem{Sirunyan:2018vjp}
\hrefCMSnoop {}{{CMS Collaboration}, ``{Search for natural and split
  supersymmetry in proton-proton collisions at $\sqrt{s}=13$ TeV in final
  states with jets and missing transverse momentum}'',} \textit{ JHEP} \textbf{
  05} (2018) 025,
  \href{http://dx.doi.org/10.1007/JHEP05(2018)025}{\doi{10.1007/JHEP05(2018)025}},
\href{http://www.arXiv.org/abs/1802.02110}{\texttt{arXiv:1802.02110}}.

\bibitem{Sirunyan:2017xse}
\hrefCMSnoop {}{{CMS Collaboration}, ``{Search for top squark pair production
  in pp collisions at $ \sqrt{s}=13 $ TeV using single lepton events}'',}
  \textit{ JHEP} \textbf{ 10} (2017) 019,
  \href{http://dx.doi.org/10.1007/JHEP10(2017)019}{\doi{10.1007/JHEP10(2017)019}},
\href{http://www.arXiv.org/abs/1706.04402}{\texttt{arXiv:1706.04402}}.

\bibitem{Sirunyan:2017kqq}
\hrefCMSnoop {}{{CMS Collaboration}, ``{Search for new phenomena with the
  $M_{\mathrm {T2}}$ variable in the all-hadronic final state produced in
  proton-proton collisions at $\sqrt{s} = 13$ TeV}'',} \textit{ Eur. Phys. J.
  C} \textbf{ 77} (2017) 710,
  \href{http://dx.doi.org/10.1140/epjc/s10052-017-5267-x}{\doi{10.1140/epjc/s10052-017-5267-x}},
\href{http://www.arXiv.org/abs/1705.04650}{\texttt{arXiv:1705.04650}}.

\bibitem{Sirunyan:2017fsj}
\hrefCMSnoop {}{{CMS Collaboration}, ``{Search for supersymmetry in pp
  collisions at $\sqrt{s}=13$ TeV in the single-lepton final state using the
  sum of masses of large-radius jets}'',} \textit{ Phys. Rev. Lett.} \textbf{
  119} (2017) 151802,
  \href{http://dx.doi.org/10.1103/PhysRevLett.119.151802}{\doi{10.1103/PhysRevLett.119.151802}},
\href{http://www.arXiv.org/abs/1705.04673}{\texttt{arXiv:1705.04673}}.

\bibitem{Sirunyan:2017cwe}
\hrefCMSnoop {}{{CMS Collaboration}, ``{Search for supersymmetry in multijet
  events with missing transverse momentum in proton-proton collisions at 13
  TeV}'',} \textit{ Phys. Rev. D} \textbf{ 96} (2017) 032003,
  \href{http://dx.doi.org/10.1103/PhysRevD.96.032003}{\doi{10.1103/PhysRevD.96.032003}},
\href{http://www.arXiv.org/abs/1704.07781}{\texttt{arXiv:1704.07781}}.

\bibitem{Sirunyan:2017uyt}
\hrefCMSnoop {}{{CMS Collaboration}, ``{Search for physics beyond the standard
  model in events with two leptons of same sign, missing transverse momentum,
  and jets in proton-proton collisions at $\sqrt{s} = 13$ TeV}'',} \textit{
  Eur. Phys. J. C} \textbf{ 77} (2017) 578,
  \href{http://dx.doi.org/10.1140/epjc/s10052-017-5079-z}{\doi{10.1140/epjc/s10052-017-5079-z}},
\href{http://www.arXiv.org/abs/1704.07323}{\texttt{arXiv:1704.07323}}.

\bibitem{Khachatryan:2017rhw}
\hrefCMSnoop {}{{CMS Collaboration}, ``{Search for supersymmetry in the
  all-hadronic final state using top quark tagging in pp collisions at
  $\sqrt{s} = 13$ TeV}'',} \textit{ Phys. Rev. D} \textbf{ 96} (2017) 012004,
  \href{http://dx.doi.org/10.1103/PhysRevD.96.012004}{\doi{10.1103/PhysRevD.96.012004}},
\href{http://www.arXiv.org/abs/1701.01954}{\texttt{arXiv:1701.01954}}.

\bibitem{Sirunyan:2017pjw}
\hrefCMSnoop {}{{CMS Collaboration}, ``{Search for supersymmetry in
  proton-proton collisions at 13 TeV using identified top quarks}'',} \textit{
  Phys. Rev. D} \textbf{ 97} (2018) 012007,
  \href{http://dx.doi.org/10.1103/PhysRevD.97.012007}{\doi{10.1103/PhysRevD.97.012007}},
\href{http://www.arXiv.org/abs/1710.11188}{\texttt{arXiv:1710.11188}}.

\bibitem{Sirunyan:2016jpr}
\hrefCMSnoop {}{{CMS Collaboration}, ``{Searches for pair production of
  third-generation squarks in $\sqrt{s}=13$ TeV pp collisions}'',} \textit{
  Eur. Phys. J. C} \textbf{ 77} (2017) 327,
  \href{http://dx.doi.org/10.1140/epjc/s10052-017-4853-2}{\doi{10.1140/epjc/s10052-017-4853-2}},
\href{http://www.arXiv.org/abs/1612.03877}{\texttt{arXiv:1612.03877}}.

\bibitem{Aaboud:2017ayj}
\hrefCMSnoop {}{{ATLAS Collaboration}, ``{Search for a scalar partner of the
  top quark in the jets plus missing transverse momentum final state at
  $\sqrt{s}=13$ TeV with the ATLAS detector}'',} \textit{ JHEP} \textbf{ 12}
  (2017) 085,
  \href{http://dx.doi.org/10.1007/JHEP12(2017)085}{\doi{10.1007/JHEP12(2017)085}},
\href{http://www.arXiv.org/abs/1709.04183}{\texttt{arXiv:1709.04183}}.

\bibitem{Aaboud:2017wqg}
\hrefCMSnoop {}{{ATLAS Collaboration}, ``{Search for supersymmetry in events
  with $b$-tagged jets and missing transverse momentum in $pp$ collisions at
  $\sqrt{s}=13$ TeV with the ATLAS detector}'',} \textit{ JHEP} \textbf{ 11}
  (2017) 195,
  \href{http://dx.doi.org/10.1007/JHEP11(2017)195}{\doi{10.1007/JHEP11(2017)195}},
\href{http://www.arXiv.org/abs/1708.09266}{\texttt{arXiv:1708.09266}}.

\bibitem{Aaboud:2017bac}
\hrefCMSnoop {}{{ATLAS Collaboration}, ``{Search for squarks and gluinos in
  events with an isolated lepton, jets, and missing transverse momentum at
  $\sqrt{s}=13$ TeV with the ATLAS detector}'',} \textit{ Phys. Rev. D}
  \textbf{ 96} (2017) 112010,
  \href{http://dx.doi.org/10.1103/PhysRevD.96.112010}{\doi{10.1103/PhysRevD.96.112010}},
\href{http://www.arXiv.org/abs/1708.08232}{\texttt{arXiv:1708.08232}}.

\bibitem{Aaboud:2017nfd}
\hrefCMSnoop {}{{ATLAS Collaboration}, ``{Search for direct top squark pair
  production in final states with two leptons in $\sqrt{s} = 13$ TeV $pp$
  collisions with the ATLAS detector}'',} \textit{ Eur. Phys. J. C} \textbf{
  77} (2017) 898,
  \href{http://dx.doi.org/10.1140/epjc/s10052-017-5445-x}{\doi{10.1140/epjc/s10052-017-5445-x}},
\href{http://www.arXiv.org/abs/1708.03247}{\texttt{arXiv:1708.03247}}.

\bibitem{Aaboud:2017hdf}
\hrefCMSnoop {}{{ATLAS Collaboration}, ``{Search for new phenomena with large
  jet multiplicities and missing transverse momentum using large-radius jets
  and flavour-tagging at ATLAS in 13 TeV $pp$ collisions}'',} \textit{ JHEP}
  \textbf{ 12} (2017) 034,
  \href{http://dx.doi.org/10.1007/JHEP12(2017)034}{\doi{10.1007/JHEP12(2017)034}},
\href{http://www.arXiv.org/abs/1708.02794}{\texttt{arXiv:1708.02794}}.

\bibitem{Aaboud:2017dmy}
\hrefCMSnoop {}{{ATLAS Collaboration}, ``{Search for supersymmetry in final
  states with two same-sign or three leptons and jets using 36 fb$^{-1}$ of
  $\sqrt{s}=13$ TeV $pp$ collision data with the ATLAS detector}'',} \textit{
  JHEP} \textbf{ 09} (2017) 084,
  \href{http://dx.doi.org/10.1007/JHEP09(2017)084}{\doi{10.1007/JHEP09(2017)084}},
\href{http://www.arXiv.org/abs/1706.03731}{\texttt{arXiv:1706.03731}}.

\bibitem{Aaboud:2017faq}
\hrefCMSnoop {}{{ATLAS Collaboration}, ``{Search for new phenomena in a lepton
  plus high jet multiplicity final state with the ATLAS experiment using
  $\sqrt{s}=13 $ TeV proton-proton collision data}'',} \textit{ JHEP} \textbf{
  09} (2017) 088,
  \href{http://dx.doi.org/10.1007/JHEP09(2017)088}{\doi{10.1007/JHEP09(2017)088}},
\href{http://www.arXiv.org/abs/1704.08493}{\texttt{arXiv:1704.08493}}.

\bibitem{bib-sms-1}
N.~Arkani-Hamed\hrefCMSnoop {}{ {et~al.}, ``{{MARMOSET}: The path from {LHC}
  data to the new standard model via on-shell effective theories}'',} (2007).
\href{http://www.arXiv.org/abs/hep-ph/0703088}{\texttt{arXiv:hep-ph/0703088}}.

\bibitem{bib-sms-2}
\hrefCMSnoop {}{J.~Alwall, P.~C. Schuster, and N.~Toro, ``Simplified models for
  a first characterization of new physics at the {LHC}'',} \textit{ Phys. Rev.
  D} \textbf{ 79} (2009) 075020,
  \href{http://dx.doi.org/10.1103/PhysRevD.79.075020}{\doi{10.1103/PhysRevD.79.075020}},
\href{http://www.arXiv.org/abs/0810.3921}{\texttt{arXiv:0810.3921}}.

\bibitem{bib-sms-3}
\hrefCMSnoop {}{J.~Alwall, M.-P. Le, M.~Lisanti, and J.~G. Wacker,
  ``{Model-independent jets plus missing energy searches}'',} \textit{ Phys.
  Rev. D} \textbf{ 79} (2009) 015005,
  \href{http://dx.doi.org/10.1103/PhysRevD.79.015005}{\doi{10.1103/PhysRevD.79.015005}},
\href{http://www.arXiv.org/abs/0809.3264}{\texttt{arXiv:0809.3264}}.

\bibitem{bib-sms-4}
D.~Alves\hrefCMSnoop {}{ {et~al.}, ``Simplified models for {LHC} new physics
  searches'',} \textit{ J. Phys. G} \textbf{ 39} (2012) 105005,
  \href{http://dx.doi.org/10.1088/0954-3899/39/10/105005}{\doi{10.1088/0954-3899/39/10/105005}},
\href{http://www.arXiv.org/abs/1105.2838}{\texttt{arXiv:1105.2838}}.

\bibitem{Adolphi:2008zzk}
\hrefCMSnoop {}{{CMS Collaboration}, ``{The CMS experiment at the CERN LHC}'',}
  \textit{ JINST} \textbf{ 3} (2008) S08004,
\href{http://dx.doi.org/10.1088/1748-0221/3/08/S08004}{\doi{10.1088/1748-0221/3/08/S08004}}.

\bibitem{Sirunyan:2017ulk}
\hrefCMSnoop {}{{CMS Collaboration}, ``{Particle-flow reconstruction and global
  event description with the CMS detector}'',} \textit{ JINST} \textbf{ 12}
  (2017) P10003,
  \href{http://dx.doi.org/10.1088/1748-0221/12/10/P10003}{\doi{10.1088/1748-0221/12/10/P10003}},
\href{http://www.arXiv.org/abs/1706.04965}{\texttt{arXiv:1706.04965}}.

\bibitem{Cacciari:2008gp}
\hrefCMSnoop {}{M.~Cacciari, G.~P. Salam, and G.~Soyez, ``The anti-\kt jet
  clustering algorithm'',} \textit{ JHEP} \textbf{ 04} (2008) 063,
  \href{http://dx.doi.org/10.1088/1126-6708/2008/04/063}{\doi{10.1088/1126-6708/2008/04/063}},
  \href{http://www.arXiv.org/abs/0802.1189}{\texttt{arXiv:0802.1189}}.

\bibitem{Cacciari:2011ma}
\hrefCMSnoop {}{M.~Cacciari, G.~P. Salam, and G.~Soyez, ``{FastJet user
  manual}'',} \textit{ Eur. Phys. J. C} \textbf{ 72} (2012) 1896,
  \href{http://dx.doi.org/10.1140/epjc/s10052-012-1896-2}{\doi{10.1140/epjc/s10052-012-1896-2}},
\href{http://www.arXiv.org/abs/1111.6097}{\texttt{arXiv:1111.6097}}.

\bibitem{Khachatryan:2016kdb}
\hrefCMSnoop {}{{CMS Collaboration}, ``{Jet energy scale and resolution in the
  CMS experiment in pp collisions at 8 TeV}'',} \textit{ JINST} \textbf{ 12}
  (2017) P02014,
  \href{http://dx.doi.org/10.1088/1748-0221/12/02/P02014}{\doi{10.1088/1748-0221/12/02/P02014}},
\href{http://www.arXiv.org/abs/1607.03663}{\texttt{arXiv:1607.03663}}.

\bibitem{CMS-PAS-JME-16-003}
\href {http://cdsweb.cern.ch/record/1279362}{{CMS Collaboration}, ``{Jet
  algorithms performance in 13 TeV data}'',} CMS Physics Analysis Summary
  CMS-PAS-JME-16-003, 2017.

\bibitem{Sirunyan:2017ezt}
\hrefCMSnoop {}{{CMS Collaboration}, ``{Identification of heavy-flavour jets
  with the CMS detector in pp collisions at 13 TeV}'',} \textit{ JINST}
  \textbf{ 13} (2018) P05011,
  \href{http://dx.doi.org/10.1088/1748-0221/13/05/P05011}{\doi{10.1088/1748-0221/13/05/P05011}},
\href{http://www.arXiv.org/abs/1712.07158}{\texttt{arXiv:1712.07158}}.

\bibitem{Thaler:2010tr}
\hrefCMSnoop {}{J.~Thaler and K.~Van~Tilburg, ``{Identifying boosted objects
  with $N$-subjettiness}'',} \textit{ JHEP} \textbf{ 03} (2011) 015,
  \href{http://dx.doi.org/10.1007/JHEP03(2011)015}{\doi{10.1007/JHEP03(2011)015}},
\href{http://www.arXiv.org/abs/1011.2268}{\texttt{arXiv:1011.2268}}.

\bibitem{Larkoski:2014wba}
\hrefCMSnoop {}{A.~J. Larkoski, S.~Marzani, G.~Soyez, and J.~Thaler, ``{Soft
  Drop}'',} \textit{ JHEP} \textbf{ 05} (2014) 146,
  \href{http://dx.doi.org/10.1007/JHEP05(2014)146}{\doi{10.1007/JHEP05(2014)146}},
\href{http://www.arXiv.org/abs/1402.2657}{\texttt{arXiv:1402.2657}}.

\bibitem{Chatrchyan:2011tn}
\hrefCMSnoop {}{{CMS Collaboration}, ``{Missing transverse energy performance
  of the CMS detector}'',} \textit{ JINST} \textbf{ 6} (2011) P09001,
  \href{http://dx.doi.org/10.1088/1748-0221/6/09/P09001}{\doi{10.1088/1748-0221/6/09/P09001}},
  \href{http://www.arXiv.org/abs/1106.5048}{\texttt{arXiv:1106.5048}}.

\bibitem{Khachatryan:2014gga}
\hrefCMSnoop {}{{CMS Collaboration}, ``{Performance of the CMS missing
  transverse momentum reconstruction in pp data at $\sqrt{s} = 8$ TeV}'',}
  \textit{ JINST} \textbf{ 10} (2015) P02006,
  \href{http://dx.doi.org/10.1088/1748-0221/10/02/P02006}{\doi{10.1088/1748-0221/10/02/P02006}},
\href{http://www.arXiv.org/abs/1411.0511}{\texttt{arXiv:1411.0511}}.

\bibitem{CMS-PAS-JME-17-001}
\href {https://cds.cern.ch/record/2628600}{{CMS Collaboration}, ``{Performance
  of missing transverse momentum in pp collisions at $\sqrt{s}=13$ TeV using
  the CMS detector}'',} CMS Physics Analysis Summary CMS-PAS-JME-17-001, 2018.

\bibitem{Khachatryan:2015hwa}
\hrefCMSnoop {}{{CMS Collaboration}, ``{Performance of electron reconstruction
  and selection with the CMS detector in proton-proton collisions at 8 TeV}'',}
  \textit{ JINST} \textbf{ 10} (2015) P06005,
  \href{http://dx.doi.org/10.1088/1748-0221/10/06/P06005}{\doi{10.1088/1748-0221/10/06/P06005}},
\href{http://www.arXiv.org/abs/1502.02701}{\texttt{arXiv:1502.02701}}.

\bibitem{Sirunyan:2018fpa}
\hrefCMSnoop {}{{CMS Collaboration}, ``{Performance of the CMS muon detector
  and muon reconstruction with proton-proton collisions at $\sqrt{s}=$ 13
  TeV}'',} \textit{ JINST} \textbf{ 13} (2018) P06015,
  \href{http://dx.doi.org/10.1088/1748-0221/13/06/P06015}{\doi{10.1088/1748-0221/13/06/P06015}},
\href{http://www.arXiv.org/abs/1804.04528}{\texttt{arXiv:1804.04528}}.

\bibitem{Khachatryan:2015dfa}
\hrefCMSnoop {}{{CMS Collaboration}, ``{Reconstruction and identification of
  $\tau$ lepton decays to hadrons and $\nu_{\tau}$ at CMS}'',} \textit{ JINST}
  \textbf{ 11} (2016) P01019,
  \href{http://dx.doi.org/10.1088/1748-0221/11/01/P01019}{\doi{10.1088/1748-0221/11/01/P01019}},
\href{http://www.arXiv.org/abs/1510.07488}{\texttt{arXiv:1510.07488}}.

\bibitem{Khachatryan:2015iwa}
\hrefCMSnoop {}{{CMS Collaboration}, ``{Performance of photon reconstruction
  and identification with the CMS detector in proton-proton collisions at
  $\sqrt{s} = 8$ TeV}'',} \textit{ JINST} \textbf{ 10} (2015) P08010,
  \href{http://dx.doi.org/10.1088/1748-0221/10/08/P08010}{\doi{10.1088/1748-0221/10/08/P08010}},
\href{http://www.arXiv.org/abs/1502.02702}{\texttt{arXiv:1502.02702}}.

\bibitem{Alwall:2011uj}
J.~Alwall\hrefCMSnoop {}{ {et~al.}, ``{MadGraph5: going beyond}'',} \textit{
  JHEP} \textbf{ 06} (2011) 128,
  \href{http://dx.doi.org/10.1007/JHEP06(2011)128}{\doi{10.1007/JHEP06(2011)128}},
  \href{http://www.arXiv.org/abs/1106.0522}{\texttt{arXiv:1106.0522}}.

\bibitem{Alwall:2014hca}
J.~Alwall\hrefCMSnoop {}{ {et~al.}, ``The automated computation of tree-level
  and next-to-leading order differential cross sections, and their matching to
  parton shower simulations'',} \textit{ JHEP} \textbf{ 07} (2014) 079,
  \href{http://dx.doi.org/10.1007/JHEP07(2014)079}{\doi{10.1007/JHEP07(2014)079}},
\href{http://www.arXiv.org/abs/1405.0301}{\texttt{arXiv:1405.0301}}.

\bibitem{Sjostrand:2014zea}
T.~Sj{\"o}strand\hrefCMSnoop {}{ {et~al.}, ``{An Introduction to PYTHIA
  8.2}'',} \textit{ Comput. Phys. Commun.} \textbf{ 191} (2015) 159,
  \href{http://dx.doi.org/10.1016/j.cpc.2015.01.024}{\doi{10.1016/j.cpc.2015.01.024}},
\href{http://www.arXiv.org/abs/1410.3012}{\texttt{arXiv:1410.3012}}.

\bibitem{Hoche:2006ph}
S.~Hoeche\hrefCMSnoop {}{ {et~al.}, ``{Matching parton showers and matrix
  elements}'',} in \textit{ {HERA and the LHC: A Workshop on the implications
  of HERA for LHC physics: Proceedings Part A}}, p.~288.
\newblock 2005.
\newblock
  \href{http://www.arXiv.org/abs/hep-ph/0602031}{\texttt{arXiv:hep-ph/0602031}}.
\newblock
\href{http://dx.doi.org/10.5170/CERN-2005-014.288}{\doi{10.5170/CERN-2005-014.288}}.

\bibitem{Alwall:2007fs}
J.~Alwall\hrefCMSnoop {}{ {et~al.}, ``{Comparative study of various algorithms
  for the merging of parton showers and matrix elements in hadronic
  collisions}'',} \textit{ Eur. Phys. J. C} \textbf{ 53} (2008) 473,
  \href{http://dx.doi.org/10.1140/epjc/s10052-007-0490-5}{\doi{10.1140/epjc/s10052-007-0490-5}},
\href{http://www.arXiv.org/abs/0706.2569}{\texttt{arXiv:0706.2569}}.

\bibitem{Khachatryan:2015pea}
\hrefCMSnoop {}{{CMS Collaboration}, ``{Event generator tunes obtained from
  underlying event and multiparton scattering measurements}'',} \textit{ Eur.
  Phys. J. C} \textbf{ 76} (2016) 155,
  \href{http://dx.doi.org/10.1140/epjc/s10052-016-3988-x}{\doi{10.1140/epjc/s10052-016-3988-x}},
\href{http://www.arXiv.org/abs/1512.00815}{\texttt{arXiv:1512.00815}}.

\bibitem{Frixione:2007nw}
\hrefCMSnoop {}{S.~Frixione, P.~Nason, and G.~Ridolfi, ``{A positive-weight
  next-to-leading-order Monte Carlo for heavy flavour hadroproduction}'',}
  \textit{ JHEP} \textbf{ 09} (2007) 126,
  \href{http://dx.doi.org/10.1088/1126-6708/2007/09/126}{\doi{10.1088/1126-6708/2007/09/126}},
\href{http://www.arXiv.org/abs/0707.3088}{\texttt{arXiv:0707.3088}}.

\bibitem{Alioli:2009je}
\hrefCMSnoop {}{S.~Alioli, P.~Nason, C.~Oleari, and E.~Re, ``{NLO single-top
  production matched with shower in POWHEG: $s$- and $t$-channel
  contributions}'',} \textit{ JHEP} \textbf{ 09} (2009) 111,
  \href{http://dx.doi.org/10.1088/1126-6708/2009/09/111}{\doi{10.1088/1126-6708/2009/09/111}},
  \href{http://www.arXiv.org/abs/0907.4076}{\texttt{arXiv:0907.4076}}.
[Erratum: \DOI{10.1007/JHEP02(2010)011}].

\bibitem{Re:2010bp}
\hrefCMSnoop {}{E.~Re, ``{Single-top Wt-channel production matched with parton
  showers using the POWHEG method}'',} \textit{ Eur. Phys. J. C} \textbf{ 71}
  (2011) 1547,
  \href{http://dx.doi.org/10.1140/epjc/s10052-011-1547-z}{\doi{10.1140/epjc/s10052-011-1547-z}},
\href{http://www.arXiv.org/abs/1009.2450}{\texttt{arXiv:1009.2450}}.

\bibitem{Ball:2014uwa}
\hrefCMSnoop {}{{NNPDF} Collaboration, ``{Parton distributions for the LHC Run
  II}'',} \textit{ JHEP} \textbf{ 04} (2015) 040,
  \href{http://dx.doi.org/10.1007/JHEP04(2015)040}{\doi{10.1007/JHEP04(2015)040}},
\href{http://www.arXiv.org/abs/1410.8849}{\texttt{arXiv:1410.8849}}.

\bibitem{geant4}
\hrefCMSnoop {}{S.~Agostinelli {et~al.}, ``Geant4 --- a simulation toolkit'',}
  \textit{ Nucl. Instrum. Meth. A} \textbf{ 506} (2003) 250,
  \href{http://dx.doi.org/10.1016/S0168-9002(03)01368-8}{\doi{10.1016/S0168-9002(03)01368-8}}.

\bibitem{FastSim}
\hrefCMSnoop {}{{CMS Collaboration}, ``The fast simulation of the {CMS}
  detector at {LHC}'',} \textit{ J. Phys.: Conf. Ser.} \textbf{ 331} (2011)
  032049,
\href{http://dx.doi.org/10.1088/1742-6596/331/3/032049}{\doi{10.1088/1742-6596/331/3/032049}}.

\bibitem{NLONLL1}
\hrefCMSnoop {}{W.~{Beenakker}, R.~{H{\"o}pker}, M.~{Spira}, and P.~M.
  {Zerwas}, ``{Squark and gluino production at hadron colliders}'',} \textit{
  Nucl. Phys. B} \textbf{ 492} (1997) 51,
  \href{http://dx.doi.org/10.1016/S0550-3213(97)80027-2}{\doi{10.1016/S0550-3213(97)80027-2}},
  \href{http://www.arXiv.org/abs/hep-ph/9610490}{\texttt{arXiv:hep-ph/9610490}}.

\bibitem{NLONLL2}
\hrefCMSnoop {}{A.~Kulesza and L.~Motyka, ``Threshold resummation for
  squark-antisquark and gluino-pair production at the {LHC}'',} \textit{ Phys.
  Rev. Lett.} \textbf{ 102} (2009) 111802,
  \href{http://dx.doi.org/10.1103/PhysRevLett.102.111802}{\doi{10.1103/PhysRevLett.102.111802}},
  \href{http://www.arXiv.org/abs/0807.2405}{\texttt{arXiv:0807.2405}}.

\bibitem{NLONLL3}
\hrefCMSnoop {}{A.~Kulesza and L.~Motyka, ``{Soft gluon resummation for the
  production of gluino-gluino and squark-antisquark pairs at the LHC}'',}
  \textit{ Phys. Rev. D} \textbf{ 80} (2009) 095004,
  \href{http://dx.doi.org/10.1103/PhysRevD.80.095004}{\doi{10.1103/PhysRevD.80.095004}},
  \href{http://www.arXiv.org/abs/0905.4749}{\texttt{arXiv:0905.4749}}.

\bibitem{NLONLL4}
W.~{Beenakker}\hrefCMSnoop {}{ {et~al.}, ``{Soft-gluon resummation for squark
  and gluino hadroproduction}'',} \textit{ JHEP} \textbf{ 12} (2009) 041,
  \href{http://dx.doi.org/10.1088/1126-6708/2009/12/041}{\doi{10.1088/1126-6708/2009/12/041}},
  \href{http://www.arXiv.org/abs/0909.4418}{\texttt{arXiv:0909.4418}}.

\bibitem{NLONLL5}
W.~{Beenakker}\hrefCMSnoop {}{ {et~al.}, ``{Squark and gluino
  hadroproduction}'',} \textit{ Int. J. Mod. Phys. A} \textbf{ 26} (2011) 2637,
  \href{http://dx.doi.org/10.1142/S0217751X11053560}{\doi{10.1142/S0217751X11053560}},
  \href{http://www.arXiv.org/abs/1105.1110}{\texttt{arXiv:1105.1110}}.

\bibitem{Borschensky:2014cia}
C.~Borschensky\hrefCMSnoop {}{ {et~al.}, ``{Squark and gluino production cross
  sections in pp collisions at $\sqrt{s}=$ 13, 14, 33 and 100 TeV}'',} \textit{
  Eur. Phys. J. C} \textbf{ 74} (2014) 3174,
  \href{http://dx.doi.org/10.1140/epjc/s10052-014-3174-y}{\doi{10.1140/epjc/s10052-014-3174-y}},
\href{http://www.arXiv.org/abs/1407.5066}{\texttt{arXiv:1407.5066}}.

\bibitem{razor2010}
\hrefCMSnoop {}{{CMS Collaboration}, ``{Inclusive search for squarks and
  gluinos in pp collisions at $\sqrt{s} = 7$ TeV}'',} \textit{ Phys. Rev. D}
  \textbf{ 85} (2012) 012004,
  \href{http://dx.doi.org/10.1103/PhysRevD.85.012004}{\doi{10.1103/PhysRevD.85.012004}},
  \href{http://www.arXiv.org/abs/1107.1279}{\texttt{arXiv:1107.1279}}.

\bibitem{Khachatryan:2015ira}
\hrefCMSnoop {}{{CMS Collaboration}, ``{Comparison of the
  Z/$\gamma^{\ast}$+jets to $\gamma$+jets cross sections in pp collisions at $
  \sqrt{s}=8$ TeV}'',} \textit{ JHEP} \textbf{ 10} (2015) 128,
  \href{http://dx.doi.org/10.1007/JHEP10(2015)128}{\doi{10.1007/JHEP10(2015)128}},
  \href{http://www.arXiv.org/abs/1505.06520}{\texttt{arXiv:1505.06520}}.
[Erratum: \DOI{10.1007/JHEP04(2016)010}].

\bibitem{Ellis:1985vn}
\hrefCMSnoop {}{S.~D. Ellis, R.~Kleiss, and W.~J. Stirling, ``{W's, Z's and
  jets}'',} \textit{ Phys. Lett. B} \textbf{ 154} (1985) 435,
\href{http://dx.doi.org/10.1016/0370-2693(85)90425-3}{\doi{10.1016/0370-2693(85)90425-3}}.

\bibitem{Berends:1989cf}
F.~A. Berends\hrefCMSnoop {}{ {et~al.}, ``{Multijet production in $\PW$, $\cPZ$
  events at $p \bar{p}$ Colliders}'',} \textit{ Phys. Lett. B} \textbf{ 224}
  (1989) 237,
\href{http://dx.doi.org/10.1016/0370-2693(89)91081-2}{\doi{10.1016/0370-2693(89)91081-2}}.

\bibitem{CMS:2017sdi}
\href {http://cdsweb.cern.ch/record/2257069}{{{CMS}} Collaboration, ``{CMS
  luminosity measurements for the 2016 data taking period}'',} CMS Physics
  Analysis Summary CMS-PAS-LUM-17-001, 2017.

\bibitem{CLS1}
\hrefCMSnoop {}{A.~L. Read, ``Presentation of search results: The
  {CL$_{\text{s}}$} technique'',} \textit{ J. Phys. G} \textbf{ 28} (2002)
  2693,
\href{http://dx.doi.org/10.1088/0954-3899/28/10/313}{\doi{10.1088/0954-3899/28/10/313}}.

\bibitem{CLS2}
\hrefCMSnoop {}{T.~Junk, ``Confidence level computation for combining searches
  with small statistics'',} \textit{ Nucl. Instrum. Meth. A} \textbf{ 434}
  (1999) 435,
  \href{http://dx.doi.org/10.1016/S0168-9002(99)00498-2}{\doi{10.1016/S0168-9002(99)00498-2}},
\href{http://www.arXiv.org/abs/hep-ex/9902006}{\texttt{arXiv:hep-ex/9902006}}.

\bibitem{CMS-NOTE-2011-005}
\href {https://cds.cern.ch/record/1379837}{{The ATLAS Collaboration, The CMS
  Collaboration, The LHC Higgs Combination Group}, ``Procedure for the {LHC}
  {Higgs} boson search combination in {Summer} 2011'',} Technical Report
  CMS-NOTE-2011-005, ATL-PHYS-PUB-2011-11, 2011.

\end{thebibliography}\endgroup

\cleardoublepage \appendix\section{The CMS Collaboration \label{app:collab}}\begin{sloppypar}\hyphenpenalty=5000\widowpenalty=500\clubpenalty=5000\vskip\cmsinstskip
\textbf{Yerevan Physics Institute, Yerevan, Armenia}\\*[0pt]
A.M.~Sirunyan, A.~Tumasyan
\vskip\cmsinstskip
\textbf{Institut f\"{u}r Hochenergiephysik, Wien, Austria}\\*[0pt]
W.~Adam, F.~Ambrogi, E.~Asilar, T.~Bergauer, J.~Brandstetter, M.~Dragicevic, J.~Er\"{o}, A.~Escalante~Del~Valle, M.~Flechl, R.~Fr\"{u}hwirth\cmsAuthorMark{1}, V.M.~Ghete, J.~Hrubec, M.~Jeitler\cmsAuthorMark{1}, N.~Krammer, I.~Kr\"{a}tschmer, D.~Liko, T.~Madlener, I.~Mikulec, N.~Rad, H.~Rohringer, J.~Schieck\cmsAuthorMark{1}, R.~Sch\"{o}fbeck, M.~Spanring, D.~Spitzbart, W.~Waltenberger, J.~Wittmann, C.-E.~Wulz\cmsAuthorMark{1}, M.~Zarucki
% ...
\vskip\cmsinstskip
% ...
1:  Also at Vienna University of Technology, Vienna, Austria\\
% ...
\end{sloppypar}
\end{document}